\documentclass[11pt]{article}
\usepackage{graphicx}      
\graphicspath{{figures/}{./}}  
\usepackage{booktabs}      
\usepackage{array}
\usepackage{amsmath,amssymb}
\usepackage[table]{xcolor}
\usepackage{multirow}
\usepackage{longtable}
\usepackage{siunitx}
\DeclareSIUnit\angstrom{\text{\AA}}
\usepackage{geometry}
\usepackage[hidelinks]{hyperref}
\usepackage{authblk}

\newcommand{\best}[1]{\textbf{#1}}     
\title{\textbf{ALIGNN 2.0: A Unified Line-Graph Neural Network Framework for
Materials Screening, Force Fields, Inverse Design, Spectroscopy, and Microscopy}}
\author[1]{Jaehyung Lee}
\author[1]{Charles Rhys Campbell}
\author[1]{Akshaya Ajith}
\author[3]{Sergei V. Kalinin}
\author[4]{Christopher Wolverton}
\author[1,2,*]{Kamal Choudhary}
\affil[1]{Department of Materials Science and Engineering, Johns Hopkins
University, Baltimore, Maryland 21218, USA}
\affil[2]{Department of Electrical and Computer Engineering, Johns Hopkins
University, Baltimore, Maryland 21218, USA}
\affil[3]{Department of Materials Science and Engineering, University of
Tennessee, Knoxville, Tennessee 37996, USA}
\affil[4]{Department of Materials Science and Engineering, Northwestern
University, Evanston, Illinois 60208, USA}
\affil[*]{Corresponding author: kchoudh2@jhu.edu}
\date{\today}

\begin{document}
\maketitle

\begin{abstract}
\noindent
Graph neural networks are central to materials property prediction and
machine-learning interatomic potentials, yet their reliance on specialized
graph libraries hampers portability and reproducibility, and property and
force-field models have historically required separate graph pipelines. We
present \textbf{ALIGNN~2.0}, a dependency-free, pure-PyTorch reimplementation of
the Atomistic Line Graph Neural Network, with the line graph and its batching
built from scratch, running on current-generation accelerators and unifying
scalar, spectral, tensorial, per-atom, and force-field prediction behind a single
graph, a combination that to our knowledge no existing framework provides. Comparing radius and $k$-nearest-neighbor (kNN) graphs, the wider kNN graph is more accurate for properties while the smoothly varying radius graph is required for energy-conserving molecular dynamics. On the JARVIS-Leaderboard, ALIGNN~2.0 leads on 26 of 30 single-property benchmarks against the original
ALIGNN, with large gains for piezoelectric and dielectric maxima,
exfoliation energy, moduli, and superconducting $T_c$, and extends to densities
of states, charges, moments, and elastic, dielectric, and piezoelectric tensors.
The LAMMPS- and OpenMM-compatible ALIGNN-FF matches leading universal potentials
on the Matbench-Discovery and CHIPS-FF benchmarks at a small fraction of their
parameters while scaling to hundred-thousand-atom cells. We further use ALIGNN
2.0 as the denoiser in a conditional crystal-diffusion model, where explicit
line-graph message passing consistently lowers structural denoising loss. We also show, as work in progress, that an independently diffused, redundant bond-angle state is learnable but does not
uniformly improve reconstruction or combine additively with the line graph.
Finally, from a single relaxed structure the same framework reconstructs infrared, Raman,
optical-dielectric, and neutron spectra in agreement with experiment and DFT, and
drives frozen-phonon electron-microscopy image simulation. \
Web app: \url{https://atomgpt.org/alignn}. \
Code: \url{https://github.com/atomgptlab/alignn}.
\end{abstract}

\section{Introduction}
Machine learning has become a routine component of computational materials
discovery, providing rapid surrogates for density functional theory (DFT) and,
increasingly, interatomic potentials capable of driving molecular dynamics~\cite{choudhary2022dlreview}.
Graph neural networks (GNNs) have been particularly influential because they
operate directly on the atomistic structure, respect its permutation symmetry,
and learn representations that transfer across chemistries; early crystal and
molecular GNNs such as CGCNN~\cite{xie2018cgcnn}, SchNet~\cite{schutt2018schnet},
and MEGNet~\cite{chen2019megnet} established the paradigm, and directional and
higher-body message passing subsequently improved it~\cite{gasteiger2020dimenet}.
Among these models, the Atomistic Line Graph Neural Network
(ALIGNN)~\cite{choudhary2021alignn} occupies a distinctive place: by augmenting
the conventional atom graph with a line graph that encodes bond angles, it
captures three-body correlations that are essential for many structural,
electronic, elastic, and vibrational properties. ALIGNN has been applied
successfully across an unusually broad task spectrum, including formation and
total energies, electronic band gaps, elastic and dielectric response,
thermoelectric descriptors~\cite{choudhary2021alignn}, superconducting critical
temperatures~\cite{choudhary2022supercon}, and, through
ALIGNN-FF~\cite{choudhary2023alignnff}, energy-force-stress interatomic
potentials for molecular dynamics. A number of variants have since built on the
line-graph idea, for example incorporating attention over the line graph to
reweight angular messages~\cite{shao2026attention}, reflecting sustained
interest in improving the ALIGNN family.

Despite this success, two practical obstacles limit how far such models can be
deployed and trusted. The first is a software-engineering problem with
scientific consequences. Most high-performing atomistic GNNs, ALIGNN included,
have historically been built on specialized graph libraries~\cite{wang2019dgl}
whose release cadence lags that of the underlying deep-learning
frameworks~\cite{paszke2019pytorch} and accelerator toolkits. As a result,
published models become progressively harder to install, to run on newer
hardware, and ultimately to reproduce; a model that was state of the art at
publication time can fail to build only a year or two later when the graph library
has not kept pace with the current compiler or driver stack. This fragility sits
uncomfortably with the community's emphasis on reproducibility, and it also
complicates the coupling of trained potentials to molecular-dynamics engines
where a heavyweight graph runtime becomes an awkward dependency. The second obstacle
is conceptual rather than technical. Property prediction and force-field
construction have traditionally been treated as separate modeling problems, each
with its own graph construction, neighbor list, and preprocessing pipeline. This
separation duplicates engineering effort, introduces subtle inconsistencies
between the graphs used for different tasks, and stands in the way of an
interactive workflow in which a single user-supplied structure yields all
quantities of interest at once.

In this work we address both obstacles with ALIGNN 2.0, a dependency-free,
pure-PyTorch reimplementation of ALIGNN. Rather than delegate graph operations
to an external library, we implement the atom graph, the angular line graph, the
neighbor search, and, critically, the batching of both graphs directly in native
PyTorch tensors. The line graph and its batching are the technically demanding
part of this exercise as batching a line graph requires consistent offsetting of
node and edge indices across structures of different sizes so that angular
messages never cross molecular boundaries. Doing this efficiently in
vectorized tensor operations, without a bespoke graph runtime, is what makes the
dependency-free design practical at training scale. The resulting model runs
unmodified on current-generation accelerators, including the newest Blackwell-class
GPUs, on which the legacy stack does not build at all, and ALIGNN 2.0 is light
enough to embed directly alongside a molecular-dynamics driver.

\begin{figure*}[t]
\centering
\includegraphics[width=\textwidth]{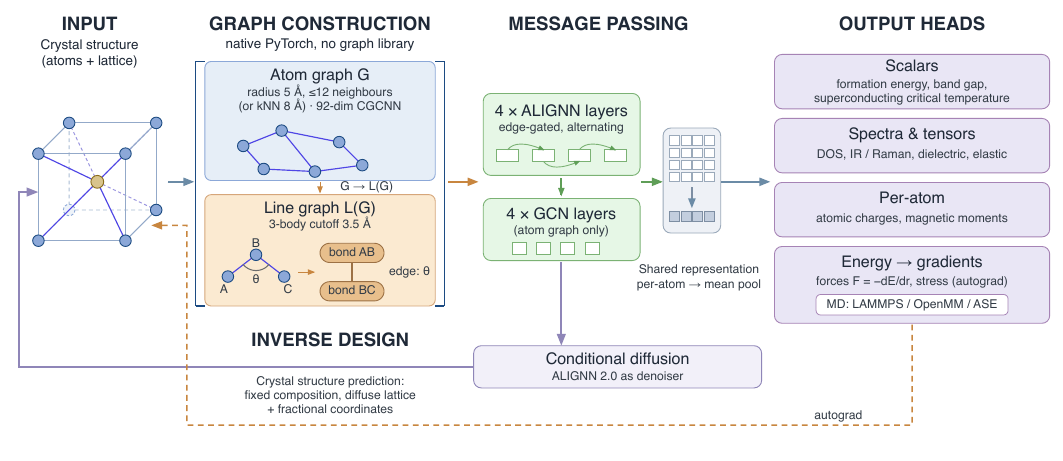}
\caption{\textbf{Overview of ALIGNN 2.0.} A crystal structure is converted in native PyTorch into an atom graph $G$ and a line graph $L(G)$, whose nodes are the bonds of $G$ and whose edges carry the angle between them. Four edge-gated ALIGNN layers alternate between the two graphs and are followed by four graph convolutions on $G$ alone, producing one shared per-atom representation. The scalar, spectral and energy heads read it after mean pooling while the per-atom heads read it unpooled. Forces and stress follow from a single backward pass of the energy through the interatomic displacement vectors, so the forces are conservative. The same representation conditions a diffusion denoiser that generates structures at fixed composition, closing the inverse-design loop.}
\label{fig:arch}
\end{figure*}

Having removed the library dependency, we exploit the freedom it affords to unify
property and force-field modeling behind a single graph construction. From one
cached graph we obtain graph-level scalar properties, full spectral outputs such
as electronic and phonon densities of states, tensorial responses, per-atom
quantities such as atomic charges and magnetic moments, and the total energy
together with its analytic forces and stress. This ``one graph, many outputs''
design (Fig.~\ref{fig:arch}) is the organizing principle of the paper. To our knowledge, no existing framework spans this range within a single model. Universal interatomic potentials~\cite{wood2025uma,batatia2022mace,deng2023chgnet,chen2022m3gnet} provide energies, forces, and stresses but not spectral, tensorial, or per-atom targets. Property-prediction models are trained one task at a time and provide no conservative force field, and generative models for structure are separate again. The design also forces the question of ``what graph should the single construction use?'' We therefore compare head-to-head and under identical training conditions a force-field-compatible radius graph and a wider k-nearest-neighbor (kNN) graph.
The comparison is consequential because the two constructions have complementary
strengths. First, we find that the kNN graph, with its fixed coordination and wider angular reach, is the more accurate choice for property prediction, but its neighbor list changes
discontinuously as atoms move across the k-th-neighbor boundary, producing energy
jumps that are incompatible with conservative dynamics. The radius graph, whose
neighbor list varies smoothly, is the construction that a force field requires,
and we show that it still remains competitive for property prediction while serving
both roles.

The remainder of the paper reports what this unified, dependency-free model
achieves and what it reveals. Using the JARVIS-Leaderboard evaluation
protocol~\cite{choudhary2024leaderboard} and its fixed train, validation, and
test splits, we benchmark ALIGNN 2.0 against the original ALIGNN and CGCNN across
thirty single-property tasks and a naive statistical baseline, and extend the
same architecture to spectral, tensorial, and per-atom targets to demonstrate the
generality of the shared graph. We evaluate the ALIGNN-FF force field against
pretrained universal potentials (UMA, M3GNet, MACE-MP-0, CHGNet) on the
Matbench-Discovery discovery set~\cite{riebesell2025framework}, the CHIPS-FF material-property suite~\cite{wines2025chipsff}, and
large-cell inference scaling. Finally, we show that the single framework reconstructs a material's
infrared, Raman, optical-dielectric, and inelastic-neutron spectra directly from
one relaxed structure, and we find that these results align closely with experiment and DFT with a speedup in computational cost. We also show that the framework couples to
frozen-phonon multislice simulation to reach scanning transmission electron
microscopy. Because the associated ALIGNN-FF potential is
LAMMPS- and OpenMM-compatible~\cite{thompson2022lammps,openmm8}, the model is immediately usable for
large-scale simulation, and its shared graph makes database-scale property
screening inexpensive.

\section{Results and Discussion}
We first benchmark accuracy on single scalar-property prediction, using the JARVIS DFT-3D
database (Table~\ref{tab:single}) and a range of additional datasets
(Table~\ref{tab:other}). We then show that the same architecture predicts many
scalar properties at once, spanning spectra, per-atom quantities, and tensors
(Table~\ref{tab:multi}). We turn next to interatomic force fields
(Table~\ref{tab:ff}), which we evaluate on the WBM (Wang-Botti-Marques) discovery set, on the CHIPS-FF
suite of defect, surface, and interface properties (Table~\ref{tab:defects}), and
on inference timing and scaling to large cells (Table~\ref{tab:scaling}). Then, ALIGNN 2.0 is integrated as a denoiser into a generative crystal diffusion model, and we benchmark the degree to which angular inductive bias improves model performance on crystal structure prediction. We close the paper with two additional applications: reconstructing vibrational spectra and
simulating electron microscopy each from a single structure. 

\subsection{Single-property prediction}
We begin with single-property prediction, the numerical results of which can be found in Table~\ref{tab:single}. Each row reports the test mean absolute error (MAE) on the JARVIS-Leaderboard fixed splits for two ALIGNN 2.0 graphs, the first on the wider \SI{8}{\angstrom} kNN graph and the second on the force-field-compatible radius graph (both trained with the improved recipe that adds exponential-moving-average weight averaging), alongside the original pre-2.0 ALIGNN~\cite{choudhary2021alignn} and coNGN~\cite{ruff2024congn}, a connectivity-optimized line-graph network reported on the JARVIS-Leaderboard and included here for comparison. The last two columns give a predict-the-mean Baseline (its MAE equals the mean absolute deviation, MAD) and the Skill $=100(1-\mathrm{MAE}/\mathrm{MAD})$. Measured against the original ALIGNN on the same splits, an ALIGNN 2.0 variant attains the lower error on 26 of the 30 JARVIS DFT-3D and superconducting tasks~\cite{choudhary2020jarvis,choudhary2022supercon}. coNGN is a strong baseline: it reaches the lowest MAE on a number of the property tasks (total energy, energy above hull, bulk and shear moduli, magnetic moment, several dielectric components, the thermoelectric Seebeck coefficient, and phonon heat capacity), while ALIGNN 2.0 leads on formation energy, both band gaps, the piezoelectric and maximum-dielectric response, exfoliation energy, carrier masses, and the electric-field gradient. The largest improvements seen by ALIGNN 2.0 against ALIGNN 1.0 appear for the tensorial
and response properties that most depend on faithful three-body information. Specifically, the
maximum piezoelectric modulus improves from around 20 in the ALIGNN 1.0
to about 12.5 in ALIGNN 2.0, the maximum dielectric response and the exfoliation
energy improve substantially, and the elastic and shear moduli, the band gaps
computed with both the OptB88vdW~\cite{klimes2011optb88} and modified
Becke-Johnson treatments, and the formation and total energies all improve by
several to tens of percent. The superconducting critical temperature is predicted
more accurately than by the original model, and the improvement persists on the
separately distributed high-pressure hydride sets~\cite{choudhary2022supercon}.
Expressed against the statistical baseline, skill scores exceed ninety percent for
the energy-like targets and remain substantial for the harder tensorial and
transport properties, quantifying how much structure the model captures beyond the
mean. Importantly, because ALIGNN 2.0 unifies the graph construction step between the force-field and property prediction functions, the property prediction algorithm is modified when compared to the original ALIGNN 1.0. We posit that this algorithmic modification is what is generating the improved performance of ALIGNN 2.0 against ALIGNN 1.0, and we discuss the specific algorithmic differences in the Methods section. 
Additionally, we benchmark property prediction using ALIGNN 2.0 on tasks derived from the QM9, QMOF, C2DB, and hMOF materials datasets to increase the statistical diversity of the benchmarks performed. The results are found in Table~\ref{tab:other}.

It is worth being explicit about where the ALIGNN 1.0 retains a narrow
advantage. On a
small number of dielectric-tensor components, maximum electric-field
gradient, average hole mass, and thermoelectric power factor, the
original model edges out ALIGNN 2.0 by margins that are, in the worst case, of order
one percent, and in several instances smaller than a thousandth of the property
value. These are tasks on which the wider property-only graph of the original
model has a slight edge that the force-field-compatible construction does not
fully recover. The fact that the gap is so small, and that the unified model
nonetheless wins the large majority of tasks, is the substantive point. CGCNN,
which lacks the angular line graph entirely, wins none of the tasks. This underlines
the value of three-body information for the property spectrum.

The comparison between the radius and kNN graphs deserves emphasis because it is
both a practical recommendation and a mechanistic observation. Holding the model,
optimizer, and training budget fixed and varying only the graph construction, we
find that the wider kNN graph, with its larger angular receptive field, is
consistently more accurate for property prediction, whereas the
force-field-compatible radius graph is slightly less accurate but shares its
single cached graph with the force field. The choice between them is dictated by
the intended use. The kNN graph is the stronger option when only property
accuracy matters, but its neighbor list changes discontinuously as atoms cross
the k-th-neighbor boundary, generating energy discontinuities that are fatal to
energy conservation in molecular dynamics. The radius graph, whose neighbor set
enters and leaves smoothly with distance, is therefore the correct default for
any workflow that must serve both property inference and dynamics, and the modest
accuracy cost relative to the kNN graph is the price of that unification.

\subsubsection{Data scale: one million structures from OQMD v1.8}
The JARVIS benchmarks above train on tens of thousands of structures. To test whether the same architecture continues to profit from data at a scale two orders of magnitude larger, we trained ALIGNN 2.0 on the OQMD v1.8 release~\cite{saal2013oqmd}, extracted directly from the February 2026 SQL distribution. Canonical entries were selected using OQMD's own duplicate marking, giving 1,259,145 structures with uncorrected PBE formation energies, split 1,133,231/62,957/62,957. Trained with the published default recipe at its stock width of 256 hidden features, the model reaches a test MAE of \SI{0.0120}{\electronvolt\per atom}, a skill of 98.6 percent against the predict-the-mean baseline and the highest skill of any model in this work, achieved with its smallest architecture.

The two databases also permit a direct test of transferability. Evaluated on a common 5,572-structure sample, the JARVIS-trained model gives a MAE of 0.0284 on its own test set and 0.1064 on OQMD, while the OQMD-trained model gives 0.0120 on its own test set and 0.0878 on JARVIS. The smaller OQMD model therefore travels better, so breadth of chemistry outweighs width of network once the domain changes. A global bias correction barely improves either figure, indicating that the PBE-to-OptB88vdW discrepancy is composition dependent rather than a constant offset, and both off-diagonal errors are comparable to a zero-shot universal potential on the same benchmark.

\subsubsection{Application: superconductor screening}
\begin{figure}[t]
\centering
\includegraphics[width=\textwidth]{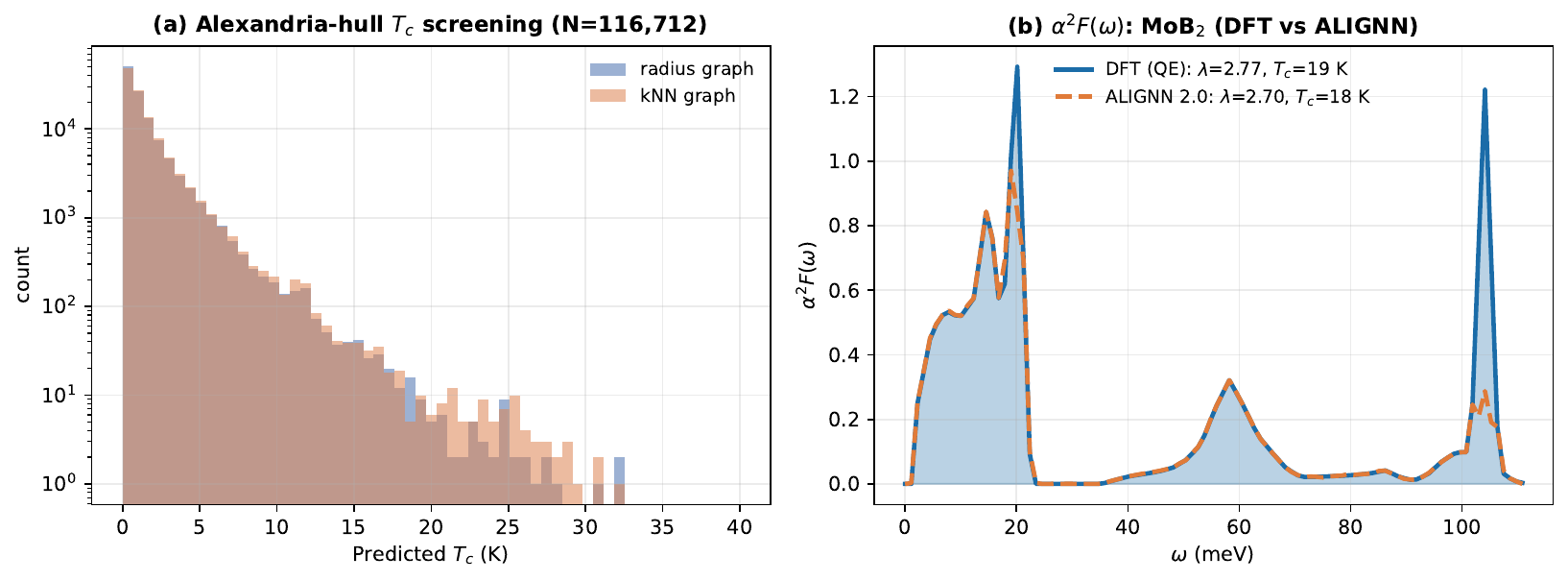}
\caption{\textbf{Superconductor screening and $\alpha^2F(\omega)$ prediction with
ALIGNN 2.0.}
(a) Predicted $T_c$ distribution over the $116{,}712$-material Alexandria-PBE hull
for the radius- and kNN-graph superconductor models. The kNN graph enhances the
high-$T_c$ tail while agreeing elsewhere.
(b) Eliashberg spectral function $\alpha^2F(\omega)$ for strong-coupling MoB$_2$: DFT
(Quantum ESPRESSO DFPT; filled blue) versus the ALIGNN 2.0 prediction (dashed
orange). The prediction tracks the DFT spectrum across the full frequency range,
recovering $\lambda=2.70$ (vs $2.77$) and an Allen--Dynes $T_c$ ($\mu^*=0.1$) of
$18$~K (vs $19$~K), so $\lambda$ and $T_c$ follow directly without electron--phonon
DFPT. Superconductor training data from~\cite{choudhary2022supercon}.}
\label{fig:supercon}
\end{figure}

Because a single ALIGNN 2.0 architecture supplies scalar $T_c$ predictors and a
force field, it can drive an end-to-end superconductor-screening workflow
(Fig.~\ref{fig:supercon}). We screened all $116{,}712$ Alexandria-PBE convex-hull
structures with two independently trained superconducting-$T_c$ models, one on a
radius graph ($5.0$\,\AA) and one on a $k$-nearest-neighbor graph ($8.0$\,\AA).
The two constructions agree across the bulk of the distribution, but the kNN graph
populates the high-$T_c$ tail more densely ($\sim$16\% more materials above
$10$\,K and roughly twice as many above $20$\,K; Fig.~\ref{fig:supercon}a). This is a
systematic, second-order effect of the graph representation on a genuine screening
task. Ranking by both models recovers known high-$T_c$ entries already present in
the Alexandria superconductor database (e.g.\ TcN, predicted $T_c=17.7$\,K versus a
DFT reference of $17.7$\,K), confirming the screen is well calibrated. Beyond a
scalar $T_c$, the same shared graph predicts the full electron--phonon Eliashberg
spectral function $\alpha^2F(\omega)$ directly (Fig.~\ref{fig:supercon}b): for
strong-coupling MoB$_2$ the ALIGNN 2.0 prediction reproduces the
density-functional perturbation theory (DFPT) $\alpha^2F(\omega)$ across the entire frequency range, recovering $\lambda=2.70$
(versus $2.77$) and an Allen--Dynes $T_c$ of $18$\,K (versus $19$\,K). The coupling
strength and $T_c$ therefore follow from a single architecture ahead of any
electron--phonon DFPT, tempering scalar-only screening with a physically grounded
spectral prediction.

\subsection{Multi-property prediction}
Table~\ref{tab:multi} demonstrates that the same
architecture, with only a change of output head, addresses targets well beyond
scalar regression. Two spectral models predict the electronic density of states
on a three-hundred-bin energy grid and the phonon density of states on a
two-hundred-bin frequency grid, in each case mapping a structure to a
fixed-length curve through a graph-level vector head trained with an averaged
absolute-error loss over bins. The electronic density of states is predicted an
order of magnitude more tightly than the phonon density of states, a difference
consistent with the smoother, broader electronic curves relative to the sharp van
Hove structure of phonon spectra. A third spectral head predicts the two-hundred-bin
Raman spectrum, where the k-nearest-neighbor graph again outperforms the radius
graph (held-out mean absolute error $0.0326$ versus $0.0378$), the wider graph
recovering roughly a third of the mean-absolute-deviation baseline (radius,
$23.9\%$; Table~\ref{tab:multi}). Two further families of targets exercise the
model's flexibility in complementary directions. Per-atom quantities, namely
atomic charges from a Bader decomposition~\cite{henkelman2006bader} and
site-projected magnetic moments, are fit with a one-dimensional per-atom head and
the graph-level loss disabled, reusing the same atomwise machinery that produces
forces. Full response tensors, including the nine-component dielectric tensor,
Born effective charges, the eighteen-component piezoelectric tensor from
density-functional perturbation theory~\cite{baroni2001dfpt}, and the
thirty-six-component elastic stiffness tensor, are fit with a vector head of the
appropriate dimension. That a single architecture spans scalars, curves, per-atom
fields, and tensors without structural change is the clearest expression of the
unified design, and it means that adding a new property to the suite is a matter
of choosing an output head rather than engineering a new pipeline.

These heads regress a target directly from geometry and do not
expose the underlying electronic structure by themselves. Questions of band character, orbital
contributions, or magnetic ordering (for instance, telling a ferromagnetic from an
antiferromagnetic state, information not contained in a single crystal structure)
lie outside this framing. We address them with a complementary tight-binding model,
SlakoNet~\cite{slakonet}, which predicts Slater-Koster electronic-structure
parameters (band structure and density of states) directly and shares the same
AtomGPT interface, so structure-property regression and electronic-structure
prediction are available side by side.

The shared architecture also lets us ask \emph{where} the network specializes
across tasks. Comparing the learned weights of all twenty-six radius property
models by the permutation-invariant Frobenius norm $\|W\|_F$ of each architecture
block (Fig.~\ref{fig:interp}), the final read-out ($\mathtt{fc}$) varies roughly
three times more across properties ($\approx$138\%) than the message-passing
backbone ($\approx$45\%). The radial and angular geometry encoders are the
most universal ($\approx$34\%). Clustering each property by its per-block weight
fingerprint recovers physically coherent families, dielectric-tensor
components, elastic moduli, and carrier effective masses by each group. Because specialization concentrates in the read-out, a single shared
backbone with lightweight per-property heads is a natural route to predicting many
properties, and, as we show next, energies and forces, from one model at a
fraction of the parameter and inference cost of independent networks.

\subsection{Unified force field}
\subsubsection{Training accuracy and generalization to discovery}
Table~\ref{tab:ff} turns to interatomic force fields, which we train
on datasets spanning several orders of magnitude in size. The ALIGNN-FF
database~\cite{choudhary2023alignnff} and the MATPES set~\cite{kaplan2025matpes}
contain of order hundreds of thousands of configurations with energies, forces,
and stresses, and two further potentials, a finite-displacement force field
trained on more than a million configurations and a Materials-Project-trajectory
model of comparable scale, probe the large-data regime. All share the unified
graph, the atom-embedding features, and the one-cycle training schedule.
The dominant design choice for these potentials is the atom representation. On the
MATPES set, replacing a scalar atomic-number embedding with the
ninety-two-dimensional atom-embedding feature vector reduces the validation energy error roughly fourfold and the force error by tens of percent at matched model width and training budget. Increasing model width alone yields only modest gains, and,
contrary to a common assumption, a smooth cutoff envelope does not help and often
hurts early training. The MATPES model reaches a force accuracy competitive with
strong contemporary universal potentials~\cite{chen2022m3gnet,deng2023chgnet,
batatia2022mace} on the same test partition. On the WBM discovery set
(Table~\ref{tab:ff}), the pretrained UMA~\cite{wood2025uma} and M3GNet~\cite{chen2022m3gnet} universal potentials are
included as external baselines: UMA attains the lowest geometry-optimization
RMSD ($0.067$\,\AA), while the ALIGNN-FF potentials remain competitive
(MATPES-PBE at $0.101$\,\AA) despite being far smaller ($0.55$\,M parameters
versus UMA's $147$\,M, a $\sim$$250\times$ reduction). The
diatomic-curve tortuosity shows that the smooth ALIGNN-FF and M3GNet potentials
($\tau\!\approx\!1.0$) yield more physical dissociation curves than UMA
($\tau=2.5$). The external baselines here are the
current MatPES/2025 checkpoints scored with our identical pipeline, not the
MP/MPtrj-trained versions on the public Matbench-Discovery leaderboard, so their
absolute F1 and DAF run lower than the public-leaderboard numbers.

\subsubsection{Molecular-dynamics stability and energy conservation}
The molecular-dynamics behavior of these potentials leads to what we regard as the
most transferable mechanistic message of the study. Energy conservation in
micro-canonical dynamics is governed primarily by training convergence, not by the
form of the cutoff. A compact model, once its one-cycle learning-rate
schedule~\cite{smith2019superconvergence} has fully annealed, drifts by only
thousandths of a milli-electronvolt per atom per picosecond for silicon,
comparable to established reference potentials. On the contrary, an under-converged
checkpoint of the identical architecture drifts by more than two orders of
magnitude more. Among the per-element potentials, close-packed metals are
essentially energy-conserving, covalent solids are stable, and the
body-centered-cubic metals exhibit a residual drift that tracks their higher force
error, so that the difficulty ordering across chemistries is preserved as training
proceeds and the drift decreases monotonically with convergence. Taken with the
atom-embedding result, this identifies two levers, learned atom features and
schedule convergence, that between them account for the difference between an
unusable and a production-quality potential. It cautions against attributing
molecular-dynamics stability to cutoff smoothing that our controlled comparisons
do not support. A natural way to reduce the residual drift of a potential trained
only on near-equilibrium finite-displacement data is to broaden the sampled
configuration space, and we are pursuing exactly this by augmenting the
finite-displacement training set with energy-volume curves and with vacancy and
surface relaxation trajectories, all computed with the same functional. That
data-scaling experiment is underway and its molecular-dynamics evaluation will be
reported when the augmented model has converged. Beyond dynamics, we probe the same
potential on point and extended to defects and heterogeneous interfaces.

\subsubsection{Material properties: defects, surfaces, and interfaces (CHIPS-FF)}
Table~\ref{tab:defects} reports monovacancy formation and surface energies (MAE vs
DFT, with the predict-the-mean MAD as baseline). Because these are far-from-equilibrium
\emph{derived} quantities, the equilibrium-trained ALIGNN force fields do not beat the
baseline on vacancies (MAE $1.90$ vs MAD $1.42$~eV), although the MATPES model is
competitive on surfaces ($0.68$ vs MAD $0.87$~J/m$^2$). The larger CHGNet~\cite{deng2023chgnet} and
MACE-MP-0~\cite{batatia2022mace} potentials lead on both. This vacancy gap is consistent with the
near-equilibrium sampling noted above and motivates the defect-augmented
finite-displacement set (Sec.~\ref{sec:methods}).
The $W_{\mathrm{ad}}$ column of Table~\ref{tab:defects} shows that with a self-consistent
split-slab evaluation in which the relaxed interface is separated into two slabs sharing its
lateral cell, the same potential predicts physically reasonable works of adhesion (of order
$-1$ to $+7$~J/m$^2$) across metal, ceramic, semiconductor, and two-dimensional
interfaces (per-interface values in Table~\ref{tab:interface}), including the
technologically important Si/SiO$_2$ contact. This replaces a
reference-cell mismatch in the default protocol that had inflated metallic adhesion
energies by two orders of magnitude.

Polymorph energetics make the provenance of a potential's physics explicit
(Fig.~\ref{fig:ev}). Energy-volume curves for competing polymorphs of Si,
SiO$_2$, Ni$_3$Al, and MoS$_2$ show that the MATPES-PBE-trained potential
recovers the correct ground state for Si and Ni$_3$Al but reproduces its
training functional's known quartz-to-cristobalite inversion (by 35~meV/atom),
and because it lacks interlayer dispersion, lets the metastable 1T stacking of
MoS$_2$ delaminate to nearly twice the physical volume, where it lands
degenerate with 2H to within 2~meV/atom. The same architecture trained on
OptB88vdW data recovers the expected ground state in all four families, as
the original ALIGNN-FF did. A potential inherits its training functional's
physics, including its flaws, and the choice of training data, not the
architecture, decides them.

\subsubsection{Inference timing and scaling}
Two further points bear on the practical value of the framework. First, because
the associated ALIGNN-FF potential is compatible with the LAMMPS and OpenMM
molecular-dynamics engines~\cite{thompson2022lammps,openmm8}, a trained model can be used directly for
large-scale simulation without a bespoke driver, closing the loop from a property
model to dynamics within a single dependency-free codebase. Second, because every
trained property model consumes the same graph, applying the full model suite to a
new structure costs one graph construction plus a forward pass per property. We
exploit this to screen the on-hull entries of the Alexandria
database~\cite{schmidt2024alexandria} in a single batched, no-gradient pass,
producing a database of GNN-predicted properties at a throughput that a per-model
pipeline could not match. Together these observations argue that the benefits of a
dependency-free, unified design are not confined to convenience: they change what
is computationally feasible, both at the scale of an interactive single-structure
query and at the scale of a database sweep, while keeping the entire workflow
reproducible on current hardware. This efficiency is intrinsic to the compact
force-field-compatible model: evaluated on a single CPU, the ALIGNN-FF potential
used here (two ALIGNN and two GCN layers, $128$ hidden features, smooth cutoff)
is roughly $4.5\times$ faster than CHGNet and MACE-MP-0 at inference and scales more
gently with system size (Fig.~\ref{fig:scaling}), so the same network that supplies
conservative dynamics also supplies the throughput a database sweep requires.
The same economy holds at the memory ceiling of a single GB10
(Table~\ref{tab:scaling}): the radius-graph property model completes
single-point evaluations of silicon supercells up to 453{,}962 atoms within 96.4 gigabytes before the unified pool is exhausted, the kNN construction reaches roughly one hundred thousand atoms, and the UMA-S universal potential~\cite{wood2025uma}, evaluated as a full energy-force-stress single point, scales almost perfectly linearly but carries the largest per-call cost over most of the size range. The corresponding force-field comparison against MACE and CHGNet is given in Table~\ref{tab:scaling}.

\begin{scriptsize}
\setlength{\tabcolsep}{2.5pt}
\begin{longtable}{>{\raggedright\arraybackslash}p{3.0cm}ccccccc}
\caption{Single-property prediction on JARVIS DFT-3D (JARVIS-Leaderboard protocol, test MAE); columns, metrics, and per-row winners are defined and discussed in the text.
Data sources: JARVIS-DFT OptB88vdW energies, band gaps, magnetic moments, transport, and $k$-point/cutoff descriptors~\cite{choudhary2020jarvis,choudhary2025jarviscms}; elastic moduli $K_V,G_V$~\cite{choudhary2018elastic}; SLME~\cite{choudhary2019solar}; spin-orbit spillage~\cite{choudhary2019spillage}; TBmBJ band gap and dielectric constants (\texttt{mbj}, \texttt{eps}, \texttt{meps})~\cite{choudhary2018diel}; DFPT piezoelectric and dielectric maxima~\cite{choudhary2020dfpt}; electric-field-gradient \texttt{max\_efg}~\cite{choudhary2020efg}.}
\label{tab:single}\\
\toprule
Task (unit) & $N_{\mathrm{tr}}/N_{\mathrm{val}}/N_{\mathrm{te}}$ & ALIGNN 2.0 & ALIGNN 2.0 & original & coNGN & Baseline & Skill \\
  &  & (kNN) & (radius) & ALIGNN & & (MAD) & (\%) \\
\midrule
\endfirsthead
\toprule
Task (unit) & $N_{\mathrm{tr}}/N_{\mathrm{val}}/N_{\mathrm{te}}$ & ALIGNN 2.0 & ALIGNN 2.0 & original & coNGN & Baseline & Skill \\
  &  & (kNN) & (radius) & ALIGNN & & (MAD) & (\%) \\
\midrule
\endhead
1) formation\_energy (eV/atom) & 44569/5572/5572 & 0.0284 & 0.0316 & 0.0331 & 0.0291 & 0.876 & 96.8 \\
2) optb88vdw\_total\_energy (eV/atom) & 44569/5572/5572 & 0.0297 & 0.0321 & 0.0367 & 0.0273 & 1.786 & 98.3 \\
3) optb88vdw\_bandgap (eV) & 44569/5572/5572 & 0.1245 & 0.1314 & 0.1423 & 0.1267 & 0.999 & 87.5 \\
4) mbj\_bandgap (eV) & 14535/1817/1815 & 0.2576 & 0.2721 & 0.3104 & 0.2719 & 1.765 & 85.4 \\
5) ehull (eV/atom) & 44290/5537/5537 & 0.0508 & 0.0576 & 0.0763 & 0.0485 & 1.148 & 95.6 \\
6) bulk\_modulus\_kv (GPa) & 15744/1968/1968 & 9.497 & 9.885 & 10.399 & 8.702 & 53.76 & 82.3 \\
7) shear\_modulus\_gv (GPa) & 15744/1968/1968 & 9.208 & 9.063 & 9.476 & 8.488 & 27.06 & 66.0 \\
8) magmom\_oszicar ($\mu_B$) & 41766/5222/5222 & 0.2622 & 0.2608 & 0.2574 & 0.2437 & 1.254 & 79.1 \\
9) slme (\%) & 7250/906/906 & 4.504 & 4.493 & 4.521 & 4.443 & 11.21 & 59.8 \\
10) spillage & 9101/1137/1137 & 0.3499 & 0.3527 & 0.3510 & 0.3463 & 0.518 & 32.5 \\
11) kpoint\_length\_unit (\AA) & 44313/5540/5539 & 9.294 & 9.699 & 9.515 & 9.346 & 17.94 & 48.2 \\
12) encut (eV) & 44308/5539/5539 & 125.46 & 131.81 & 133.80 & 129.83 & 262.6 & 52.2 \\
13) epsx & 35592/4449/4449 & 19.853 & 20.705 & 20.394 & 18.574 & 57.45 & 65.4 \\
14) epsy & 35592/4449/4449 & 19.352 & 20.088 & 19.999 & 18.592 & 57.32 & 66.2 \\
15) epsz & 35592/4449/4449 & 19.503 & 19.633 & 19.568 & 17.810 & 55.79 & 65.0 \\
16) mepsx & 13447/1681/1681 & 24.148 & 24.646 & 24.046 & 18.574 & 63.39 & 61.9 \\
17) mepsy & 13447/1681/1681 & 23.840 & 23.823 & 23.648 & 18.592 & 63.68 & 62.6 \\
18) mepsz & 13447/1681/1681 & 23.572 & 23.247 & 23.731 & 17.810 & 60.71 & 61.2 \\
19) dfpt\_piezo\_max\_dij (pC/N) & 2677/334/334 & 13.426 & 12.603 & 20.570 & 13.887 & 22.69 & 40.8 \\
20) dfpt\_piezo\_max\_dielectric & 3764/470/470 & 25.175 & 26.823 & 28.151 & 25.555 & 43.91 & 42.7 \\
21) exfoliation\_energy (meV/atom) & 650/81/81 & 39.350 & 40.272 & 52.703 & 46.272 & 61.03 & 35.5 \\
22) max\_efg ($10^{21}$V/m$^2$) & 9493/1186/1186 & 18.834 & 19.802 & 19.121 & 19.549 & 44.46 & 57.6 \\
23) avg\_elec\_mass ($m_e$) & 14114/1764/1764 & 0.0797 & 0.0837 & 0.0853 & 0.0876 & 0.225 & 64.6 \\
24) avg\_hole\_mass ($m_e$) & 14114/1764/1764 & 0.1196 & 0.1299 & 0.1239 & 0.1285 & 0.399 & 70.0 \\
25) n\_Seebeck ($\mu$V/K) & 18568/2321/2321 & 41.454 & 41.524 & 40.921 & 40.098 & 111.5 & 62.8 \\
26) n\_powerfact ($\mu$W/mK$^2$) & 18568/2321/2321 & 482.84 & 469.07 & 442.30 & 456.61 & 709.2 & 31.9 \\
27) ph\_heat\_capacity (J/mol/K) & 9644/1205/1205 & 8.468 & 9.577 & 9.606 & 7.813 & 40.16 & 78.9 \\
\bottomrule
\end{longtable}
\end{scriptsize}

Beyond the JARVIS-DFT JARVIS DFT-3D database of Table~\ref{tab:single}, the same
architecture transfers, without modification, to a range of other materials
databases. Table~\ref{tab:other} collects single-property results on molecular
(QM9), metal--organic-framework (QMOF, hMOF), superconductor, two-dimensional
(C2DB, MXene, \texttt{twod\_matpd}), and polymer datasets: ALIGNN 2.0 again removes
most of the naive-baseline error relative to the predict-the-mean baseline.
The ALIGNN 2.0 column reports the best-performing graph on each dataset, which is
the base kNN model for most of these smaller transfer sets, the radius graph for
QM9, and the improved training recipe for QMOF and the OQMD lattice thermal
conductivity; the full radius/kNN/recipe ablation is in
Table~\ref{tab:other_ablation}. For comparison we also list the original pre-2.0
ALIGNN wherever a JARVIS-Leaderboard entry exists on the identical benchmark (a
dash marks the datasets, such as \texttt{alex\_supercon} and the OQMD thermal
conductivity, for which no such entry is available). ALIGNN 2.0 attains the lower
error on all of these benchmarks except the hydride-plus-bulk $T_c$ set and the
QM9 gap, where the original model is marginally ahead, and the QMOF gap, where the
two are tied to the reported precision. Baseline (MAD) and Skill are defined as in
Table~\ref{tab:single}.

\begin{scriptsize}
\setlength{\tabcolsep}{2.5pt}
\begin{longtable}{>{\raggedright\arraybackslash}p{3.5cm}ccccc}
\caption{Single-property prediction on additional datasets beyond JARVIS DFT-3D (JARVIS-Leaderboard protocol, test MAE); columns and per-row winners are discussed in the text.
Data sources: bulk superconductor $T_c$~\cite{choudhary2022supercon}; hydride $T_c$~\cite{choudhary2024hydride,choudhary2024hydrideafm}; the remaining datasets (QM9, QMOF, MXene, polymer, C2DB, 2D, OMDB, hMOF) via the JARVIS-Leaderboard~\cite{choudhary2024leaderboard,choudhary2025jarviscms}; lattice thermal conductivity on OQMD structures and formation energies from the OQMD v1.8 SQL release~\cite{saal2013oqmd}. The OQMD v1.8 row (row 18) uses the default recipe at 256 hidden features rather than the wider recipe applied elsewhere.}
\label{tab:other}\\
\toprule
Task (unit) & $N_{\mathrm{tr}}/N_{\mathrm{val}}/N_{\mathrm{te}}$ & ALIGNN 2.0 & original & Baseline & Skill \\
  &  & (kNN) & ALIGNN & (MAD) & (\%) \\
\midrule
\endfirsthead
\toprule
Task (unit) & $N_{\mathrm{tr}}/N_{\mathrm{val}}/N_{\mathrm{te}}$ & ALIGNN 2.0 & original & Baseline & Skill \\
  &  & (kNN) & ALIGNN & (MAD) & (\%) \\
\midrule
\endhead
1) Thermal conductivity (log$_{10}\kappa_L$, OQMD) & 3227/403/404 & 0.194 & -- & 0.597 & 67.5 \\
2) QMOF bandgap (eV) & 16{,}340/2042/2042 & 0.202 & 0.202 & 0.946 & 78.6 \\
3) Tc\_supercon (K) & 556/30/30 & 1.490 & 2.032 & 2.723 & 45.3 \\
4) Tc\_supercon\_hydride (K) & 763/95/95 & 9.425 & 12.310 & 33.56 & 71.9 \\
5) Tc\_supercon\_\allowbreak hydride\_plus\_bulk (K) & 1595/199/199 & 8.407 & 7.962 & 22.33 & 62.4 \\
6) alex\_supercon Tc (K) & 6592/824/825 & 0.864 & -- & 2.818 & 69.3 \\
7) alex\_supercon $N(E_F)$ (states/eV) & 6592/824/825 & 0.791 & -- & 1.559 & 49.3 \\
8) alex\_supercon $\theta_D$ (K) & 6592/824/825 & 10.68 & -- & 80.30 & 86.7 \\
9) alex\_supercon $\lambda$ & 6592/824/825 & 0.0679 & -- & 0.194 & 65.0 \\
10) alex\_supercon $\omega_{\log}$ (K) & 6592/824/825 & 20.08 & -- & 55.37 & 63.7 \\
11) mxene275, formation energy (eV/atom) & 220/27/27 & 0.0343 & -- & 0.732 & 95.3 \\
12) polymer\_genome, GGA gap (eV) & 858/107/108 & 0.2273 & -- & 1.101 & 79.4 \\
13) c2db, band gap (eV) & 2816/352/352 & 0.0802 & -- & 0.641 & 87.5 \\
14) twod\_matpd, band gap (eV) & 5080/635/636 & 0.3660 & -- & 1.234 & 70.3 \\
15) omdb, band gap (eV) & 10000/1250/1250 & 0.2411 & -- & 0.783 & 69.2 \\
16) hMOF, CO$_2$ uptake (mol/kg) & 110{,}121/13{,}765/13{,}766 & 0.4687 & -- & 2.232 & 79.0 \\
17) QM9 HOMO--LUMO gap (eV) & 110{,}000/10{,}000/10{,}829 & 0.031 & 0.030 & 0.834 & 96.3 \\
18) OQMD, formation energy (eV/atom) & 1{,}133{,}231/62{,}957/62{,}957 & 0.0120 & -- & 0.830 & 98.6 \\
\bottomrule
\end{longtable}
\end{scriptsize}

\begin{scriptsize}
\setlength{\tabcolsep}{4pt}
\begin{longtable}{>{\raggedright\arraybackslash}p{5.2cm}cccc}
\caption{Multi-property outputs on the shared ALIGNN~2.0 graph (radius graph, held-out MAE; $D$ is the output dimension): spectra, per-atom, and tensor targets. Baseline (MAD) and Skill $=100(1-\mathrm{MAE}/\mathrm{MAD})$ as in Table~\ref{tab:single}; ``--'' where the baseline is ill-defined (e.g.\ Born/net charge).
Data sources: electronic/phonon DOS, atomic charges, and magnetic moments from JARVIS-DFT~\cite{choudhary2020jarvis}; Raman spectra~\cite{choudhary2023ramandb}; IR, Born effective charges, dielectric and piezoelectric tensors from DFPT~\cite{choudhary2020dfpt}; elastic tensor~\cite{choudhary2018elastic}.}
\label{tab:multi}\\
\toprule
Task (unit) & $N_{\mathrm{tr}}/N_{\mathrm{val}}/N_{\mathrm{te}}$ & ALIGNN 2.0 & Baseline & Skill \\
 & & (MAE) & (MAD) & (\%) \\
\midrule
\endfirsthead
\toprule
Task (unit) & $N_{\mathrm{tr}}/N_{\mathrm{val}}/N_{\mathrm{te}}$ & ALIGNN 2.0 & Baseline & Skill \\
 & & (MAE) & (MAD) & (\%) \\
\midrule
\endhead
1) eDOS, electronic DOS ($D{=}300$) & 4103/227/229 & 0.0138 & 0.0213 & 35.2 \\
2) pDOS, phonon DOS ($D{=}200$) & 4103/227/229 & 0.0819 & 0.117 & 29.8 \\
3) Raman spectrum ($D{=}200$) & 4059/507/508 & 0.0378 & 0.0497 & 23.9 \\
4) IR spectrum ($D{=}200$) & 3844/480/481 & 0.0232 & 0.0400 & 42.4 \\
5) Bader charge, per atom ($e$) & 75{,}028/3000/3000 & 0.0192 & 2.124 & 99.1 \\
6) Net charge, per atom ($e$) & 75{,}033/3000/3000 & 0.0167 & 2.980 & 99.4 \\
7) Magnetic moment, per atom ($\mu_B$) & 89{,}231/3000/3000 & 0.0256 & 2.063 & 98.8 \\
8) Dielectric tensor ($D{=}9$) & 4103/227/229 & 1.690 & 3.401 & 50.3 \\
9) Born effective charge ($e$) & 4472/248/249 & 0.234 & 0.746 & 68.6 \\
10) Piezoelectric tensor, C/m$^2$ ($D{=}18$) & 4513/250/252 & 0.077 & 0.089 & 13.9 \\
11) Elastic $C_{ij}$ tensor, GPa ($D{=}36$) & 15{,}936/885/886 & 5.593 & 18.73 & 70.1 \\
\bottomrule
\end{longtable}
\end{scriptsize}

\begin{table}[htbp]\centering\scriptsize
\setlength{\tabcolsep}{2.5pt}
\caption{\textbf{Universal force-field report card.} Training energy/force MAE and
Matbench-Discovery WBM metrics: stability F1, discovery acceleration factor DAF
($>1$ beats random), geometry-optimization RMSD, diatomic tortuosity $\tau$
(ideal $1$), and trainable parameters (units in the header). The five ALIGNN-FF
(AFF) variants share one architecture at two widths; UMA, M3GNet, MACE-MP-0 and
CHGNet are external baselines (current MatPES/2025 checkpoints, our identical
pipeline; see text). F1/DAF use per-model self-consistent elemental references
over the full $256{,}963$-material WBM set; coverage caveats and formation-energy
MAE are in the SI (Table~\ref{tab:wbm_eform}). Training data:
MATPES~\cite{kaplan2025matpes}, JARVIS-FF, MPtrj; metrics follow~\cite{riebesell2025matbench};
baselines UMA~\cite{wood2025uma}, M3GNet~\cite{chen2022m3gnet}, MACE-MP-0~\cite{batatia2022mace}, CHGNet~\cite{deng2023chgnet}.
\best{Bold}: column best.}
\label{tab:ff}
\begin{tabular}{l c c c c c c c c}
\toprule
Model & Params (M) & $N_{\mathrm{tr}}/N_{\mathrm{val}}/N_{\mathrm{te}}$ & $E$ MAE & $F$ MAE & F1 & DAF & WBM RMSD & $\tau$ \\
\midrule
AFF MATPES-R2SCAN & 0.55 & 347{,}889/19{,}327/19{,}328 & 0.0487 & 0.163 & 0.380 & 1.51 & 0.102 & 1.02 \\
AFF MATPES-PBE & 0.55 & 391{,}241/21{,}735/21{,}736 & 0.0404 & 0.148 & 0.388 & 2.02 & 0.101 & \best{1.01} \\
AFF JV-DFT-DB1 & 0.55 & 276{,}401/15{,}355/15{,}355 & 0.0324 & 0.0564 & 0.389 & 1.66 & 0.143 & 1.04 \\
AFF JV-DFT-DB2 & 4.04 & 1{,}176{,}773/60{,}957/60{,}958 & \best{0.0289} & \best{0.0445} & 0.375 & 1.60 & 0.148 & 1.23 \\
AFF MPtrj & 0.55 & 1{,}376{,}739/76{,}485/76{,}485 & 0.0579 & 0.0721 & 0.453 & 1.86 & 0.118 & 1.03 \\
\midrule
UMA (uma-s-1p1, baseline) & 146.6 & -- & -- & -- & \best{0.575} & \best{3.58} & \best{0.067} & 2.53 \\
M3GNet (MatPES-PBE, baseline) & 0.29 & -- & -- & -- & 0.401 & 1.81 & 0.101 & 1.05 \\
MACE-MP-0 (baseline) & 4.69 & -- & -- & -- & 0.430 & 2.08 & 0.097 & 1.60 \\
CHGNet (baseline) & 0.41 & -- & -- & -- & 0.459 & 1.84 & 0.101 & 9.63 \\
\bottomrule
\end{tabular}
\end{table}

\begin{table*}[t]
\centering
\caption{\textbf{CHIPS-FF material-property benchmark} (MAE vs JARVIS-DFT; single
relax-and-evaluate protocol, \texttt{FrechetCellFilter}). Lattice $a,c$; formation
$E_f$; elastic $C_{11},C_{44},K_V$; monovacancy (Vac) and surface (Surf) energies;
interface work of adhesion $W_{\mathrm{ad}}$; amorphous-Si density
$\rho_{\mathrm{aSi}}$; phonon-band MAE $\omega_{\mathrm{ph}}$ (units in the header).
UMA-S~\cite{wood2025uma}, M3GNet~\cite{chen2022m3gnet}, MACE-MP-0~\cite{batatia2022mace},
and CHGNet~\cite{deng2023chgnet} are external universal-potential baselines (identical
pipeline; their divergent $c$/$K_V$ relaxations excluded). ``AFF''=ALIGNN-FF,
``MPS''=MATPES; protocol, system counts, and self-consistent chemical potentials in
Methods. References: JARVIS-DFT~\cite{choudhary2020jarvis,choudhary2018elastic,choudhary2023vacancy},
InterMat~\cite{choudhary2024intermat}, CHIPS-FF~\cite{wines2025chipsff}. \best{Bold}: column best.}
\label{tab:defects}
\setlength{\tabcolsep}{4.5pt}
\begin{scriptsize}
\resizebox{\textwidth}{!}{%
\begin{tabular}{l r ccccccccccc}
\toprule
Model & $N_{\mathrm{tr}}$ & $a$ & $c$ & $E_f$ & $C_{11}$ & $C_{44}$ & $K_V$ & Vac & Surf & $W_{\mathrm{ad}}$ & $\rho_{\mathrm{aSi}}$ & $\omega_{\mathrm{ph}}$ \\
 & struct. & \AA & \AA & eV/at & GPa & GPa & GPa & eV & J/m$^2$ & J/m$^2$ & g/cm$^3$ & cm$^{-1}$ \\
\midrule
AFF MPS-R2SCAN & 347{,}889 & 0.056 & 0.094 & 0.159 & 41.5 & 33.9 & 128.1 & 0.790 & 0.418 & 1.742 & 2.324 & 62.6 \\
AFF MPS-PBE & 391{,}241 & \best{0.025} & 0.063 & 0.081 & 51.5 & 33.9 & 95.7 & 1.036 & 0.618 & 1.657 & \best{2.281} & 45.2 \\
AFF JV-DFT-DB1 & 276{,}401 & 0.041 & 0.058 & 0.031 & 154.1 & 64.0 & 51.3 & 1.926 & 1.313 & 1.595 & 2.314 & 198.3 \\
AFF JV-DFT-DB2 & 1{,}176{,}773 & \best{0.025} & \best{0.029} & \best{0.027} & 144.9 & 50.0 & \best{50.2} & 1.899 & 1.194 & \best{0.963} & 2.381 & 207.6 \\
AFF MPtrj & 1{,}376{,}739 & 0.031 & 0.047 & 0.190 & 68.9 & 46.9 & 105.4 & 1.630 & 0.948 & 1.226 & 2.975 & 161.4 \\
MACE-MP-0 & 1{,}580{,}395 & 0.028 & 0.084 & 0.104 & 42.4 & 38.0 & 94.2 & 1.012 & 0.359 & 1.623 & 2.226 & 55.1 \\
CHGNet & 1{,}580{,}395 & 0.035 & 0.083 & 0.142 & 59.3 & 46.9 & 85.6 & 1.235 & 0.666 & 1.369 & 2.198 & 72.7 \\
UMA-S & $\sim$100\,M & 0.026 & \best{0.029} & 0.076 & \best{29.7} & 35.2 & 85.3 & \best{0.464} & \best{0.122} & 1.910 & 2.202 & \best{39.1} \\
M3GNet-MatPES-PBE & 434{,}712 & 0.038 & 0.049 & 0.148 & 49.3 & \best{33.2} & 83.5 & 0.928 & 0.297 & 1.667 & 2.944 & 94.8 \\
\bottomrule
\end{tabular}%
}
\end{scriptsize}
\end{table*}

\begin{table}[htbp]\centering\small
\caption{Force-field inference scaling on one NVIDIA GB10 (Grace--Blackwell,
\SI{121}{\giga\byte} unified memory): median wall time for one
energy\,+\,force evaluation of cubic Si supercells, the largest system reached
before out-of-memory (OOM), and the limiting resource. Each model run
GPU-exclusive. ALIGNN-FF (AFF) is fastest with a high memory ceiling, UMA reaches the
largest system but is slowest per step, and MACE/CHGNet are memory-bound.}
\label{tab:scaling}
\begin{tabular}{l r r r r l}
\toprule
Model & max atoms & mem there & $t$ @ 21{,}952 & OOM at & bound by \\
\midrule
AFF-r2SCAN & 54{,}872 & \SI{107}{GB} & \SI{2.0}{s} & $\sim$64{,}000 & balanced (fastest) \\
UMA-S & 140{,}608 & \SI{104.1}{GB} & \SI{19.1}{s} & 175{,}616 & compute (slowest) \\
MACE-MP-0 & 21{,}952 & \SI{79.2}{GB} & \SI{7.4}{s} & 27{,}000 & memory \\
CHGNet & 21{,}952 & \SI{101.8}{GB} & \SI{5.6}{s} & 27{,}000 & memory \\
M3GNet-MatPES-PBE & 32{,}768 & \SI{99.7}{GB} & \SI{2.3}{s} & 39{,}304 & balanced (fast) \\
\bottomrule
\end{tabular}
\end{table}

\paragraph{Direct predictors versus the force field.}
The two model families in this work are complementary rather than competing. A
direct property model, trained end to end on labels for a single target, is the
more accurate and cheaper choice wherever such labels exist: on formation energy it
reaches \SI{0.0288}{eV/atom} (Table~\ref{tab:single}), against
\SI{0.226}{eV/atom} for the same quantity obtained from relaxed force-field
energies over the WBM set (Table~\ref{tab:ff}). The force field trades this
per-property accuracy for breadth and physical consistency: from one energy surface
it yields elastic constants, phonons, thermal conductivity, defect and surface
energetics, molecular dynamics, and formation energies for structures that carry no
direct label, all from a single set of weights. Because errors propagate through the
intervening relaxation or lattice-dynamics calculation, the practical guidance is to
use a direct model when a well-populated label set is available. The force field is recommended
when a property is tied to the potential energy surface, when many properties are
needed at once, or when the target structures lie outside any labelled set. The two
therefore occupy distinct roles in a screening pipeline: the force field is the
generalist that relaxes and filters arbitrary candidates, and the direct models are
the specialists that assign precise values once a labelled target is in reach.

\paragraph{Metal-ceramic interfaces.}
Heterogeneous metal-ceramic contacts are a demanding test, historically requiring
charge-optimized many-body (COMB3) potentials fitted specifically to aluminium oxide
and nitride~\cite{choudhary2015comb,choudhary2016comb}. The same ALIGNN 2.0 stack
reaches this setting without any interface-specific training
(Fig.~\ref{fig:interface}, Supplementary Information): Al(111)/Al$_2$O$_3$(0001) and Al(111)/AlN(0001) junctions
are assembled with InterMat~\cite{choudhary2024intermat}, which enumerates
lattice-matched supercells, terminations, and in-plane registries. The lowest-energy
stacking is relaxed with the MatPES-PBE force field, and a per-atom Born effective
charge $Z^*$ is then read off the relaxed geometry with the ALIGNN 2.0 Born-tensor
model. The predicted charge state is the physically expected one: the metallic Al slab
stays near neutral ($Z^*\!\approx\!0$), while in the ceramic the cation Al is strongly
positive ($Z^*\!\approx\!+1.5$ to $+3$) and the anions carry the compensating charge
(O $\approx\!-2.0$, N $\approx\!-2.6$), with the transition confined to the first one or
two atomic planes on either side of the contact. A single graph thus supplies the
geometry, the energetics, and the local charge state that a dedicated reactive potential
had to be parameterised for, recovering the qualitative picture of the COMB3 studies at a
fraction of the fitting effort.

\begin{figure}[t]
\centering
\includegraphics[width=0.85\textwidth]{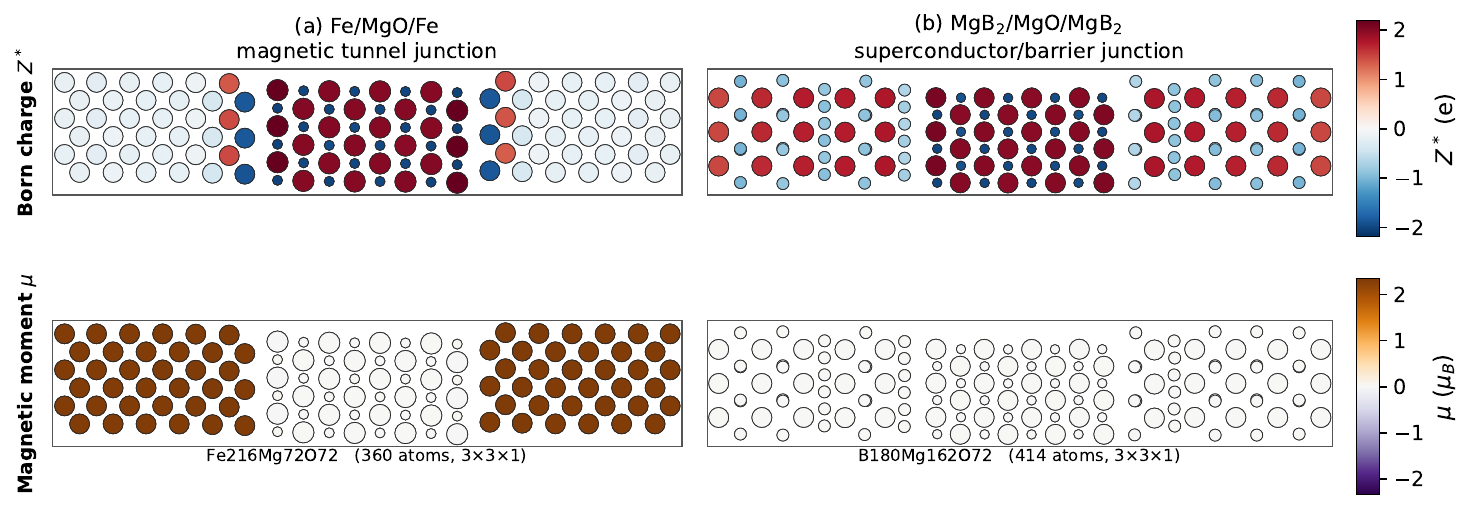}
\caption{\textbf{Three-material device stacks with two per-atom fields from ALIGNN 2.0.}
Fe/MgO/Fe (a magnetic tunnel junction) and MgB$_2$/MgO/MgB$_2$ (a superconductor-barrier
junction), built as symmetric trilayers with InterMat~\cite{choudhary2024intermat}, expanded to
$3{\times}3{\times}1$ supercells (360 and 414 atoms), relaxed with the ALIGNN-FF, and
coloured by the predicted per-atom Born effective charge $Z^{*}$ (top row) and magnetic
moment $\mu$ (bottom row). Charge marks the ionic MgO
barrier in both stacks; magnetism appears only on the Fe electrodes of the tunnel
junction, with the MgO spacer non-magnetic. All fields come from public pretrained
ALIGNN 2.0 models with no stack-specific fitting.}
\label{fig:multilayer}
\end{figure}

The same components extend from a single contact to the three-material stacks that
define real devices, and to a second per-atom field. Figure~\ref{fig:multilayer} shows
two symmetric trilayers assembled by mirroring the InterMat contact about its central
layer: an Fe/MgO/Fe magnetic tunnel junction and an MgB$_2$/MgO/MgB$_2$
superconductor-barrier junction, each coloured by both the Born effective charge and the
per-atom magnetic moment predicted by the corresponding ALIGNN 2.0 models. The two fields
report complementary physics from the same geometry. The charge field recovers the
$+2/-2$ Mg/O dipole of the MgO barrier in both stacks, while the moment field is nonzero
only in the tunnel junction. On either side of a magnetically dead MgO spacer, the two Fe electrodes carry $\approx\!2.3\,\mu_B$ per
atom, in line with bulk bcc iron, which
is precisely the spin-filtering geometry the device relies on. That a charge model, a
magnetic-moment model, and a force field, all sharing one graph, can be read off the same
relaxed multilayer illustrates the practical payoff of the multi-task design for device
modelling.

\subsection{Generative inverse design}

Every application above maps a crystal structure to one or more properties. We next ask whether the same ALIGNN 2.0 representation can be used in the inverse direction, with composition and a target property fixed while the crystal geometry is generated. We refer to this model as ALIGNN Crystal Structure Prediction (ALIGNN-CSP). Crystal reconstruction is evaluated with AtomBench~\cite{campbell2026atombench} on JARVIS Supercon-3D and Alexandria DS-A/B, containing 103 and 825 test structures, respectively.

The experiment tests two questions motivated by the ALIGNN line graph. First, does explicit message passing over three-body geometry improve denoising of the lattice and fractional coordinates? Second, does independently corrupting and denoising bond angles improve reconstruction beyond ordinary ALIGNN angle features or direct supervision of the angular error? We compare six models. The angular treatment is either absent, directly supervised, or independently diffused, and each case is tested with and without the line graph.

Across all three angular treatments and both datasets, adding the line graph lowers the validation structural loss. With no angular objective, the reduction is 12.2\% on JARVIS and 25.2\% on Alexandria. With direct angular supervision, the reductions are 12.9\% and 21.9\%. With independent angle diffusion, they are 7.6\% and 10.4\%. Since the paired models use the same number of convolution stages and have closely matched parameter counts, the improvement is not explained by greater network depth or size.

This provides an inverse-design counterpart to the role of the line graph in forward ALIGNN~\cite{choudhary2021alignn}. Explicit message passing over bonded atom triplets improves prediction of the noisy lattice and coordinates even though the structural loss contains no direct bond-angle term.

The improvement in denoising does not consistently increase the final reconstruction match rate. On JARVIS, adding the line graph to the model with no angular objective lowers the structural loss from 7.149 to 6.279 while changing the match rate from 0.476 to 0.447. On Alexandria, the loss decreases from 5.718 to 4.278 while the match rate changes from 0.589 to 0.566. This distinction is expected because the validation loss measures the denoising model directly, whereas the reconstruction metrics are measured after stochastic generation, energy ranking, structural relaxation, and crystallographic post-processing.

\paragraph{Independent bond-angle diffusion.}

Directly supervising the angular displacement caused by coordinate and lattice corruption has little effect on the structural loss. Relative to the model with no angular objective, the change is +1.0\% without the line graph and +0.2\% with it on JARVIS, and -2.2\% and +2.1\% on Alexandria. The angular loss also reaches an early plateau near 0.28. Directly predicting the bond-angle error therefore does not consistently improve lattice and coordinate denoising.

Independent angle diffusion behaves differently. Without the line graph, it lowers the JARVIS structural loss from 7.149 to 6.682, a 6.5\% reduction. On Alexandria, the change is negligible, from 5.718 to 5.712. The independently corrupted angular variables are nevertheless learned. Across these experiments, the fitted angular score loss reaches approximately 0.29--0.62 times the loss obtained by always predicting zero.

The benefit becomes smaller or disappears when the line graph is present. On JARVIS, independent angle diffusion reduces the structural loss by only 1.7\% once the line graph has already been added. On Alexandria, it instead increases the loss from 4.278 to 5.118, or 19.6\%. One possible explanation is that independently diffused angles provide useful three-body information when the main network contains only pair connections, whereas the same information can become redundant or conflicting when the line graph already propagates bonded-triplet geometry.

\paragraph{Reconstruction fidelity.}

On JARVIS, the independent-angle model without the line graph gives the lowest matched-coordinate RMSD among the six models, 0.0276\,\AA{}, compared with 0.0502\,\AA{} for the corresponding model with no angular objective. Its ccRMSD decreases from 0.540 to 0.492, while the number of matched structures changes only from 49 of 103 to 48 of 103. Independent angle diffusion therefore improves the coordinate accuracy of matched structures much more strongly than the number of structures satisfying the matching criterion.

Direct angular supervision also lowers the JARVIS RMSD to 0.0322\,\AA{} without the line graph, despite leaving the structural loss nearly unchanged. Direct supervision and independent diffusion therefore affect the model differently even though both introduce explicit angular information.

On the larger Alexandria test set, the independent-angle model without the line graph gives the highest match rate and the lowest RMSD, lattice-length MAE, and KLD among the six models. The differences from the corresponding no-angle model are, however, extremely small. The match rate differs by one structure out of 825, the RMSD differs by 0.0018\,\AA{}, and the mean KLD differs by 0.0001. We therefore do not interpret these results as evidence that independent angle diffusion improves Alexandria reconstruction. They instead show that the additional angular state can be introduced without meaningful degradation when the line graph is absent.

The model combining independent angle diffusion with the ALIGNN line graph does not improve the post-relaxation metrics. Its match rates are 0.447 on JARVIS and 0.552 on Alexandria, with RMSDs of 0.0490\,\AA{} and 0.0301\,\AA{}, respectively. Together with its increased Alexandria structural loss, this result indicates that the two ways of providing angular information do not combine constructively in all settings.

\begin{table}[t]
\centering
\scriptsize
\setlength{\tabcolsep}{3.5pt}
\caption{\textbf{Angular representation and line-graph ablation for ALIGNN-CSP.}
Validation structural loss is $\mathcal{L}_{\mathrm{struct}}$ and is directly comparable across all six models. Reconstruction metrics are evaluated after generation of 32 candidates, ALIGNN-FF energy prescreening and relaxation, and crystallographic symmetrization. RMSD is evaluated only for structures satisfying the AtomBench structure matcher. All models use seed 0.}
\label{tab:inverse_ablation}
\begin{tabular}{llrrrrrrr}
\toprule
Angular state & LG &
$\mathcal{L}_{\mathrm{struct}}$ &
Match &
RMSD (\AA)  &
ccRMSD  &
MAE $abc$ (\AA)  &
MAE $\alpha\beta\gamma$ ($^\circ$)  &
KLD  \\
\midrule
\multicolumn{9}{l}{\emph{JARVIS Supercon-3D} (103 test targets)}\\
none        & no  & 7.1485 & 0.4757 & 0.0502 & 0.5398 & 0.5316 & 10.23 & 0.0234 \\
none        & yes & 6.2792 & 0.4466 & 0.0436 & 0.5429 & \best{0.4888} & 11.12 & 0.0233 \\
direct      & no  & 7.2223 & \best{0.4757} & 0.0322 & 0.5095 & 0.4886 & \best{9.67} & \best{0.0199} \\
direct      & yes & 6.2919 & 0.4660 & 0.0403 & 0.5662 & 0.5102 & 11.24 & 0.0241 \\
independent & no  & 6.6815 & 0.4660 & \best{0.0276} & \best{0.4922} & 0.5019 & 10.73 & 0.0240 \\
independent & yes & \best{6.1718} & 0.4466 & 0.0490 & 0.5573 & 0.5208 & 10.41 & 0.0231 \\
\midrule
\multicolumn{9}{l}{\emph{Alexandria DS-A/B} (825 test targets)}\\
none        & no  & 5.7178 & 0.5891 & 0.0266 & 0.3038 & 0.4845 & \best{8.28} & 0.0200 \\
none        & yes & \best{4.2778} & 0.5661 & 0.0253 & 0.3177 & 0.5332 & 8.82 & 0.0217 \\
direct      & no  & 5.5913 & 0.5721 & 0.0304 & \best{0.2998} & 0.4897 & 8.70 & 0.0206 \\
direct      & yes & 4.3665 & 0.5758 & 0.0312 & 0.3118 & 0.5240 & 8.62 & 0.0211 \\
independent & no  & 5.7121 & \best{0.5903} & \best{0.0248} & 0.3071 & \best{0.4782} & 8.51 & \best{0.0199} \\
independent & yes & 5.1182 & 0.5515 & 0.0301 & 0.3193 & 0.5131 & 9.47 & 0.0215 \\
\bottomrule
\end{tabular}
\end{table}

\paragraph{Crystallographic post-processing.}

The symmetry idealization used by AtomBench mainly changes the reported lattice geometry rather than the Cartesian atomic positions. Averaged over the six models, symmetrization lowers the lattice-angle MAE by 37.7\% on JARVIS and 35.9\% on Alexandria, and lowers the mean lattice-parameter KLD by 29.1\% and 24.2\%, respectively. In comparison, lattice-length MAE changes by less than one percent and ccRMSD by approximately one-half of one percent on both datasets.

This is consistent with the symmetry procedure described in Methods. Small symmetry-breaking distortions can cause discrete changes in the reduced lattice under Niggli reduction, substantially changing the reported lattice angles without a comparable change in Cartesian geometry. We therefore use the symmetrized metrics as the primary crystallographic evaluation and retain the unsymmetrized values as a check.

\paragraph{Computational cost.}

The six models differ in parameter count by only 2.8\%, although they were not matched in GPU time. Independent angle diffusion increases total GPU time from 0.67 to 0.75 hours without the line graph and from 0.81 to 0.94 hours with it on JARVIS. On Alexandria, the corresponding changes are 2.40 to 2.86 and 3.81 to 4.08 GPU-hours. The line graph contributes the larger computational cost because message passing is performed over atom triplets in addition to bonded pairs.

\paragraph{Interpretation.}

The ablation supports three main conclusions. The ALIGNN line graph consistently improves lattice and coordinate denoising. Directly supervising the bond-angle error does not reproduce the effect of independently corrupting and denoising the angles. Independent angle diffusion can improve reconstruction when the line graph is absent, most clearly on JARVIS, but the benefit is not consistent across datasets and does not combine favorably with the line graph on Alexandria.

A useful interpretation is that an independently corrupted angle can provide an additional measurement of local three-atom geometry when the network otherwise operates primarily on bonded pairs. Once the line graph already represents three-body geometry explicitly, a separately corrupted representation of the same geometry need not provide an additional benefit.

\paragraph{Comparison with contemporary crystal generators.}

We compare the complete ALIGNN-CSP reconstruction workflow with the four generative model families evaluated in AtomBench~\cite{campbell2026atombench}: AtomGPT~\cite{choudhary2023atomgpt}, CDVAE~\cite{xie2022cdvae}, FlowMM~\cite{miller2024flowmm}, and MatterGen~\cite{zeni2025mattergen}. The JARVIS Supercon-3D and Alexandria DS-A/B test sets used here are the same as those used in AtomBench, and the reconstruction metrics follow the same AtomBench definitions. For ALIGNN-CSP, we use the standard line-graph model without an auxiliary angular objective as the representative configuration. Because ALIGNN-CSP conditions on composition and the target scalar property, the $T_c$-conditioned AtomGPT and MatterGen variants provide the closest comparison in terms of information available at inference. FlowMM is conditioned on composition only, whereas CDVAE reconstructs from a latent representation of the complete target structure, so comparisons involving those models also reflect differences in conditioning information.

\begin{table}[t]
\centering
\scriptsize
\setlength{\tabcolsep}{4pt}
\caption{\textbf{Crystal reconstruction comparison with contemporary models.}
ALIGNN-CSP values correspond to the complete 32-candidate generation, ALIGNN-FF ranking and relaxation, and crystallographic post-processing workflow. AtomGPT, CDVAE, FlowMM, and MatterGen values, as well as the statistical evaluation procedure, are taken from AtomBench~\cite{campbell2026atombench}. Mean RMSD is evaluated over structures accepted by the AtomBench StructureMatcher criterion, while ccRMSD is evaluated over the full test set. Mean absolute error (MAE) is reported for lattice parameters and angles, and the mean Kullback-Leibler divergence (KLD) is given for ground truth and predicted histograms of lattice parameters.}
\label{tab:inverse_external}
\begin{tabular}{llrrrrrr}
\toprule
Dataset & Model & Match & Mean RMSD (\AA) & ccRMSD & MAE $abc$ (\AA) & MAE $\alpha\beta\gamma$ ($^\circ$) & Mean KLD \\
\midrule
\multirow{5}{*}{Alexandria}
& ALIGNN-CSP workflow & 0.5661 & 0.0253 & 0.3177 & 0.5332 & 8.82 & 0.0217 \\
& AtomGPT $T_c$ & 0.4939 & 0.0347 & 0.5228 & 0.541 & 9.262 & 0.0221 \\
& MatterGen $T_c$ & 0.6352 & 0.0182 & 0.2440 & 0.400 & 11.851 & 0.0243 \\
& CDVAE & 0.3564 & 0.4181 & 1.2218 & 0.177 & 8.351 & 0.0118 \\
& FlowMM & 0.0897 & 0.3810 & 1.3998 & 1.066 & 17.883 & 0.0380 \\
\midrule
\multirow{5}{*}{JARVIS}
& ALIGNN-CSP workflow & 0.4466 & 0.0436 & 0.5429 & 0.4888 & 11.12 & 0.0233 \\
& AtomGPT $T_c$ & 0.4706 & 0.0376 & 0.7665 & 0.565 & 7.283 & 0.0235 \\
& MatterGen $T_c$ & 0.4951 & 0.0446 & 0.4988 & 0.587 & 12.648 & 0.0312 \\
& CDVAE & 0.3592 & 0.4083 & 1.1555 & 0.348 & 7.128 & 0.0110 \\
& FlowMM & 0.0291 & 0.4077 & 1.5392 & 0.936 & 17.427 & 0.0356 \\
\bottomrule
\end{tabular}
\end{table}

On Alexandria, the complete ALIGNN-CSP workflow attains a higher match rate and lower ccRMSD than AtomGPT $T_c$, while MatterGen $T_c$ retains the higher match rate and lower coordinate error. ALIGNN-CSP also gives the lowest mean lattice-angle error among these three composition-and-property-conditioned workflows. On JARVIS, its match rate is slightly lower than those of AtomGPT $T_c$ and MatterGen $T_c$, but its ccRMSD is substantially lower than that of AtomGPT $T_c$ and close to that of MatterGen $T_c$. CDVAE gives the lowest lattice-distribution and lattice-length errors on both datasets, consistent with its access to a latent embedding of the target structure, while FlowMM has substantially larger reconstruction errors under the present benchmark.

The ALIGNN-CSP numbers should be interpreted as performance of the reconstruction workflow as a whole rather than as a controlled comparison of raw generator outputs. For each target, ALIGNN-CSP generates 32 candidates, ranks them using ALIGNN-FF energies, relaxes the four lowest-energy candidates, and retains the lowest-energy relaxed structure before crystallographic post-processing. This ranking and relaxation procedure is likely to improve reconstruction fidelity relative to an unrelaxed sample, whereas the AtomBench values characterize the corresponding standalone model reconstructions without the same ALIGNN-FF refinement procedure. We nevertheless report the comparison because the generator, energetic prescreening, and relaxation stages are intended to operate together as the deployed ALIGNN-CSP workflow.

\subsection{Spectroscopy}
\begin{figure}[t]
\centering
\includegraphics[width=\textwidth]{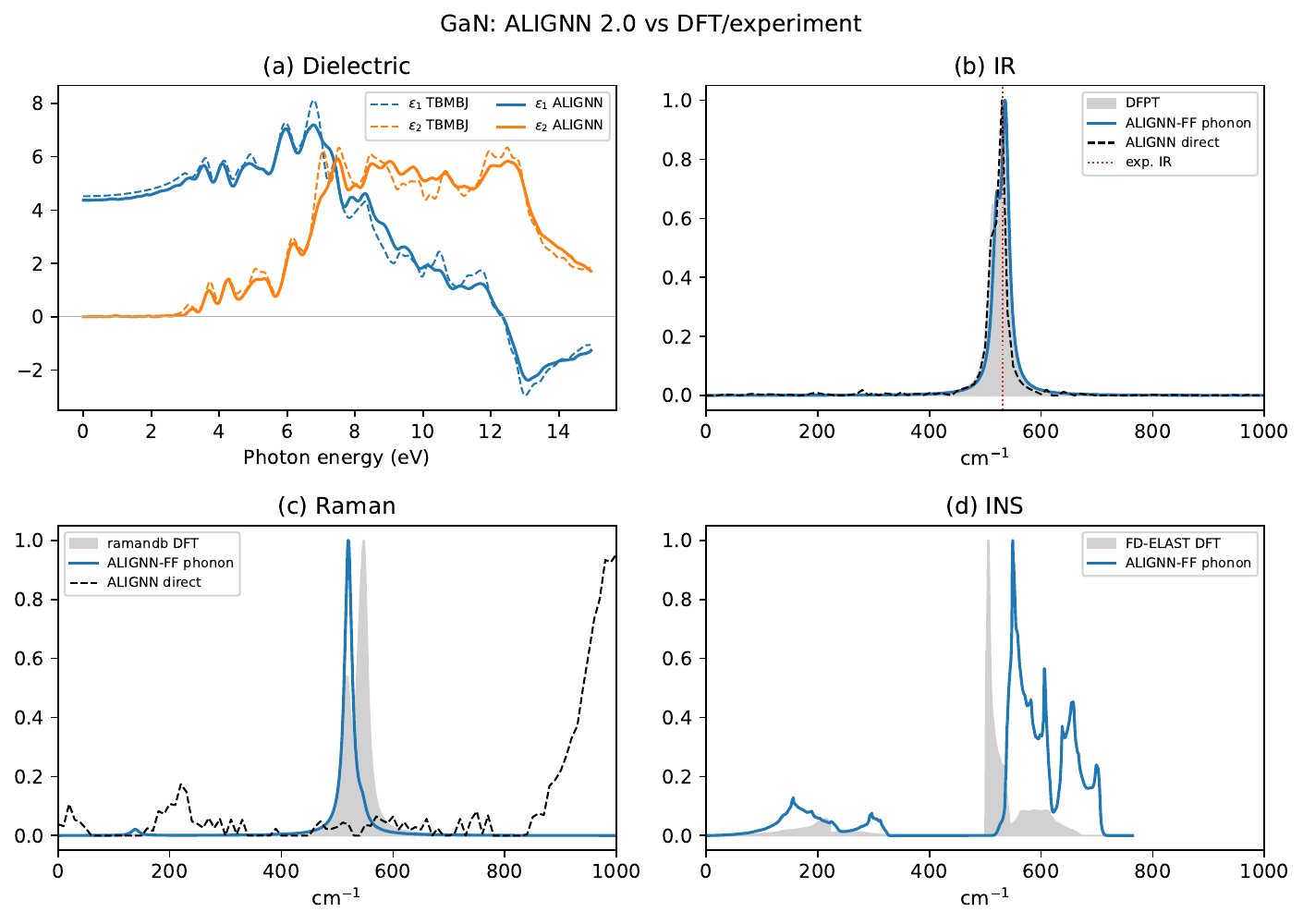}
\caption{\textbf{Spectroscopy of wurtzite GaN (JVASP-30\,/\,mp-804) from a single
relaxed structure, versus experiment and DFT.} ALIGNN 2.0 predictions (solid;
direct end-to-end predictor dashed) against reference data (shaded/dashed).
(a) Optical dielectric $\varepsilon_1,\varepsilon_2$ vs TBmBJ.
(b) Infrared (ALIGNN-FF phonons) vs DFPT and the experimental line, coinciding near
$531$~cm$^{-1}$.
(c) Raman vs the JARVIS DFT Raman database, reproducing the dominant $E_2$/$A_1$
band near $520$~cm$^{-1}$.
(d) Inelastic neutron scattering vs the finite-displacement DFT reference.
Reference data: TBmBJ dielectric~\cite{choudhary2018diel}, DFPT infrared~\cite{choudhary2020dfpt}, JARVIS Raman database~\cite{choudhary2023ramandb}.}
\label{fig:spectra}
\end{figure}

Beyond triaging a screening hit, the same set of ALIGNN 2.0 models reproduces the
principal \emph{spectroscopic} observables of a material directly from its relaxed
structure, and, because the probes are independent and have independent ground
truths, the predictions can be validated against experiment and DFT at once. We
assembled a validation set of seven crystals that appear in \emph{both} the JARVIS
density-functional-perturbation-theory infrared database~\cite{choudhary2020jarvis},
which tabulates experimental IR frequencies, and the Raman
database~\cite{choudhary2024leaderboard} of DFT Raman activities: ZnO, AlN, GaN,
SnS, SnSe, LiNbO$_3$, and GeS. Figure~\ref{fig:spectra} shows the representative
case of wurtzite GaN (JVASP-30\,/\,mp-804). The remaining six materials and the
rutile TiO$_2$ reference are collected in the Supplementary Information. The
wurtzite semiconductors ZnO and AlN (Figs.~\ref{fig:spec_zno},~\ref{fig:spec_aln})
follow the same pattern as GaN, with predicted infrared bands within a few
wavenumbers of experiment. The orthorhombic (\emph{Pnma}) chalcogenides SnS, SnSe,
and GeS (Figs.~\ref{fig:spec_sns},~\ref{fig:spec_snse},~\ref{fig:spec_ges}) show
richer multi-line spectra that the phonon-derived route reproduces mode by mode;
and the ferroelectric oxide LiNbO$_3$ (Fig.~\ref{fig:spec_linbo3}) and rutile
TiO$_2$ (Fig.~\ref{fig:spec_tio2}) complete the set across polar chemistries.

For each material, four independent probes are predicted from the single relaxed
structure: the optical dielectric function $\varepsilon(\omega)$, with
$\varepsilon_2(\omega)$ from the TBmBJ dielectric model and $\varepsilon_1(\omega)$
from a Kramers--Kronig transform, together with the infrared, Raman, and inelastic
neutron scattering (INS) responses. The IR and INS spectra are built from the
r2SCAN ALIGNN-FF $\Gamma$-point and Brillouin-zone phonons (IR intensities weighted
by the predicted Born effective charges, INS by the neutron-weighted phonon density
of states). The Raman activities follow from the derivatives of the predicted
dielectric tensor along each mode, and for infrared and Raman we additionally
overlay a \emph{direct} end-to-end ALIGNN 2.0 spectrum predictor. For GaN the
predicted optical dielectric tracks the TBmBJ reference across the full
$0$--$15$~eV window. The infrared response peaks at the correct transverse-optical
mode near $530$~cm$^{-1}$, coincident with both the DFPT calculation and the
experimental line at $531$~cm$^{-1}$; and the phonon-derived Raman activity
reproduces the dominant $E_2$/$A_1$ band near $520$~cm$^{-1}$ recorded in the DFT
Raman database. A single framework thus delivers optical, infrared, Raman, and
neutron spectra that are each directly comparable to experiment and DFT.

\subsection{Microscopy}
\begin{figure}[t]
\centering
\includegraphics[width=\textwidth]{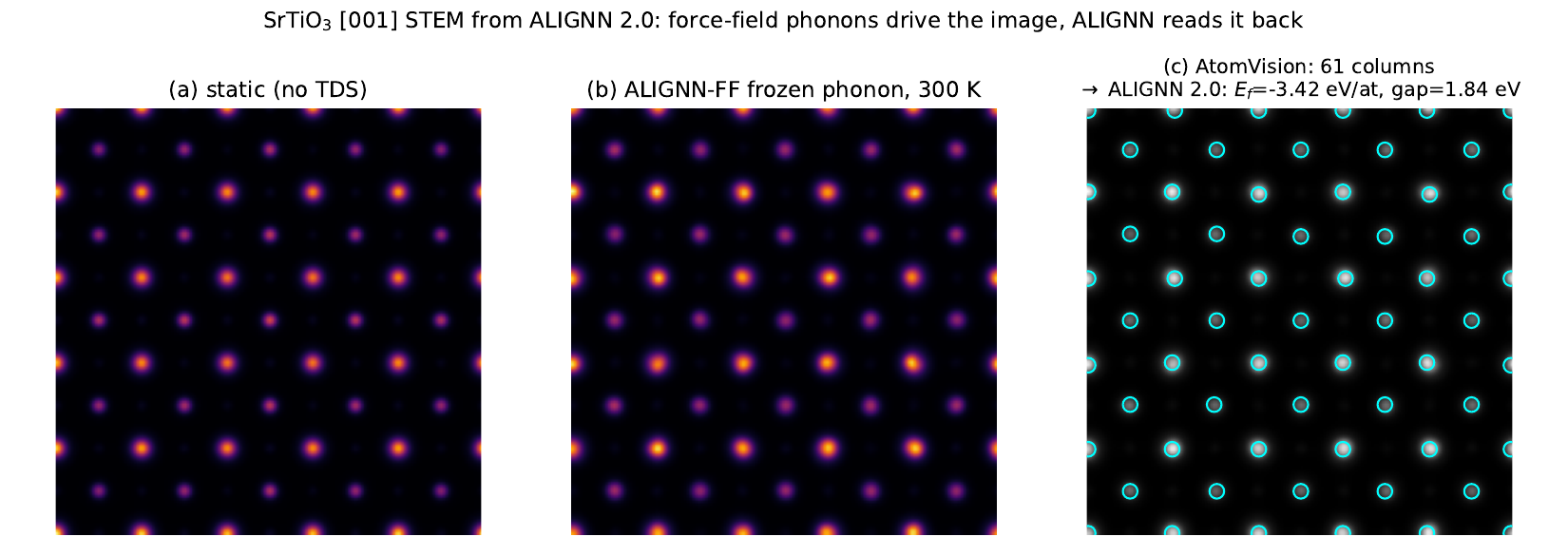}
\caption{\textbf{Force-field-driven STEM with ALIGNN 2.0}
(SrTiO$_3$ [001]; structure$\,\rightarrow\,$image$\,\rightarrow\,$structure$\,\rightarrow\,$property).
(a) Static HAADF multislice.
(b) Frozen-phonon HAADF using $300$\,K thermal amplitudes from the ALIGNN-FF phonons;
thermal-diffuse scattering brightens the heavy Sr columns.
(c) AtomVision detects the $61$ atomic columns, from which ALIGNN 2.0 predicts
$E_f=-3.42$\,eV/atom and band gap $1.84$\,eV.}
\label{fig:microscopy}
\end{figure}

The same framework reaches a fourth experimental modality, scanning transmission
electron microscopy (STEM), by coupling ALIGNN 2.0 to image simulation and
analysis (Fig.~\ref{fig:microscopy}). Quantitative high-angle annular dark-field
(HAADF) imaging depends on the thermal-diffuse scattering (TDS) produced by atomic
vibrations, for which one normally supplies tabulated Debye--Waller factors. Here
the anisotropic thermal-displacement amplitudes come directly from the ALIGNN-FF
phonon spectrum. For SrTiO$_3$ the ALIGNN-FF phonons at $300$\,K give per-element
root-mean-square displacements of $0.11$, $0.07$, and $0.12$\,\AA{} for Sr, Ti and
O, correctly ordered, with the light O and the loosely bound Sr ``rattler''
vibrating most and the stiff Ti least. A frozen-phonon multislice
simulation~\cite{madsen2021abtem} driven by these amplitudes reproduces the
characteristic TDS brightening of the heavy Sr columns that a static calculation
misses (Fig.~\ref{fig:microscopy}a versus b). The loop then closes on the analysis
side: AtomVision~\cite{choudhary2023atomvision}, running dependency-free within the
same ALIGNN ecosystem, localizes the $61$ atomic columns, and ALIGNN 2.0 predicts
the material's formation energy ($-3.42$\,eV/atom) and band gap ($1.84$\,eV) from
the recovered structure (Fig.~\ref{fig:microscopy}c). A single graph thus carries a
structure to a simulated micrograph and back to a structure and its properties; the
same pipeline extends across bonding types (covalent, 2D, metallic, and perovskite)
in Fig.~\ref{fig:microscopy_grid}.

\paragraph{Outlook.}
A single line-graph network, built once and run on current accelerators, has here
supplied direct property prediction, a transferable force field, spectra and
microscopy from a relaxed structure, and a conditional generator, with the same
angular channel earning its place in every mode. The value is not that any one of
these beats a specialised model everywhere, but that they share one representation
and compose: the force field relaxes and scores what the generator proposes, while
the property models assign precise numbers once a labelled target is in reach. The
open directions follow the same logic, pretraining the generator at scale,
extending the force field to the reactive and finite-temperature regimes its
spectroscopy already probes, and folding experimental conditioning, diffraction and
micrographs, more tightly into the inverse-design loop.

\section{Methods}\label{sec:methods}
ALIGNN 2.0 retains the alternating atom-graph and line-graph update structure of
ALIGNN~\cite{choudhary2021alignn} but implements every graph operation in native
PyTorch tensors~\cite{paszke2019pytorch} under a neighbor strategy we refer to as
pure-torch, with no external graph runtime. Atoms are represented by
ninety-two-dimensional atom-embedding vectors~\cite{xie2018cgcnn}; edges carry
radial-basis-function distance features; and triplets, which are the nodes of the
line graph, carry bond-angle features expanded in a radial basis. The model
alternates edge-gated graph convolutions on the atom graph with convolutions on
the line graph, so that angular information computed on the line graph modulates
the messages passed on the atom graph. The line graph is constructed on the fly
from the atom graph's edges, and, because training proceeds over batches of
structures of different sizes, both graphs are batched by concatenation with
consistent index offsetting, so that atoms, edges, and triplets belonging to
different structures never exchange messages. This batched line-graph
construction, expressed entirely in vectorized tensor operations, is what allows
the dependency-free implementation to train efficiently at the scale of more than
a million configurations. Two model families are used throughout: a compact configuration with two
atom-graph and two line-graph layers, at a hidden width of sixty-four for the
controlled architecture comparisons and one hundred and twenty-eight for the
deployed force field, and a production configuration with four atom-graph and
four line-graph layers and a hidden width of two hundred and fifty-six.

\subsection{Line-graph message passing}
Each atom $i$ carries a $92$-dimensional atom-embedding feature
vector~\cite{xie2018cgcnn} projected to the hidden width $d$, while edges and
triplets (the line-graph nodes) carry radial-basis (RBF) expansions of the bond
length and of the bond-angle cosine,
\begin{equation}
\mathbf{h}_i^{(0)} = \mathbf{W}_\mathrm{emb}\,\mathbf{x}_i,\quad
\mathbf{e}_{ij}^{(0)} = \mathrm{RBF}\big(\lVert\mathbf{r}_{ij}\rVert\big),\quad
\mathbf{t}_{ijk}^{(0)} = \mathrm{RBF}\big(\cos\theta_{ijk}\big),\quad
\cos\theta_{ijk} = \frac{\mathbf{r}_{ij}\cdot\mathbf{r}_{ik}}
{\lVert\mathbf{r}_{ij}\rVert\,\lVert\mathbf{r}_{ik}\rVert},
\end{equation}
with $\mathbf{r}_{ij}=\mathbf{r}_j-\mathbf{r}_i+\mathbf{T}_{ij}\mathbf{L}$ the
minimum-image bond vector ($\mathbf{T}_{ij}$ the integer cell offset, $\mathbf{L}$
the lattice matrix). The core operation is the edge-gated graph
convolution~\cite{choudhary2021alignn},
\begin{align}
\mathbf{e}_{ij} &\leftarrow \mathbf{e}_{ij} + \mathrm{SiLU}\!\big(
\mathrm{LN}(\mathbf{A}\mathbf{h}_i+\mathbf{B}\mathbf{h}_j+\mathbf{C}\mathbf{e}_{ij})
\big),\qquad
\hat{\boldsymbol{\sigma}}_{ij} = \frac{\sigma(\mathbf{e}_{ij})}
{\sum_{k\in\mathcal{N}(i)}\sigma(\mathbf{e}_{ik})+\epsilon}, \\
\mathbf{h}_i &\leftarrow \mathbf{h}_i + \mathrm{SiLU}\!\Big(
\mathrm{LN}\big(\mathbf{D}\mathbf{h}_i + \textstyle\sum_{j\in\mathcal{N}(i)}
\hat{\boldsymbol{\sigma}}_{ij}\odot\mathbf{E}\mathbf{h}_j\big)\Big),
\end{align}
where $\sigma$ is the logistic gate, $\odot$ the Hadamard product, $\mathrm{LN}$
layer normalization, and $\mathbf{A}$--$\mathbf{E}$ learned weights. An ALIGNN
layer applies this convolution first on the line graph, so updated triplet
features modulate the bond features, and then on the atom graph,
\begin{equation}
(\mathbf{h},\mathbf{e}) \leftarrow
\mathrm{EGGC}_{\mathrm{atom}}\!\big(\mathbf{h},\,
\mathrm{EGGC}_{\mathrm{line}}(\mathbf{e},\mathbf{t})\big).
\end{equation}
After $L$ such layers, mean pooling and a linear head give the graph-level output,
\begin{equation}
\mathbf{y} = \mathbf{W}_\mathrm{out}\,\frac{1}{N}\sum_{i=1}^{N}\mathbf{h}_i^{(L)},
\end{equation}
whose dimension is set per task ($1$ for a scalar, $D$ for a $D$-bin spectrum or a
flattened tensor); per-atom targets read out the node features
$\mathbf{h}_i^{(L)}$ directly.

Two graph constructions are compared under otherwise identical settings. The
force-field-compatible radius graph uses a five-angstrom cutoff, retains at most
twelve nearest neighbors, and builds the line graph from a three-and-a-half
angstrom three-body cutoff; its neighbor set varies continuously with atomic
displacement, which is the property required for conservative dynamics. The
alternative kNN graph uses an eight-angstrom cutoff with twelve nearest neighbors
and the full angular line graph, providing a wider angular receptive field at the
cost of a neighbor list that changes discontinuously as atoms move. All other
architectural and training settings are held fixed between the two constructions
so that any difference in accuracy is attributable to the graph alone. For
force-field training the total energy is predicted as a per-atom energy multiplied
by the number of atoms. Atomic forces are obtained analytically as the negative
gradient of the predicted energy with respect to atomic positions through
automatic differentiation; and, where trained, the stress is obtained from the
corresponding virial. The training loss combines a graph-level term for energy
with a gradient term for forces and, when enabled, a stress term, with fixed
relative weights. The same per-atom head that produces forces is reused, with the
gradient disabled, to fit non-gradient per-atom targets such as atomic charges and
magnetic moments, and a graph-level vector head of the appropriate output
dimension is used for spectral and tensorial targets.

\subsection{Force-field energies, forces, and training objective}
The potential predicts a per-atom energy $\epsilon_\theta(i)$ summed to the total
energy; forces and the (optional) stress follow analytically by automatic
differentiation,
\begin{equation}
E = \sum_{i=1}^{N}\epsilon_\theta(i),\qquad
\mathbf{F}_i = -\frac{\partial E}{\partial \mathbf{r}_i},\qquad
\sigma_{\alpha\beta} = \frac{1}{V}\frac{\partial E}{\partial \eta_{\alpha\beta}},
\end{equation}
where $V$ is the cell volume and $\eta_{\alpha\beta}$ the symmetric strain. Because
$\mathbf{F}_i$ and $\boldsymbol{\sigma}$ are exact gradients of the single scalar
$E$, the potential is conservative by construction, the prerequisite for
energy-conserving dynamics. Training minimizes a weighted energy--force--stress
objective,
\begin{equation}
\mathcal{L} = w_E\Big(\frac{E-E^\mathrm{DFT}}{N}\Big)^2
+ \frac{w_F}{3N}\sum_{i=1}^{N}\big\lVert\mathbf{F}_i-\mathbf{F}_i^\mathrm{DFT}\big\rVert^2
+ w_\sigma\big\lVert\boldsymbol{\sigma}-\boldsymbol{\sigma}^\mathrm{DFT}\big\rVert^2,
\end{equation}
with fixed relative weights $(w_E,w_F,w_\sigma)$. Accuracy on the leaderboard is the
test-set mean absolute error, and skill relative to the mean-predicting baseline
(MAD) is $\mathrm{Skill}=100\,(1-\mathrm{MAE}/\mathrm{MAD})$.

The property datasets are drawn from JARVIS-DFT~\cite{choudhary2020jarvis,garrity2021wannier,choudhary2025jarviscms}, in
which the underlying quantities are computed with the projector-augmented-wave
method~\cite{kresse1996vasp} using, for the JARVIS DFT-3D properties, the
OptB88vdW van der Waals density functional~\cite{klimes2011optb88}. Formation and
total energies, band gaps, and derived electronic descriptors follow from
self-consistent calculations. Elastic stiffness tensors are obtained from
finite strain-stress relations. Dielectric, Born-effective-charge, and
piezoelectric tensors are obtained from density-functional perturbation
theory~\cite{baroni2001dfpt}. Thermoelectric Seebeck coefficients and power
factors follow from semiclassical Boltzmann transport. The lattice thermal
conductivity, reported on a base-ten logarithmic scale, quantifies phonon
transport. Electronic densities of states are taken directly from the
self-consistent calculations, while phonon densities of states derive from
force constants evaluated by the finite-displacement method and post-processed
with phonopy~\cite{togo2015phonopy}. Atomic charges are obtained by Bader
decomposition of the charge density~\cite{henkelman2006bader}, and site magnetic
moments from spin-polarized calculations. Superconducting critical temperatures,
including the separately distributed high-pressure hydride sets, follow the
electron-phonon-coupling screening workflow of Ref.~\cite{choudhary2022supercon}.
For force fields, the ALIGNN-FF database aggregates JARVIS-DFT
relaxation trajectories~\cite{choudhary2023alignnff}, and the MATPES set provides
of order four hundred thousand structures carefully sampled from a very large pool
of molecular-dynamics snapshots to span equilibrium and near-equilibrium
configurations~\cite{kaplan2025matpes}. To broaden the configuration space sampled
by a finite-displacement potential, we additionally harvest OptB88vdW
configurations from energy-volume curves and from vacancy and surface relaxation
trajectories generated in JARVIS-DFT workflows, parsing every ionic step of each
converged calculation into an energy-per-atom-and-forces record consistent with
the finite-displacement data's reference, and merge these with the
finite-displacement set while holding the original test partition fixed so that
any change in accuracy is attributable to the added training data alone. All
property and force-field evaluations follow the JARVIS-Leaderboard protocol and
its fixed train, validation, and test splits and report the mean absolute error on
the test partition~\cite{choudhary2024leaderboard}. The force-field discovery
generalization is measured on the WBM set with the Matbench-Discovery
protocol~\cite{riebesell2025matbench}: each candidate is relaxed and scored by its
formation-energy error, a convex-hull stability F1, and the geometry-optimization
root-mean-square displacement to the DFT-relaxed structure.

\subsection{Material-property benchmarks}
The CHIPS-FF suite~\cite{wines2025chipsff} evaluates every potential under one
relax-and-evaluate protocol against JARVIS-DFT references. The monovacancy
formation energy for removing a species $X$ from an $N$-atom supercell is
\begin{equation}
E_f^{\mathrm{vac}} = E_{N-1}^{\mathrm{def}} - E_N^{\mathrm{bulk}} + \mu_X,
\end{equation}
with $E_N^{\mathrm{bulk}}$ the relaxed perfect-cell energy and $\mu_X$ the chemical
potential of the removed species; the surface energy of a slab of area
$A$ and $N_s$ atoms, and the interface work of adhesion (\textsc{InterMat}
protocol~\cite{choudhary2024intermat}), are
\begin{equation}
\gamma = \frac{E^{\mathrm{slab}}_{N_s} - N_s\,E^{\mathrm{bulk}}_{\mathrm{atom}}}{2A},
\qquad
W_{\mathrm{ad}} = \frac{E_{\mathrm{slab},1} + E_{\mathrm{slab},2}
- E_{\mathrm{interface}}}{A},
\end{equation}
where the two free slabs are obtained by a self-consistent split of the relaxed
interface (identical lateral cell and atom set). Elastic constants follow from the
stress--strain response, $C_{ijkl}=\partial\sigma_{ij}/\partial\eta_{kl}$, with the
Voigt bulk modulus $K_V=\tfrac{1}{9}(C_{11}+C_{22}+C_{33})+\tfrac{2}{9}(C_{12}+C_{13}+C_{23})$;
The benchmark covers, per property, $104$ systems for the lattice
constants and elastic moduli, $49$ vacancies, $82$ surfaces, and $15$ interface
work-of-adhesion pairs. formation energies and vacancies use per-model self-consistent references $\mu_X$ so
each potential is scored against its own elemental states.

\subsection{Spectroscopy from a single structure}
Phonon frequencies $\omega_m$ and eigenvectors $\mathbf{e}^m_{\kappa}$ (atom
$\kappa$, mass $M_\kappa$) are obtained from finite-displacement force constants on
the relaxed structure. Infrared intensities weight each mode by the predicted Born
effective charges $\mathbf{Z}^*_\kappa$, and Raman activities by the derivative of
the predicted electronic dielectric tensor along the mode (the Raman tensor
$\mathbf{R}^m$),
\begin{equation}
I^{\mathrm{IR}}(\omega_m) \propto \sum_{\alpha}\Big|\sum_{\kappa\beta}
Z^*_{\kappa,\alpha\beta}\,\frac{e^m_{\kappa\beta}}{\sqrt{M_\kappa}}\Big|^2,
\qquad
R^m_{\alpha\beta} = \sum_{\kappa\gamma}\frac{\partial\varepsilon^{\infty}_{\alpha\beta}}
{\partial r_{\kappa\gamma}}\,\frac{e^m_{\kappa\gamma}}{\sqrt{M_\kappa}},
\end{equation}
the latter evaluated either by central finite differences of
$\varepsilon^{\infty}(\mathbf{r})$ or analytically by back-propagation through the
dielectric head. The powder Raman activity uses the rotational invariants
\begin{equation}
I^{\mathrm{Raman}}_m \propto 45\,\bar a_m^2 + 7\,\gamma_m^2,\qquad
\bar a_m = \tfrac{1}{3}\mathrm{Tr}\,\mathbf{R}^m,\quad
\gamma_m^2 = \tfrac{1}{2}\sum_{\alpha\neq\beta}\big[(R^m_{\alpha\alpha}-R^m_{\beta\beta})^2
+ 3\,(R^m_{\alpha\beta})^2\big].
\end{equation}
The inelastic-neutron response is the neutron-weighted phonon density of states, and
the optical dielectric function is predicted as $\varepsilon_2(\omega)$ (TBmBJ
target) with the real part recovered by the Kramers--Kronig transform,
\begin{equation}
G(\omega) \propto \sum_m\sum_\kappa \frac{\sigma_\kappa}{M_\kappa}\,
\big|\mathbf{e}^m_\kappa\big|^2\,\delta(\omega-\omega_m),
\qquad
\varepsilon_1(\omega) = 1 + \frac{2}{\pi}\,\mathcal{P}\!\int_0^\infty
\frac{\omega'\,\varepsilon_2(\omega')}{\omega'^2-\omega^2}\,\mathrm{d}\omega',
\end{equation}
with $\sigma_\kappa$ the neutron cross-section and $\mathcal{P}$ the Cauchy
principal value.

\subsection{Superconductor screening}
Candidate critical temperatures from the direct $T_c$ predictors are refined, where
indicated, through the Eliashberg spectral function $\alpha^2F(\omega)$, from which
the electron--phonon coupling and logarithmic phonon frequency,
\begin{equation}
\lambda = 2\int_0^\infty \frac{\alpha^2F(\omega)}{\omega}\,\mathrm{d}\omega,
\qquad
\omega_{\mathrm{ln}} = \exp\!\left[\frac{2}{\lambda}\int_0^\infty
\frac{\ln\omega}{\omega}\,\alpha^2F(\omega)\,\mathrm{d}\omega\right],
\end{equation}
give the Allen--Dynes critical temperature
\begin{equation}
T_c = \frac{\omega_{\mathrm{ln}}}{1.2}\,\exp\!\left[
-\frac{1.04\,(1+\lambda)}{\lambda-\mu^{*}(1+0.62\,\lambda)}\right],
\end{equation}
with $\mu^{*}$ the Morel--Anderson Coulomb pseudopotential.

Alongside the model comparisons we report, for each single-property task where it
is well defined, the mean absolute error of a baseline that predicts the mean of
the training targets, evaluated on the test partition, and a skill score defined
as one minus the ratio of the model error to this baseline error. The skill score
is positive when the model outperforms the naive predictor and equals the fraction
of the baseline error removed. For a small number of targets whose per-atom or
reference conventions are still being reconciled, the corresponding baseline or
skill entry is omitted rather than reported provisionally.

Models are trained with a one-cycle learning-rate
schedule~\cite{smith2019superconvergence} and the AdamW
optimizer~\cite{loshchilov2019adamw}. Because the schedule anneals the learning
rate to near zero only at the final epoch, all results are reported at full
convergence, with property models trained for one hundred and fifty to three
hundred epochs and force fields for up to several tens to a few hundred epochs
depending on dataset size. Warm starting from a converged checkpoint is used when
continuing a potential's training on an augmented dataset. Training was performed
on current-generation Blackwell-class GPUs and on Grace-Blackwell hardware; the
dependency-free implementation runs on these accelerators without modification.
Molecular-dynamics stability is assessed by micro-canonical integration with a
one-femtosecond time step, using structures handled through the atomic simulation
environment~\cite{larsen2017ase}, initializing supercells at finite temperature
and reporting the linear drift of the total energy per atom in
milli-electronvolts per atom per picosecond together with its maximum absolute
deviation. The pure-PyTorch ALIGNN implementation, the trained models, and the
associated LAMMPS-compatible~\cite{thompson2022lammps} ALIGNN-FF potential are
openly available at \url{https://github.com/atomgptlab/alignn}, with an
interactive web application at \url{https://atomgpt.org/alignn}~\cite{lee2026agapi}; evaluations follow the JARVIS-Leaderboard protocol~\cite{choudhary2024leaderboard} and the contributions will be provided under the corresponding leaderboard entries.

\subsection{Generative inverse design}\label{sec:methods_inverse}

ALIGNN-CSP is a conditional denoising diffusion model over the lattice,
fractional coordinates, and a redundant bond-angle field. For a crystal with a
fixed set of $N$ atoms, where $N$ is the number of atoms in the generated
cell, let $\mathbf{L}\in\mathbb{R}^{3\times3}$ denote the lattice matrix and
$\mathbf{F}\in[0,1)^{N\times3}$ denote the matrix of fractional coordinates.
Here $\mathbb{R}$ denotes the real numbers, the rows of $\mathbf{L}$ are the
three lattice vectors, and the $i$th row $\mathbf{f}_i$ of $\mathbf{F}$ is the
fractional position of atom $i$, with $i\in\{1,\ldots,N\}$. Joint diffusion
of the lattice and fractional coordinates follows established crystal
diffusion formulations \cite{jiao2023diffcsp,zeni2025mattergen}.

We augment these variables with an independently diffused bond-angle field.
Unlike ordinary ALIGNN, where bond angles are computed from the current atomic
geometry, this angular state is corrupted through its own forward process and
therefore need not equal the angles reconstructed from the noisy lattice and
coordinates at the same timestep. It provides a redundant representation of
three-body structure rather than an additional physical degree of freedom and
is discarded after sampling.

The complete stochastic state at diffusion timestep $t$ is
\begin{equation}
\mathbf{X}_t
=
\left(
\mathbf{x}_t,
\mathbf{F}_t,
\boldsymbol{\Phi}_t
\right),
\end{equation}
where $\mathbf{X}_t$ denotes the complete noisy state,
$\mathbf{x}_t\in\mathbb{R}^{6}$ is the rotationally invariant lattice state,
$\mathbf{F}_t\in[0,1)^{N\times3}$ contains the noisy fractional coordinates,
and $\boldsymbol{\Phi}_t$ denotes the collection of noisy bond-angle
variables. The integer $t\in\{0,\ldots,T\}$ is the diffusion timestep, with
$t=0$ denoting the clean state and $T$ denoting the total number of diffusion
steps. All three channels are denoised jointly by one ALIGNN-CSP network.

\paragraph{Lattice process.}
A lattice is defined only up to global Cartesian rotation:
$\mathbf{L}\rightarrow\mathbf{L}\mathbf{R}$, where
$\mathbf{R}\in\mathbb{R}^{3\times3}$ is a rotation matrix satisfying
$\mathbf{R}^{\mathsf T}\mathbf{R}=\mathbf{I}$. Here the superscript
$\mathsf T$ denotes matrix transpose and $\mathbf{I}$ is the
$3\times3$ identity matrix. This rotation leaves the represented crystal
unchanged. We remove this degree of freedom using
\begin{equation}
\mathbf{S}
=
\left(
\mathbf{L}\mathbf{L}^{\mathsf T}
\right)^{1/2},
\end{equation}
where $\mathbf{S}$ is the symmetric positive-definite lattice factor and
$(\cdot)^{1/2}$ denotes the positive-definite matrix square root, consistent
with invariant lattice representations used in crystal diffusion
\cite{jiao2024space,zeni2025mattergen}.

The clean lattice variable is
\begin{equation}
\mathbf{x}_0
=
\mathrm{vec}_6
\left[
\log\left(
N^{-1/3}\mathbf{S}
\right)
\right]
\in\mathbb{R}^{6},
\end{equation}
where $\mathbf{x}_0$ is the six-component clean lattice representation,
$\log$ denotes the matrix logarithm, and $N^{-1/3}$ removes the leading
dependence of linear cell dimensions on the number of atoms. The map
$\mathrm{vec}_6$ converts a symmetric $3\times3$ matrix into a
six-component vector containing its three diagonal entries and its three
off-diagonal entries multiplied by $\sqrt{2}$. This scaling preserves the
Frobenius inner product under vectorization. The resulting six components are
standardized using means and standard deviations computed from the training
set.

The matrix logarithm maps the symmetric positive-definite matrices to the
unconstrained vector space of symmetric matrices \cite{jiao2024space}, and
the matrix exponential performs the inverse map when the lattice is decoded.
After decoding, we use $\mathbf{L}=\mathbf{S}$ as the representative lattice.
A variance-preserving diffusion process \cite{ho2020ddpm} with a cosine
cumulative noise schedule \cite{nichol2021improved,jiao2023diffcsp} gives
\begin{equation}
\mathbf{x}_t
=
\sqrt{\bar{\alpha}_t}\mathbf{x}_0
+
\sqrt{1-\bar{\alpha}_t}\boldsymbol{\epsilon},
\qquad
\boldsymbol{\epsilon}
\sim
\mathcal{N}(\mathbf{0},\mathbf{I}_6),
\end{equation}
where $\mathbf{x}_t$ is the noisy lattice state at timestep $t$,
$\boldsymbol{\epsilon}\in\mathbb{R}^{6}$ is the Gaussian noise added to the
lattice state, $\mathcal{N}$ denotes a normal distribution, $\mathbf{0}$ is
the six-component zero vector, and $\mathbf{I}_6$ is the $6\times6$ identity
matrix. The cumulative signal-retention coefficient is
\begin{equation}
\bar{\alpha}_t
=
\prod_{s=1}^{t}\alpha_s,
\end{equation}
where $\alpha_s$ is the variance-preserving diffusion coefficient at step
$s$, $s$ is an integer indexing diffusion steps inside the product, and
$\prod$ denotes multiplication over those steps. The lattice output head of
the network predicts $\boldsymbol{\epsilon}$, with prediction denoted
$\hat{\boldsymbol{\epsilon}}$.

\paragraph{Fractional-coordinate process.}
Fractional coordinates are periodic and occupy a three-dimensional torus for
each atom. We use wrapped-normal score matching
\cite{jiao2023diffcsp,zeni2025mattergen} with a geometrically spaced
variance-exploding noise ladder
\begin{equation}
\sigma_t\in[0.005,0.5],
\end{equation}
where $\sigma_t$ is the standard deviation of the coordinate noise at
timestep $t$. The coordinate forward process is
\begin{equation}
\mathbf{F}_t
=
w\left(
\mathbf{F}_0+\sigma_t\mathbf{Z}^{F}
\right),
\qquad
Z^{F}_{ij}
\overset{\mathrm{iid}}{\sim}
\mathcal{N}(0,1),
\end{equation}
where $\mathbf{F}_0$ is the clean fractional-coordinate matrix,
$\mathbf{F}_t$ is the noisy coordinate matrix, and
$\mathbf{Z}^{F}\in\mathbb{R}^{N\times3}$ is a matrix of independent
standard-normal noise samples. The element $Z^{F}_{ij}$ is the noise applied
to coordinate component $j$ of atom $i$, with $j\in\{1,2,3\}$ in this
equation. The notation $\overset{\mathrm{iid}}{\sim}$ means independently
and identically distributed according to the indicated probability
distribution.

The wrapping function is
\begin{equation}
w(u)=u-\lfloor u\rfloor,
\end{equation}
where $u$ is a real-valued coordinate and $\lfloor u\rfloor$ is its floor,
or greatest integer not exceeding $u$. Thus $w$ maps each component onto the
unit-period interval $[0,1)$.

For one fractional-coordinate component,
\begin{equation}
\delta
=
w_{\pm}(f_t-f_0)
\in
\left[-\frac{1}{2},\frac{1}{2}\right)
\end{equation}
is the minimum signed periodic displacement between its noisy value $f_t$ and
clean value $f_0$. Here $w_{\pm}$ denotes wrapping onto the centered
unit-period interval $[-1/2,1/2)$, and $\delta$ denotes the resulting signed
displacement.

The training target for this component is the $\sigma_t$-scaled
wrapped-normal score
\begin{equation}
S^{F}_t
=
\sigma_t
\frac{\partial}{\partial\delta}
\log
\sum_{n\in\mathbb{Z}}
\mathcal{N}
\left(
\delta+n;
0,\sigma_t^2
\right),
\end{equation}
where $S^{F}_t$ is the target score for the selected coordinate component,
$\partial/\partial\delta$ denotes differentiation with respect to the
periodic displacement, $\log$ is the natural logarithm, $n\in\mathbb{Z}$ is
an integer periodic-image index, and $\mathbb{Z}$ denotes the integers. The
quantity
$\mathcal{N}(\delta+n;0,\sigma_t^2)$ is the normal probability density
evaluated at $\delta+n$ with mean zero and variance $\sigma_t^2$. The infinite
periodic-image sum is evaluated numerically using a finite number of image
terms \cite{jiao2023diffcsp,song2021score}.

Collecting the target scores for all atoms and coordinate components gives
$\mathbf{S}^{F}_t\in\mathbb{R}^{N\times3}$. The corresponding network
prediction is denoted $\hat{\mathbf{S}}^{F}_t$.

\paragraph{Independent bond-angle process.}
Direct diffusion in internal angular coordinates has precedent in molecular
and protein generative models
\cite{jing2022torsional,wu2024foldingdiff}. For a triplet of atoms
$(i,j,k)$ centered on atom $j$, the clean bond angle is
\begin{equation}
\theta_{ijk,0}
=
\arccos
\left[
\frac{
\mathbf{r}_{ji,0}\cdot\mathbf{r}_{jk,0}
}{
\lVert\mathbf{r}_{ji,0}\rVert
\lVert\mathbf{r}_{jk,0}\rVert
}
\right],
\end{equation}
where $i$, $j$, and $k$ are atom indices,
$\theta_{ijk,0}$ is the clean angle formed by atoms $i$-$j$-$k$,
$\mathbf{r}_{ji,0}$ is the clean minimum-image Cartesian displacement vector
from atom $j$ to atom $i$, and $\mathbf{r}_{jk,0}$ is the corresponding
vector from atom $j$ to atom $k$. The centered dot denotes the Euclidean dot
product, $\lVert\cdot\rVert$ denotes the Euclidean vector norm, and
$\arccos$ is the inverse cosine.

We represent each angle in turns,
\begin{equation}
\phi_{ijk,0}
=
\frac{\theta_{ijk,0}}{2\pi},
\end{equation}
where $\phi_{ijk,0}$ is the clean angle expressed as a fraction of one full
rotation and $\pi$ is the usual circle constant. This places the angular
variable on the same unit-period domain as the fractional coordinates.

The angular forward process is
\begin{equation}
\phi_{ijk,t}
=
w\left(
\phi_{ijk,0}
+
\sigma_t Z^{\Theta}_{ijk}
\right),
\qquad
Z^{\Theta}_{ijk}
\overset{\mathrm{iid}}{\sim}
\mathcal{N}(0,1),
\end{equation}
where $\phi_{ijk,t}$ is the noisy angular variable for triplet $(i,j,k)$ at
timestep $t$, $Z^{\Theta}_{ijk}$ is its independent standard-normal noise
sample, and the superscript $\Theta$ labels quantities belonging to the
angular channel. The angular noise scale uses
\begin{equation}
\sigma_t\in[0.005,0.5].
\end{equation}
The angular and fractional-coordinate channels therefore use the same values
of $\sigma_t$ but independent Gaussian noise draws.

The complete forward transition probability is
\begin{equation}
q\left(
\mathbf{x}_t,
\mathbf{F}_t,
\boldsymbol{\Phi}_t
\mid
\mathbf{x}_0,
\mathbf{F}_0,
\boldsymbol{\Phi}_0
\right)
=
q_L(\mathbf{x}_t\mid\mathbf{x}_0)
q_F(\mathbf{F}_t\mid\mathbf{F}_0)
q_{\Theta}
\left(
\boldsymbol{\Phi}_t
\mid
\boldsymbol{\Phi}_0
\right),
\end{equation}
where $q$ denotes the joint forward corruption distribution,
$q_L$ denotes the lattice forward distribution,
$q_F$ denotes the fractional-coordinate forward distribution, and
$q_{\Theta}$ denotes the angular forward distribution. The vertical bar
denotes conditioning on the clean state. The factorization follows from the
independent noise draws used by the three forward processes.

The collection $\boldsymbol{\Phi}(\mathbf{F},\mathbf{L})$ denotes the bond
angles obtained deterministically from a lattice $\mathbf{L}$ and fractional
coordinates $\mathbf{F}$. Since the angular channel is corrupted
independently,
\begin{equation}
\boldsymbol{\Phi}_t
\neq
\boldsymbol{\Phi}
\left(
\mathbf{F}_t,\mathbf{L}_t
\right)
\end{equation}
in general, where $\mathbf{L}_t$ is the lattice decoded from
$\mathbf{x}_t$.

The angular training target is also a wrapped-normal score. Define
\begin{equation}
\delta^\Theta_{ijk}
=
w_{\pm}
\left(
\phi_{ijk,t}
-
\phi_{ijk,0}
\right),
\end{equation}
where $\delta^\Theta_{ijk}$ is the minimum signed periodic displacement
between the noisy and clean angular variables for triplet $(i,j,k)$. The
target angular score is
\begin{equation}
S^\Theta_{ijk,t}
=
\sigma_t
\frac{\partial}{\partial\delta^\Theta_{ijk}}
\log
\sum_{n\in\mathbb{Z}}
\mathcal{N}
\left(
\delta^\Theta_{ijk}+n;
0,\sigma_t^2
\right),
\end{equation}
where $S^\Theta_{ijk,t}$ is the target score for triplet $(i,j,k)$ at
timestep $t$. The corresponding prediction of the angular output head is
$\hat{S}^{\Theta}_{ijk,t}$. The collection of angular score predictions over
all triplets is denoted $\hat{\mathbf{S}}^\Theta_t$.

\paragraph{Persistent triplet field.}
Independent angular diffusion requires each angular variable to retain the
same atom-triplet identity throughout the forward and reverse trajectories.
Rebuilding the triplet list from a distance cutoff or a
$k$-nearest-neighbor ranking would violate this requirement when neighbors
enter, leave, or exchange rank. Here $k$ in $k$-nearest-neighbor denotes the
number of neighbors retained for each atom. We therefore use a fixed dense
triplet set, denoted $\mathcal{T}$, constructed from the ordered atom-pair
graph. Each element $(i,j,k)\in\mathcal{T}$ identifies one persistent
triplet centered on atom $j$.

For $N$ atoms, the number of ordered atom pairs scales as
$\mathcal{O}(N^2)$ and the number of triplets as $\mathcal{O}(N^3)$, where
$\mathcal{O}$ denotes asymptotic computational scaling. On a representative
64-crystal JARVIS batch, the persistent construction contains 8722 triplets
compared with 8124 for the geometry-derived construction, a factor of $1.07$,
with approximately $8.5$ MiB of angular-state activations, where MiB denotes
mebibytes.

Each ordered pair receives a continuous geometric relevance
\begin{equation}
s_{ij,t}
=
u(r_{ij,t};r_c),
\end{equation}
where $s_{ij,t}$ is the relevance weight of pair $(i,j)$ at timestep $t$,
$r_{ij,t}$ is the minimum-image Cartesian distance between atoms $i$ and $j$
at that timestep, $r_c$ is the radial cutoff distance, and
$u(r;r_c)$ is the polynomial cutoff envelope used in DimeNet
\cite{gasteiger2020dimenet}. We use
\begin{equation}
r_c=5\,\text{\AA},
\end{equation}
where \AA{} denotes the angstrom, with envelope exponent five. The relevance
of triplet $(i,j,k)$ is
\begin{equation}
s_{ijk,t}
=
s_{ji,t}s_{jk,t},
\end{equation}
where $s_{ijk,t}$ is the product of the relevance weights of the two bonds
meeting at central atom $j$. The angular variable remains defined when
$s_{ijk,t}=0$; this weight only controls its contribution to message passing
and to the angular loss.

\paragraph{Angular representation.}
Geometric bond angles used as ordinary ALIGNN features are expanded with a
40-bin cosine radial basis. The independently diffused angular state instead
uses the periodic harmonic features
\begin{equation}
\left\{
\sin(2\pi k\phi_{ijk,t}),
\cos(2\pi k\phi_{ijk,t})
\right\}_{k=1}^{8},
\end{equation}
where $\sin$ and $\cos$ are the sine and cosine functions and, in this
equation, $k$ is the harmonic order rather than an atom index. The eight
harmonic orders provide sixteen scalar angular features and respect the
unit-period identification of $\phi_{ijk,t}$.

\paragraph{Joint ALIGNN denoiser.}
All stochastic channels are denoised in one network pass,
\begin{equation}
D_{\boldsymbol{\psi}}
\left(
\mathbf{x}_t,
\mathbf{F}_t,
\boldsymbol{\Phi}_t,
t,
\mathbf{c}
\right)
\longrightarrow
\left(
\hat{\boldsymbol{\epsilon}},
\hat{\mathbf{S}}^{F}_t,
\hat{\mathbf{S}}^{\Theta}_t
\right),
\end{equation}
where $D_{\boldsymbol{\psi}}$ denotes the ALIGNN-CSP denoising network,
$\boldsymbol{\psi}$ denotes all trainable network parameters, and
$\mathbf{c}$ denotes the conditioning information. The three outputs are the
predicted lattice noise $\hat{\boldsymbol{\epsilon}}$, the predicted
fractional-coordinate score $\hat{\mathbf{S}}^{F}_t$, and the predicted
angular score $\hat{\mathbf{S}}^{\Theta}_t$.

Each atom receives learned embeddings of its chemical species, the diffusion
timestep $t$, the conditioning variables $\mathbf{c}$, and the current
lattice state $\mathbf{x}_t$. Pair features include radial basis functions of
interatomic distance and periodic Fourier features of the fractional
displacement,
\begin{equation}
\left\{
\sin\left(2\pi k\Delta\mathbf{f}_{ij}\right),
\cos\left(2\pi k\Delta\mathbf{f}_{ij}\right)
\right\}_{k=1}^{10},
\end{equation}
where $\Delta\mathbf{f}_{ij}$ is the signed periodic fractional displacement
vector between atoms $i$ and $j$. Here $k\in\{1,\ldots,10\}$ is the Fourier
harmonic order, and sine and cosine are applied componentwise to the
three-component displacement vector.

For models containing the line graph, the independently diffused angular
state initializes a triplet hidden representation
$\mathbf{z}_{ijk}$. ALIGNN then uses the standard
triplet-to-pair-to-atom message-passing sequence
\cite{choudhary2021alignn},
\begin{equation}
\boldsymbol{\Phi}_t
\rightarrow
\mathbf{z}_{ijk}
\rightarrow
\mathbf{y}_{ij}
\rightarrow
\mathbf{h}_i,
\end{equation}
where $\mathbf{z}_{ijk}$ is the learned hidden representation of triplet
$(i,j,k)$, $\mathbf{y}_{ij}$ is the learned hidden representation of ordered
pair $(i,j)$, and $\mathbf{h}_i$ is the learned hidden representation of atom
$i$. The arrows indicate the direction in which information is passed between
these representations. The angular output head also reads the hidden pair
representations belonging to the triplet, allowing information from the
structural and angular channels to influence one another.

Models without a line graph use nine pair-graph convolutions in place of the
three ALIGNN layers and three subsequent pair convolutions. Their triplet
representation for the angular output head is constructed after pair message
passing as
\begin{equation}
\mathbf{z}_{ijk}
=
\mathrm{Emb}
\left(
\phi_{ijk,t}
\right)
+
\mathbf{y}_{ji}
+
\mathbf{y}_{jk},
\end{equation}
where $\mathrm{Emb}(\phi_{ijk,t})$ denotes the learned embedding of the noisy
angular variable and $\mathbf{y}_{ji}$ and $\mathbf{y}_{jk}$ are the hidden
representations of the two ordered pairs forming the triplet. These models
have no forward
angle$\rightarrow$pair$\rightarrow$atom message path, although the angular
loss still updates the shared pair-processing parameters through
backpropagation.

The pooled atomic representation used for lattice prediction is
\begin{equation}
\bar{\mathbf{h}}
=
\frac{1}{N}
\sum_{i=1}^{N}
\mathbf{h}_i,
\end{equation}
where $\bar{\mathbf{h}}$ is the mean of the $N$ atom representations and
$\sum$ denotes summation over atoms. A multilayer perceptron maps
$\bar{\mathbf{h}}$ to
$\hat{\boldsymbol{\epsilon}}\in\mathbb{R}^{6}$.

The fractional-coordinate score uses learned scalar coefficients and signed
fractional displacements. For atom $i$,
\begin{equation}
\hat{\mathbf{s}}_i
=
\sum_{c=1}^{C}
a_c
\sum_j
s_{ij,t}
w_{ij}^{(c)}
\left(
\mathbf{h}_i,
\mathbf{h}_j,
\mathbf{y}_{ij}
\right)
\Delta\mathbf{f}_{ij},
\qquad
C=32,
\end{equation}
where $\hat{\mathbf{s}}_i\in\mathbb{R}^{3}$ is the predicted coordinate-score
vector for atom $i$, and the collection of these vectors over all atoms forms
$\hat{\mathbf{S}}^{F}_t$. The integer $c\in\{1,\ldots,C\}$ indexes coordinate
score channels and $C=32$ is the number of such channels. The coefficient
$a_c$ is a learned scalar associated with channel $c$.
The function
$w_{ij}^{(c)}(\mathbf{h}_i,\mathbf{h}_j,\mathbf{y}_{ij})$ is a learned scalar
weight for pair $(i,j)$ and channel $c$, computed from the two atomic hidden
representations and their pair representation. The quantity $s_{ij,t}$ is the
geometric relevance weight defined above and $\Delta\mathbf{f}_{ij}$ supplies
the signed direction of the predicted coordinate score.

No explicit consistency loss or projection is imposed between the
independently diffused angular state $\boldsymbol{\Phi}_t$ and the angles
$\boldsymbol{\Phi}(\mathbf{F}_t,\mathbf{L}_t)$ reconstructed from the noisy
structure. Their agreement is used only as a diagnostic.

\paragraph{Training objectives.}
The model predicts the noise or score associated with each diffused state.
For the lattice state, the target is the Gaussian noise
$\boldsymbol{\epsilon}$ used to construct $\mathbf{x}_t$, and the network
prediction is $\hat{\boldsymbol{\epsilon}}$. The lattice loss is
\begin{equation}
\mathcal{L}_{L}
=
\left\|
\hat{\boldsymbol{\epsilon}}
-
\boldsymbol{\epsilon}
\right\|_2^2,
\end{equation}
where $\mathcal{L}_L$ denotes the lattice loss and
$\lVert\cdot\rVert_2$ denotes the Euclidean norm over the six lattice
components.

For the fractional-coordinate state, the target is the wrapped-normal score
$\mathbf{S}^{F}_t$ and the prediction is
$\hat{\mathbf{S}}^{F}_t$. The fractional-coordinate loss is
\begin{equation}
\mathcal{L}_{F}
=
\left\|
\hat{\mathbf{S}}^{F}_t
-
\mathbf{S}^{F}_t
\right\|_F^2,
\end{equation}
where $\mathcal{L}_F$ denotes the coordinate loss and
$\lVert\cdot\rVert_F$ denotes the Frobenius norm over all atoms and all three
fractional-coordinate components.

For the independently diffused angular state,
$S^\Theta_{ijk,t}$ is the target wrapped-normal score and
$\hat{S}^\Theta_{ijk,t}$ is its network prediction. The relevance-weighted
angular loss is
\begin{equation}
\mathcal{L}_{\Theta}^{\mathrm{ind}}
=
\frac{
\displaystyle
\sum_{(i,j,k)\in\mathcal{T}}
s_{ijk,t}
\left(
\hat{S}^{\Theta}_{ijk,t}
-
S^{\Theta}_{ijk,t}
\right)^2
}{
\displaystyle
\sum_{(i,j,k)\in\mathcal{T}}
s_{ijk,t}
},
\end{equation}
where $\mathcal{L}_{\Theta}^{\mathrm{ind}}$ denotes the angular loss for the
independent-angle model, $\mathcal{T}$ is the persistent triplet set, and
$s_{ijk,t}$ is the geometric relevance weight of triplet $(i,j,k)$. The score
residual itself is not wrapped because periodicity has already entered through
the wrapped-normal target.

The complete loss for the independent-angle model is
\begin{equation}
\mathcal{L}^{\mathrm{ind}}
=
w_L\mathcal{L}_L
+
w_F\mathcal{L}_F
+
w_{\Theta}\mathcal{L}_{\Theta}^{\mathrm{ind}},
\end{equation}
where $\mathcal{L}^{\mathrm{ind}}$ is the total independent-angle training
loss, $w_L$ is the lattice-loss weight, $w_F$ is the fractional-coordinate
loss weight, and $w_{\Theta}$ is the angular-loss weight. We use
\begin{equation}
(w_L,w_F,w_{\Theta})
=
(1,10,1).
\end{equation}

For models without an angular objective, the total loss is
\begin{equation}
\mathcal{L}^{\mathrm{none}}
=
w_L\mathcal{L}_L
+
w_F\mathcal{L}_F,
\end{equation}
where $\mathcal{L}^{\mathrm{none}}$ denotes the total loss for the no-angle
model.

\paragraph{Direct angular-supervision comparator.}
To distinguish independent angular diffusion from direct angular supervision,
we also train a model with no stochastic angular state. For triplet $(i,j,k)$,
let $\theta_{ijk,t}$ denote the angle reconstructed directly from the noisy
lattice and fractional coordinates at timestep $t$. The target angular
displacement is
\begin{equation}
\Delta\theta_{ijk,t}
=
w_{2\pi}
\left(
\theta_{ijk,t}
-
\theta_{ijk,0}^{(t)}
\right),
\end{equation}
where $\Delta\theta_{ijk,t}$ is the signed target angular displacement and
$\theta_{ijk,0}^{(t)}$ is the clean reference angle evaluated using the same
periodic-image identity as the corresponding noisy triplet at timestep $t$.
The angular wrapping function is
\begin{equation}
w_{2\pi}(x)
=
\mathrm{mod}(x+\pi,2\pi)-\pi,
\end{equation}
where $x$ is an angular residual in radians and $\mathrm{mod}(a,b)$ denotes
$a$ modulo $b$. Thus $w_{2\pi}$ maps an angular residual onto
$[-\pi,\pi)$.

The network prediction of the target displacement is denoted
$\widehat{\Delta\theta}_{ijk,t}$. The direct angular loss is
\begin{equation}
\mathcal{L}_{\Theta}^{\mathrm{dir}}
=
\frac{
\displaystyle
\sum_{(i,j,k)\in\mathcal{T}}
s_{ijk,t}
\,
\mathrm{SmoothL1}
\left[
w_{2\pi}
\left(
\widehat{\Delta\theta}_{ijk,t}
-
\Delta\theta_{ijk,t}
\right)
\right]
}{
\displaystyle
\sum_{(i,j,k)\in\mathcal{T}}
s_{ijk,t}
},
\end{equation}
where $\mathcal{L}_{\Theta}^{\mathrm{dir}}$ denotes the direct angular loss
and $\mathrm{SmoothL1}$ denotes the standard piecewise quadratic and linear
SmoothL1 regression loss applied to the wrapped prediction residual.

The complete loss for the direct angular-supervision model is
\begin{equation}
\mathcal{L}^{\mathrm{dir}}
=
w_L\mathcal{L}_L
+
w_F\mathcal{L}_F
+
w_{\Theta}\mathcal{L}_{\Theta}^{\mathrm{dir}},
\end{equation}
where $\mathcal{L}^{\mathrm{dir}}$ denotes the total direct-supervision loss.
The same weights
$(w_L,w_F,w_{\Theta})=(1,10,1)$ are used.

\paragraph{Conditioning.}
The experiments condition on composition and one target scalar property.
Composition is represented by a 118-dimensional element-count vector, with
one component for each chemical element represented by atomic number. The
target scalar property is passed through a continuous learned embedding to
the common hidden dimension of the network. These composition and property
representations are added to the timestep and lattice embeddings.

Classifier-free guidance is implemented by replacing the composition and
property conditions with learned null embeddings during training
\cite{ho2022classifierfree}. The composition condition is dropped with
probability $0.10$ and the target-property condition with probability $0.15$.
At inference, all available diffusion channels use classifier-free guidance
with
\begin{equation}
\gamma=2.0,
\end{equation}
where $\gamma$ is the classifier-free guidance scale.

\paragraph{Sampling.}
Sampling begins from a uniform distribution over fractional coordinates, a
Gaussian distribution over the standardized lattice state, and, for the
independent-angle model, a uniform distribution over the unit-period angular
variables.

Fractional coordinates use the predictor--corrector formulation for
variance-exploding score models
\cite{song2021score,jiao2023diffcsp}. Each reverse predictor step is followed
by one annealed Langevin corrector step with
\begin{equation}
\mathrm{step\_lr}
=
10^{-5},
\end{equation}
where $\mathrm{step\_lr}$ is the step-size parameter used by the Langevin
corrector. The independently diffused angular state uses the same
predictor--corrector construction on its own wrapped-normal noise ladder.
Each predictor or corrector substep evaluates the shared denoising network
once to obtain all available score predictions.

The lattice uses the ancestral reverse update of its variance-preserving DDPM
process \cite{ho2020ddpm}. Before the posterior mean is formed, each component
of the reconstructed standardized clean lattice state is clipped according to
\begin{equation}
\left|
\hat{x}_{0,i}
\right|
\leq 4,
\end{equation}
where $\hat{x}_{0,i}$ is the reconstructed value of component $i$ of the
six-component clean lattice vector $\hat{\mathbf{x}}_0$. Here $i$ indexes a
lattice-vector component rather than an atom, and $|\cdot|$ denotes absolute
value.

For the independent-angle model, the joint state
\begin{equation}
\left(
\mathbf{x}_t,
\mathbf{F}_t,
\boldsymbol{\Phi}_t
\right)
\end{equation}
is propagated through all $T=1000$ diffusion steps. After the final step,
$\boldsymbol{\Phi}_0$ is discarded and only the generated lattice,
fractional coordinates, atomic species, and atom count are retained.

\paragraph{Candidate selection and relaxation.}
For each conditioning target, 32 candidate crystal structures are generated
and assigned single-point energies by ALIGNN-FF
\cite{choudhary2023alignnff}. The four candidates with the lowest predicted
energies are relaxed over both atomic positions and lattice parameters for at
most 200 optimization steps. Relaxation terminates when the maximum residual
atomic force satisfies
\begin{equation}
f_{\max}
=
0.05\,\mathrm{eV}\,\text{\AA}^{-1},
\end{equation}
where $f_{\max}$ is the maximum force magnitude over all atoms,
$\mathrm{eV}$ denotes electronvolts, and \AA{} denotes angstroms. The relaxed
candidate with the lowest energy per atom is retained.

\paragraph{Crystallographic post-processing and metrics.}
The primary reconstruction pipeline applies crystallographic idealization
using spglib \cite{togo2024spglib} with
\begin{equation}
\mathrm{symprec}=0.1\,\text{\AA},
\end{equation}
where $\mathrm{symprec}$ is the Cartesian distance tolerance used by spglib
when identifying crystallographic symmetry. Primitive-cell reduction is then
applied. This tolerance was not tuned on the validation set. We also report an
unsymmetrized evaluation performed directly after generation and relaxation.

Lattice comparisons are performed after Niggli reduction
\cite{grossekunstleve2004niggli}. Because Niggli reduction selects a
canonical reduced lattice basis through discrete conditions, small
symmetry-breaking perturbations can select different reduced bases and
produce comparatively large changes in the reported lattice angles.

Reconstruction is evaluated using AtomBench
\cite{campbell2026atombench}. The structure-match rate uses the AtomBench
StructureMatcher criterion with
$\mathrm{stol}=0.5$, where $\mathrm{stol}$ is the dimensionless site-position
tolerance used by the structure matcher. Additional metrics are Cartesian
root-mean-square displacement over matched structures, normalized
root-mean-square displacement over matched structures, continuous corrected
RMSE using average-minimum-distance descriptors with $k=100$, mean absolute
error of the three lattice lengths, mean absolute error of the three lattice
angles, and the mean Kullback--Leibler divergence of the six
lattice-parameter distributions. Here $k=100$ denotes the number of
average-minimum-distance descriptor components retained for the continuous
corrected RMSE calculation.

\paragraph{Ablations.}
Six models are compared. They are defined by the combination of one of three
angular training objectives and either the presence or absence of a line
graph. The three angular objectives are as follows:
\begin{enumerate}
    \item \emph{none}: no angular objective; when a line graph is present,
    geometric bond angles appear only as ordinary ALIGNN input features;
    \item \emph{direct}: supervised angular displacement without an
    independently diffused angular state; and
    \item \emph{independent}: an independently diffused bond-angle state with
    wrapped-normal forward and reverse processes.
\end{enumerate}

Models without a line graph use nine pair-graph convolutions. Models with a
line graph use three ALIGNN layers, corresponding to six edge-gated graph
convolutions across the atom and line graphs, followed by three pair-graph
convolutions. All models therefore contain nine convolution operations.
Parameter counts range from $3.751$ to $3.856$ million, a spread of $2.8\%$.
The direct and independent models are parameter matched between their
line-graph and no-line-graph versions, while the two no-angle models differ by
approximately $1\%$. Compute cost is not matched because constructing and
propagating the line graph introduces additional triplet-level operations.

The angular models use the smooth radius-based topology with
$r_c=5$\,\AA{} and the polynomial relevance envelope defined above, whereas
the no-angle models use a hard $k$-nearest-neighbor graph with $k=12$. Here
$k=12$ means that twelve neighbors are retained for each atom. The smooth
topology is required because discontinuous changes in neighbor identity would
make the supervised triplet field discontinuous. Comparisons between no-angle
and angular models therefore include both the change in angular treatment and
the associated change in graph topology. The direct-versus-independent
comparison holds the topology fixed and isolates the effect of independently
diffusing the angular state. The effect of smooth topology alone is measured
separately.

\paragraph{Training protocol.}
The JARVIS-DFT Supercon-3D dataset contains 847 training, 105 validation, and
103 test structures. Alexandria DS-A/DS-B contains 6603 training, 825
validation, and 825 test structures. Both use the AtomBench splits with split
seed 123 and are stored in the primitive Niggli scoring basis.

All denoisers use hidden dimension 256 and $T=1000$ diffusion steps. Pair
distances are expanded using 64 radial basis functions extending to
$10$\,\AA{}. Fractional displacements use Fourier orders
$k=1,\ldots,10$, where $k$ again denotes Fourier harmonic order. Geometric
angles in the direct model use 40 cosine radial basis functions. The
independent angular state uses harmonic orders $k=1,\ldots,8$. The
fractional-coordinate score uses $C=32$ learned score channels.

Optimization uses AdamW \cite{loshchilov2019adamw} with peak learning rate
$10^{-3}$, weight decay $10^{-5}$, batch size 64, and gradient clipping at
$1.0$. A one-cycle learning-rate schedule
\cite{smith2019superconvergence} is used with a $5\%$ warm-up fraction. An
exponential moving average of the network parameters with decay $0.999$ is
used for sampling. No structural data augmentation is applied.

JARVIS models are trained for 3000 epochs and Alexandria models for 1000
epochs. Because only four of the six model configurations contain an angular
objective, all checkpoints are selected using the validation structural loss
\begin{equation}
\mathcal{L}_{\mathrm{struct}}
=
\mathcal{L}_L
+
10\mathcal{L}_F,
\end{equation}
where $\mathcal{L}_{\mathrm{struct}}$ is the checkpoint-selection loss,
$\mathcal{L}_L$ is the lattice loss defined above, and $\mathcal{L}_F$ is the
fractional-coordinate loss defined above. This quantity is available for all
six models regardless of their angular objective.

All twelve models across the two datasets are trained from random seed 0 on
one NVIDIA GB10 GPU on the same compute node. The results therefore provide a
controlled comparison at one training seed rather than an estimate of
run-to-run variance. This limitation is most important for the 103-structure
JARVIS test set; the 825-structure Alexandria test set provides greater
resolution for reconstruction metrics.

Rare sampling trajectories that leave the real-valued numerical domain are
removed individually. Approximately one trajectory in $10^4$ diverges
numerically, where $10^4$ denotes ten thousand trajectories. Five candidates
out of 26,400 were removed in the affected model configuration.

All reported reconstruction metrics are measured after the complete
32-candidate generation, ALIGNN-FF prescreening, and relaxation procedure and
therefore characterize the combined generator--force-field search procedure.

\bigskip
\noindent\textit{Use of AI tools.} During the preparation of this work, the author(s) used large language models (LLMs) to assist with drafting and editing portions of this manuscript. After using this tool/service, the author(s) reviewed and edited the content as needed and take(s) full responsibility for the content of the published article.

\section*{Acknowledgements}
This work was supported by the National Science Foundation under Award
No.~2607469 and the U.S.\ Department of Energy under Award No.~DE-SC0026725.

\bibliographystyle{unsrt}
\bibliography{references}

\clearpage
\section*{Supplementary Information}
\setcounter{figure}{0}
\renewcommand{\thefigure}{S\arabic{figure}}
\setcounter{table}{0}
\renewcommand{\thetable}{S\arabic{table}}

\begin{scriptsize}
\setlength{\tabcolsep}{3pt}
\begin{longtable}{>{\raggedright\arraybackslash}p{3.6cm}cccc}
\caption{\textbf{Graph-and-recipe ablation for JARVIS DFT-3D single-property
prediction} (test MAE), expanding the ALIGNN~2.0 column of Table~\ref{tab:single}.
ALIGNN~2.0 on the radius graph, on the \SI{8}{\angstrom} kNN graph, and the improved recipe
(EMA weight averaging) on the kNN graph, against the original
ALIGNN. \best{Bold}: row best among the three ALIGNN~2.0 variants. ``--'' marks a variant
not run.}
\label{tab:single_ablation}\\
\toprule
Task (unit) & 2.0 (radius) & 2.0 (kNN) & 2.0 (kNN, improved) & orig.\ ALIGNN \\
\midrule
\endfirsthead
\toprule
Task (unit) & 2.0 (radius) & 2.0 (kNN) & 2.0 (kNN, improved) & orig.\ ALIGNN \\
\midrule
\endhead
1) formation\_energy (eV/atom) & 0.0316 & 0.0307 & \best{0.0284} & 0.0331 \\
2) optb88vdw\_total\_energy (eV/atom) & 0.0321 & 0.0314 & \best{0.0297} & 0.0367 \\
3) optb88vdw\_bandgap (eV) & 0.1314 & 0.1306 & \best{0.1245} & 0.1423 \\
4) mbj\_bandgap (eV) & 0.2721 & 0.2730 & \best{0.2576} & 0.3104 \\
5) ehull (eV/atom) & 0.0576 & 0.0590 & \best{0.0508} & 0.0763 \\
6) bulk\_modulus\_kv (GPa) & 9.885 & \best{9.302} & 9.497 & 10.399 \\
7) shear\_modulus\_gv (GPa) & 9.063 & \best{8.825} & 9.208 & 9.476 \\
8) magmom\_oszicar ($\mu_B$) & 0.2608 & \best{0.2567} & 0.2622 & 0.2574 \\
9) slme (\%) & 4.493 & \best{4.447} & 4.504 & 4.521 \\
10) spillage & 0.3527 & \best{0.3456} & 0.3499 & 0.3510 \\
11) kpoint\_length\_unit (\AA) & 9.699 & 9.342 & \best{9.294} & 9.515 \\
12) encut (eV) & 131.81 & 128.08 & \best{125.46} & 133.80 \\
13) epsx & 20.705 & 20.139 & \best{19.853} & 20.394 \\
14) epsy & 20.088 & 19.829 & \best{19.352} & 19.999 \\
15) epsz & 19.633 & \best{19.453} & 19.503 & 19.568 \\
16) mepsx & 24.646 & \best{23.847} & 24.148 & 24.046 \\
17) mepsy & \best{23.823} & 24.044 & 23.840 & 23.648 \\
18) mepsz & \best{23.247} & 23.531 & 23.572 & 23.731 \\
19) dfpt\_piezo\_max\_dij (pC/N) & 12.603 & \best{12.498} & 13.426 & 20.570 \\
20) dfpt\_piezo\_max\_dielectric & 26.823 & \best{24.305} & 25.175 & 28.151 \\
21) exfoliation\_energy (meV/atom) & 40.272 & \best{37.628} & 39.350 & 52.703 \\
22) max\_efg ($10^{21}$V/m$^2$) & 19.802 & 19.248 & \best{18.834} & 19.121 \\
23) avg\_elec\_mass ($m_e$) & 0.0837 & 0.0810 & \best{0.0797} & 0.0853 \\
24) avg\_hole\_mass ($m_e$) & 0.1299 & 0.1240 & \best{0.1196} & 0.1239 \\
25) n\_Seebeck ($\mu$V/K) & 41.524 & \best{40.346} & 41.454 & 40.921 \\
26) n\_powerfact ($\mu$W/mK$^2$) & 469.07 & \best{451.90} & 482.84 & 442.30 \\
27) ph\_heat\_capacity (J/mol/K) & 9.577 & -- & \best{8.468} & 9.606 \\
\bottomrule
\end{longtable}
\end{scriptsize}

\begin{scriptsize}
\setlength{\tabcolsep}{3pt}
\begin{longtable}{>{\raggedright\arraybackslash}p{4.2cm}ccc}
\caption{\textbf{Graph-and-recipe ablation for the additional datasets} (test MAE),
expanding the ALIGNN~2.0 column of Table~\ref{tab:other}. \best{Bold}: row best among
the three variants; ``--'' marks a variant not run. On these smaller and more diverse
datasets the base kNN model is generally strongest, whereas the improved training recipe helps most on
the large JARVIS DFT-3D tasks of Table~\ref{tab:single_ablation}.}
\label{tab:other_ablation}\\
\toprule
Task (unit) & 2.0 (radius) & 2.0 (kNN) & 2.0 (kNN, improved) \\
\midrule
\endfirsthead
\toprule
Task (unit) & 2.0 (radius) & 2.0 (kNN) & 2.0 (kNN, improved) \\
\midrule
\endhead
1) Thermal conductivity (log$_{10}\kappa_L$, OQMD) & 0.386 & 0.375 & \best{0.194} \\
2) QMOF bandgap (eV) & 0.208 & -- & \best{0.202} \\
3) Tc\_supercon (K) & 1.637 & \best{1.490} & 1.997 \\
4) Tc\_supercon\_hydride (K) & 9.937 & \best{9.425} & -- \\
5) Tc\_supercon\_\allowbreak hydride\_plus\_bulk (K) & 8.670 & \best{8.407} & -- \\
6) alex\_supercon Tc (K) & 0.883 & 0.864 & \best{0.850} \\
7) alex\_supercon $N(E_F)$ (states/eV) & 0.821 & \best{0.791} & 0.800 \\
8) alex\_supercon $\theta_D$ (K) & 11.33 & \best{10.68} & -- \\
9) alex\_supercon $\lambda$ & 0.0707 & \best{0.0679} & 0.0693 \\
10) alex\_supercon $\omega_{\log}$ (K) & 20.31 & \best{20.08} & -- \\
11) mxene275, formation energy (eV/atom) & 0.0348 & 0.0343 & \best{0.0312} \\
12) polymer\_genome, GGA gap (eV) & 0.2274 & \best{0.2273} & -- \\
13) c2db, band gap (eV) & 0.0971 & \best{0.0802} & 0.0980 \\
14) twod\_matpd, band gap (eV) & 0.3802 & \best{0.3660} & -- \\
15) omdb, band gap (eV) & 0.2428 & \best{0.2411} & -- \\
16) hMOF, CO$_2$ uptake (mol/kg) & -- & \best{0.4687} & -- \\
17) QM9 HOMO--LUMO gap (eV) & \best{0.031} & -- & 0.0345 \\
\bottomrule
\end{longtable}
\end{scriptsize}

\begin{table}[htbp]\centering\small
\caption{\textbf{WBM formation-energy MAE (eV/atom), full coverage.} Formation-energy
MAE for every force field in Table~\ref{tab:ff}, scored with per-model
self-consistent elemental references. ``84-element subset'' is the common
$\sim$222{,}000-material set that excludes the five heavy/$f$-electron/alkali-earth
elements Sr, Cs, Gd, Th, Np (whose model elemental references complete the periodic
table); ``full set'' is the complete $256{,}963$-material WBM set (identical to the
public Matbench-Discovery coverage). The external baselines are essentially
unchanged between the two, whereas the ALIGNN-FF potentials degrade on the five
added elements, which are under-represented in their MATPES/JARVIS-FF training data.
This is a coverage limitation, not a scoring artifact (the subset numbers reproduce
exactly, and the stability F1/DAF of Table~\ref{tab:ff} remain robust).}
\label{tab:wbm_eform}
\begin{tabular}{l c c}
\toprule
Model & $E_f$ MAE (84-element subset) & $E_f$ MAE (full set) \\
\midrule
AFF MATPES-R2SCAN & 0.163 & 0.243 \\
AFF MATPES-PBE    & 0.147 & 0.183 \\
AFF JV-DFT-DB1  & 0.126 & 0.163 \\
AFF JV-DFT-DB2  & 0.147 & 0.182 \\
AFF MPtrj               & 0.131 & 0.174 \\
\midrule
UMA (uma-s-1p1)         & 0.120 & 0.125 \\
M3GNet (MatPES-PBE)     & 0.128 & 0.129 \\
MACE-MP-0               & 0.139 & 0.143 \\
CHGNet                  & 0.097 & 0.098 \\
\bottomrule
\end{tabular}
\end{table}

\begin{figure}[htbp]
\centering
\includegraphics[width=\textwidth]{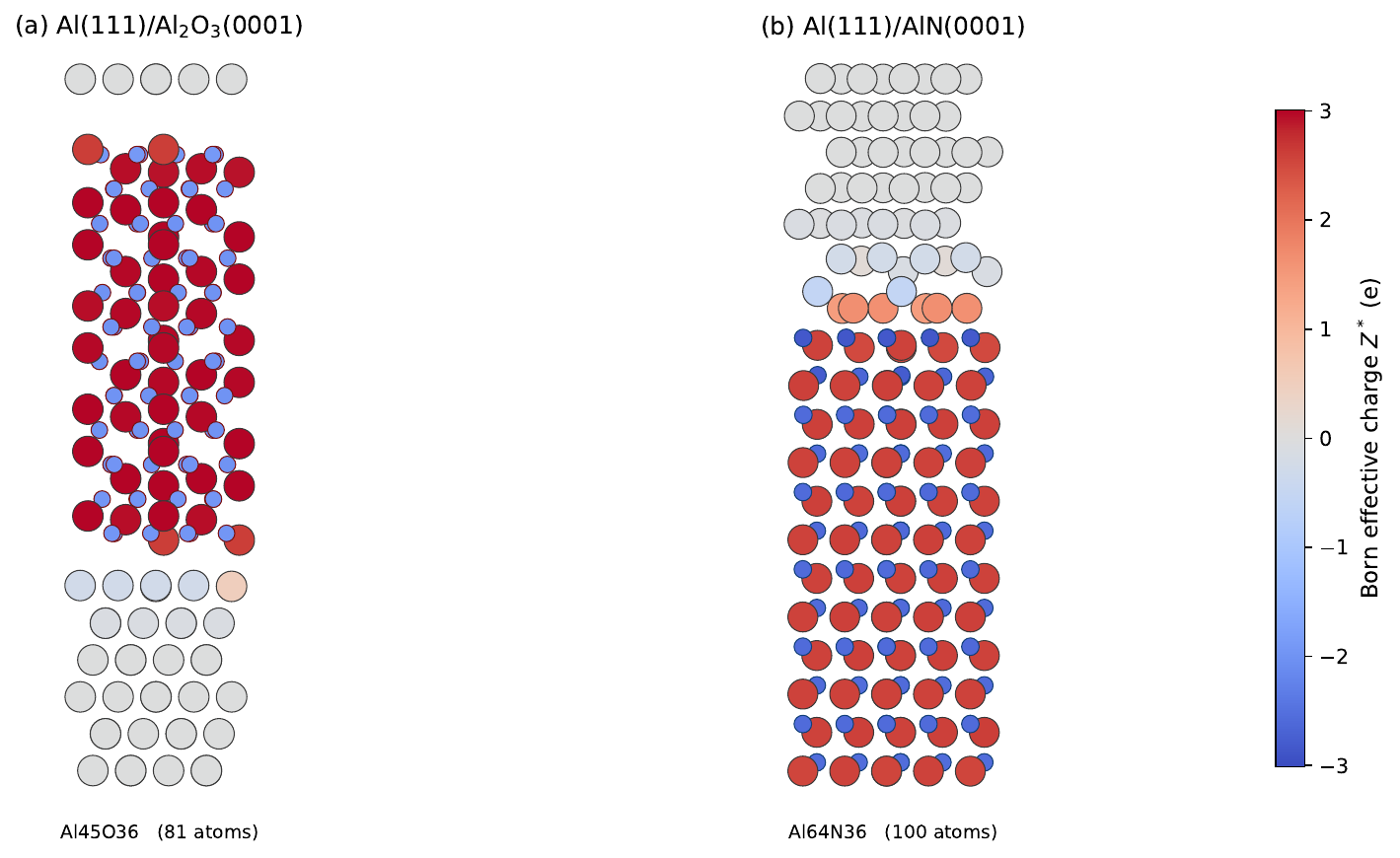}
\caption{\textbf{Metal-ceramic interfaces from ALIGNN-FF, coloured by predicted Born
effective charge.} (a) Al(111)/Al$_2$O$_3$(0001) and (b) Al(111)/AlN(0001), assembled
with InterMat~\cite{choudhary2024intermat}, relaxed with the MatPES-PBE ALIGNN-FF, and
coloured by the per-atom dynamical charge $Z^{*}=\mathrm{tr}(\mathbf{Z}^{*})/3$ from the
ALIGNN 2.0 Born-tensor model (shared colour scale). The metallic Al slab is near-neutral;
the ceramic cation Al is positive and the O/N anions negative, with charge transfer
localised to the interfacial planes. No interface-specific fitting was used; models are
loaded from the public pretrained registry.}
\label{fig:interface}
\end{figure}

\begin{figure}[htbp]
\centering
\includegraphics[width=\textwidth]{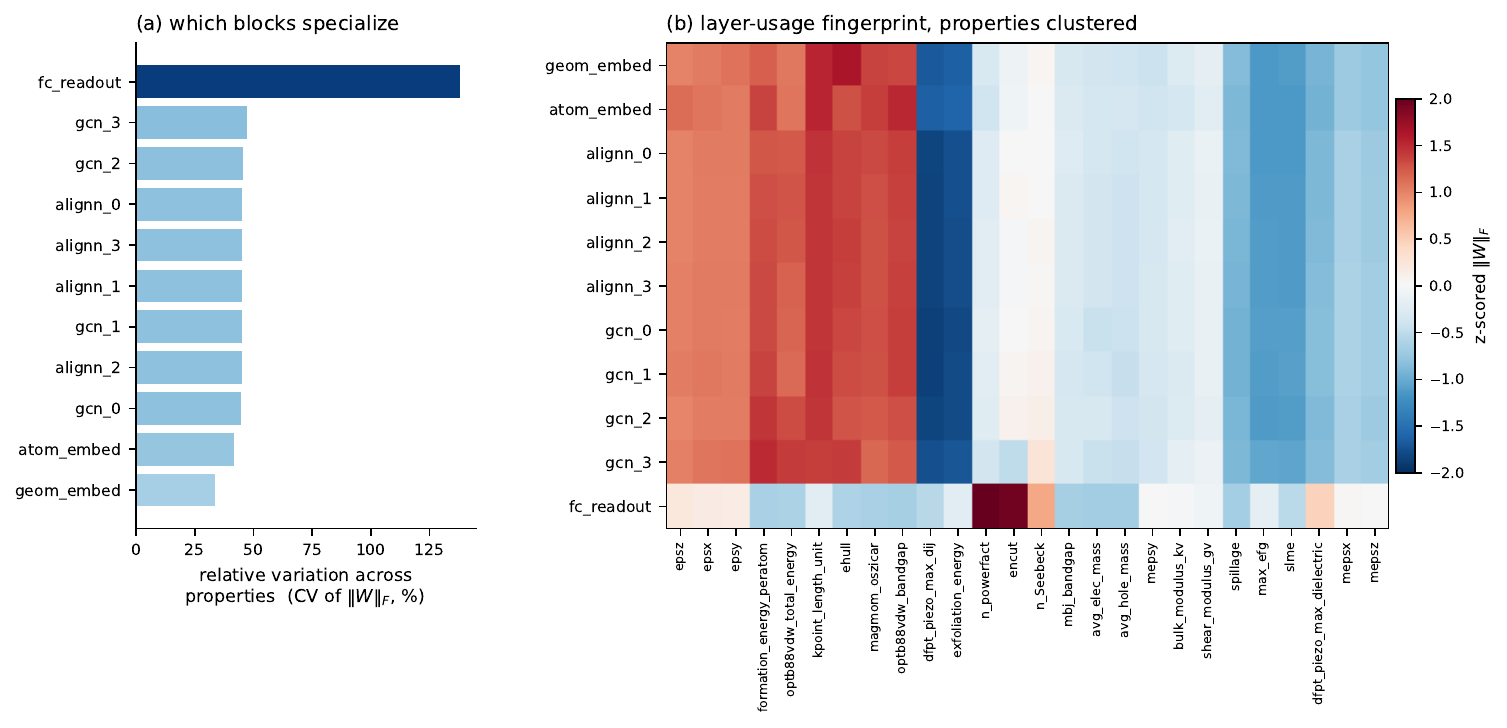}
\caption{\textbf{Where ALIGNN 2.0 specializes across properties.}
(a) Relative variation (coefficient of variation of the permutation-invariant
Frobenius norm $\|W\|_F$) of each architecture block across the twenty-six radius
property models: the read-out layer adapts far more than the message-passing
backbone, and the radial/angular geometry encoders are the most universal.
(b) Per-block weight fingerprint ($z$-scored $\|W\|_F$) with properties ordered by
hierarchical clustering. Physically related properties, dielectric-tensor
components, elastic moduli, and carrier effective masses, share similar
fingerprints.}
\label{fig:interp}
\end{figure}

\begin{figure}[htbp]
\centering
\includegraphics[width=\textwidth]{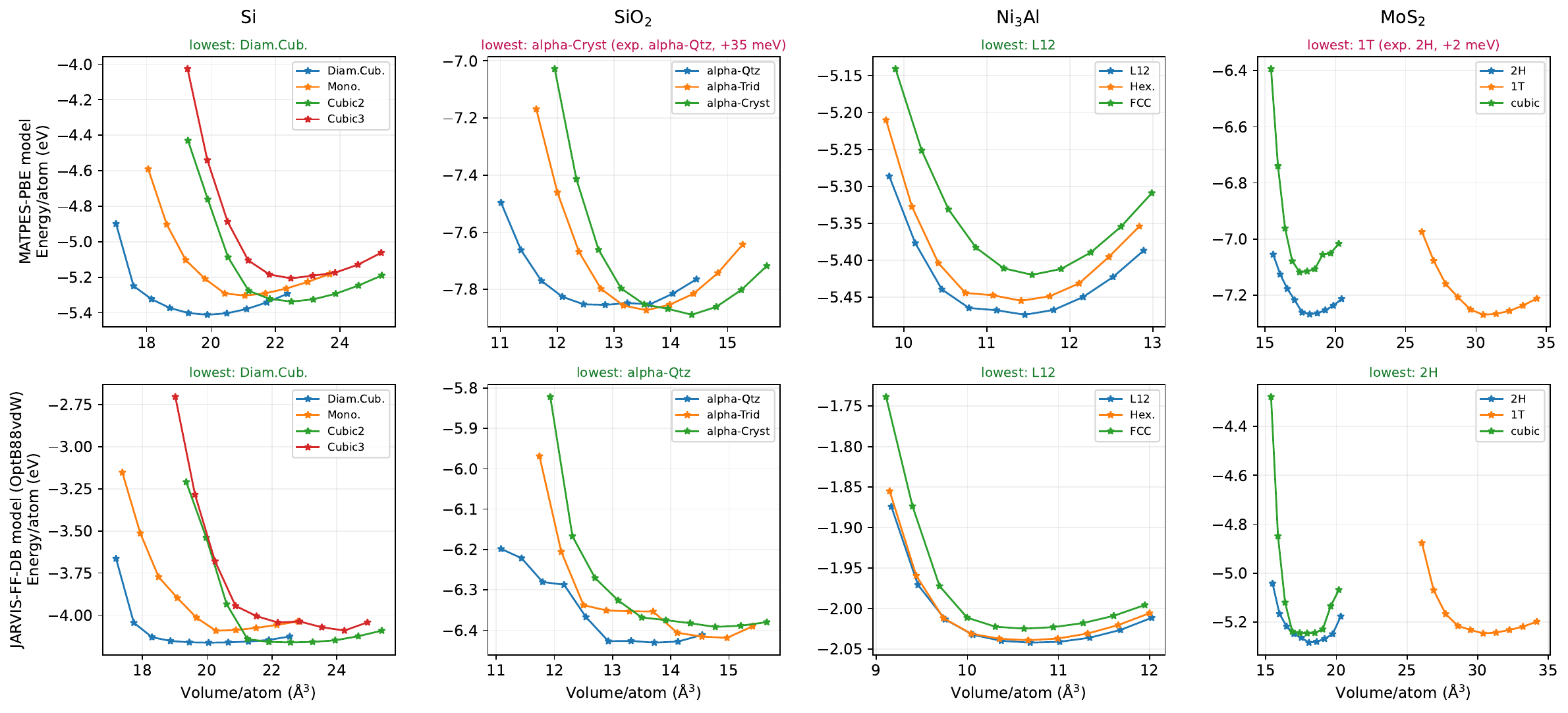}
\caption{Energy--volume curves for competing polymorphs of four materials under two
ALIGNN 2.0 force fields (rows). Each panel labels the lowest-lying polymorph and the
margin where it differs from the expected ground state. The MATPES-PBE potential
(top) reproduces PBE's quartz--cristobalite inversion and a near-degenerate,
delaminating 1T-MoS$_2$ (missing interlayer dispersion). The OptB88vdW potential
(bottom) recovers all four ground states. Per-row energy references differ;
within-family orderings do not.}
\label{fig:ev}
\end{figure}

\begin{figure}[htbp]
\centering
\includegraphics[width=0.62\linewidth]{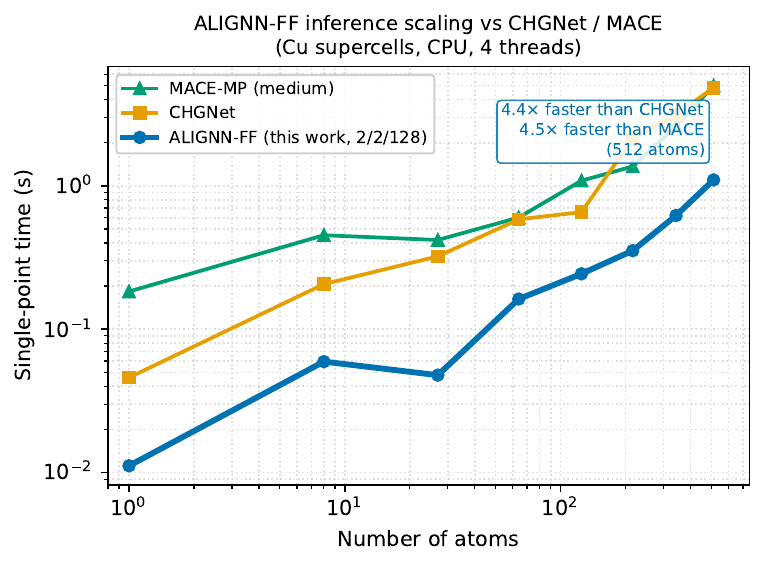}
\caption{Inference cost of the ALIGNN 2.0 force field versus CHGNet and MACE-MP-0.
Single-point energy-and-force evaluation wall-time on cubic Cu supercells
($1$--$512$ atoms, CPU, $4$ threads). The compact force-field-compatible model
(two ALIGNN and two GCN layers, $128$ hidden features, smooth cutoff, trained on
MATPES) is the fastest at every size and ${\sim}4.5\times$ faster than both
baselines at $512$ atoms.}
\label{fig:scaling}
\end{figure}

\begin{figure*}[htbp]
\centering
\includegraphics[width=\textwidth]{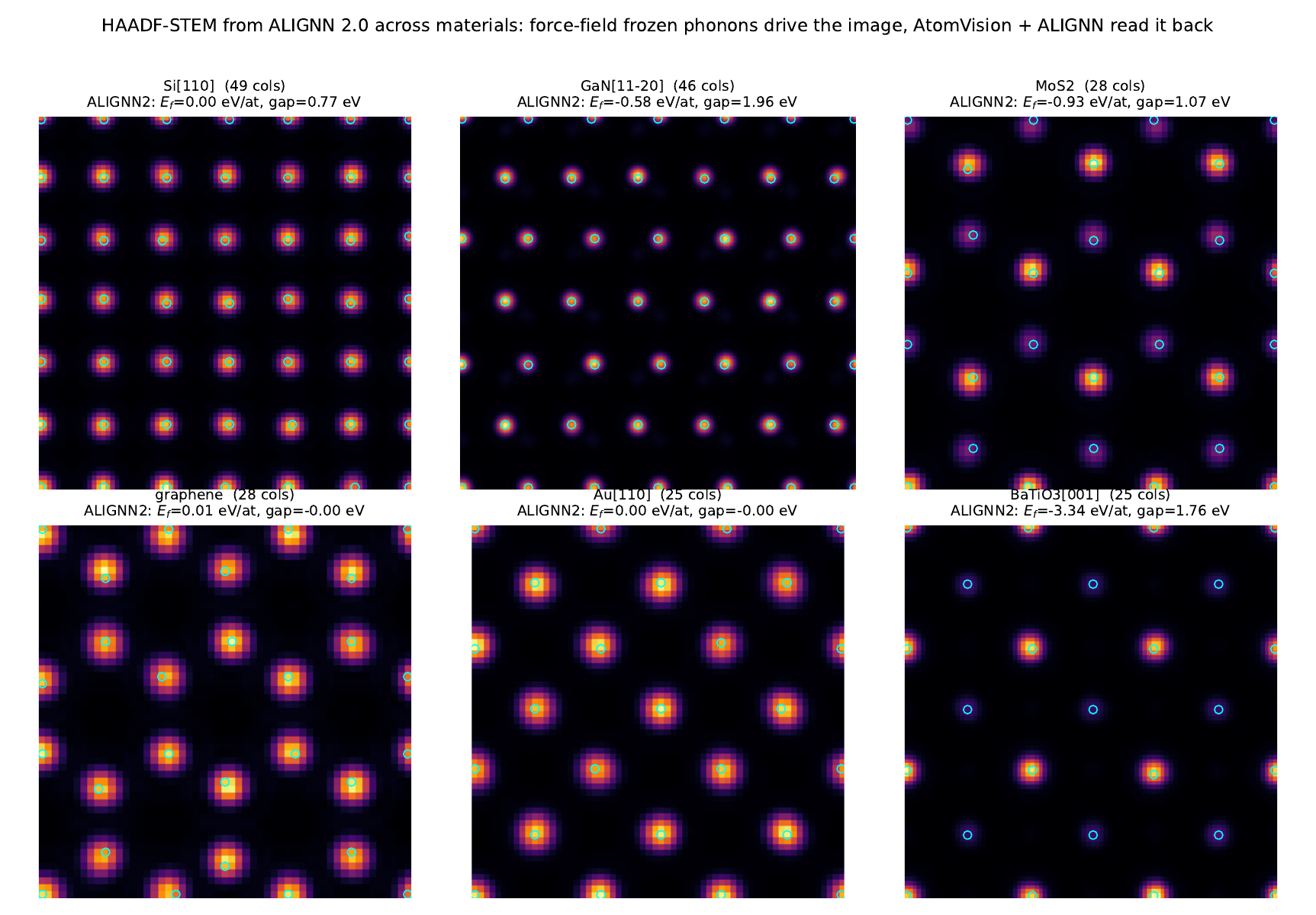}
\caption{\textbf{Force-field-driven HAADF-STEM across materials with ALIGNN 2.0.}
The single-framework microscopy pipeline of Fig.~\ref{fig:microscopy}, applied to six
materials spanning covalent (Si$[110]$), wide-gap semiconductor (GaN$[11\bar{2}0]$),
2D (monolayer MoS$_2$, graphene), close-packed metal (Au$[110]$), and perovskite
(BaTiO$_3[001]$) structures. For each, the ALIGNN-FF (MATPES-R2SCAN) relaxes the cell
and supplies per-element $300$\,K thermal (frozen-phonon) displacements that drive an
abTEM multislice HAADF-STEM simulation. AtomVision then detects the atomic columns
(cyan markers, count in each title) and ALIGNN 2.0 predicts the formation energy and
OptB88vdW band gap (title) from the same relaxed structure. One model set closes
the simulate$\rightarrow$image$\rightarrow$read-back loop across bonding types.}
\label{fig:microscopy_grid}
\end{figure*}

\begin{figure}[htbp]
\centering
\includegraphics[width=\textwidth]{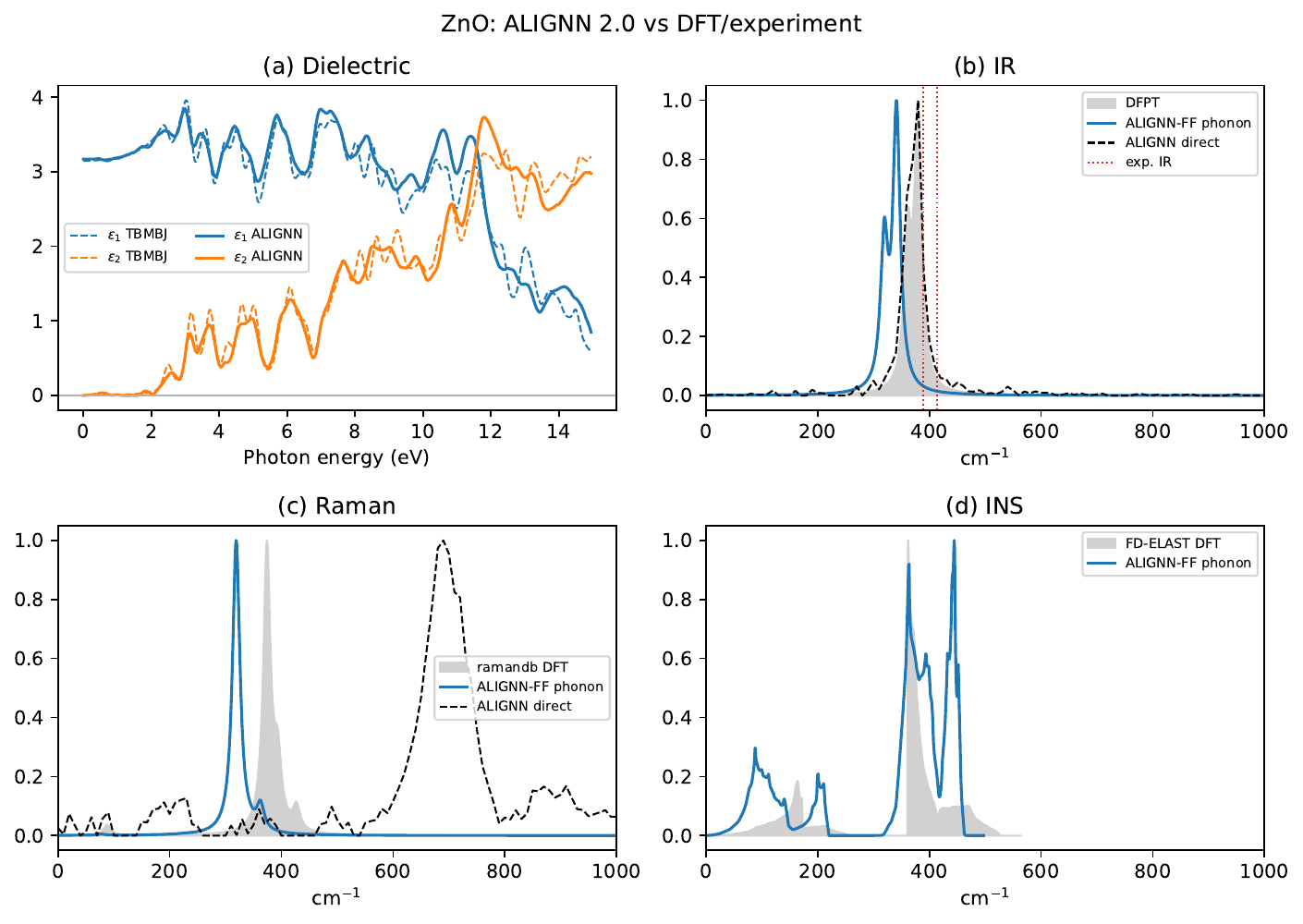}
\caption{Spectroscopic validation for ZnO (wurtzite; IR JVASP-1195, Raman mp-2133),
panels as in Fig.~\ref{fig:spectra}: (a) optical dielectric vs TBmBJ, (b) IR vs DFPT
and experiment, (c) Raman vs the JARVIS DFT Raman database, (d) INS vs
finite-displacement DFT, all from a single relaxed structure with ALIGNN 2.0.}
\label{fig:spec_zno}
\end{figure}

\begin{figure}[htbp]
\centering
\includegraphics[width=\textwidth]{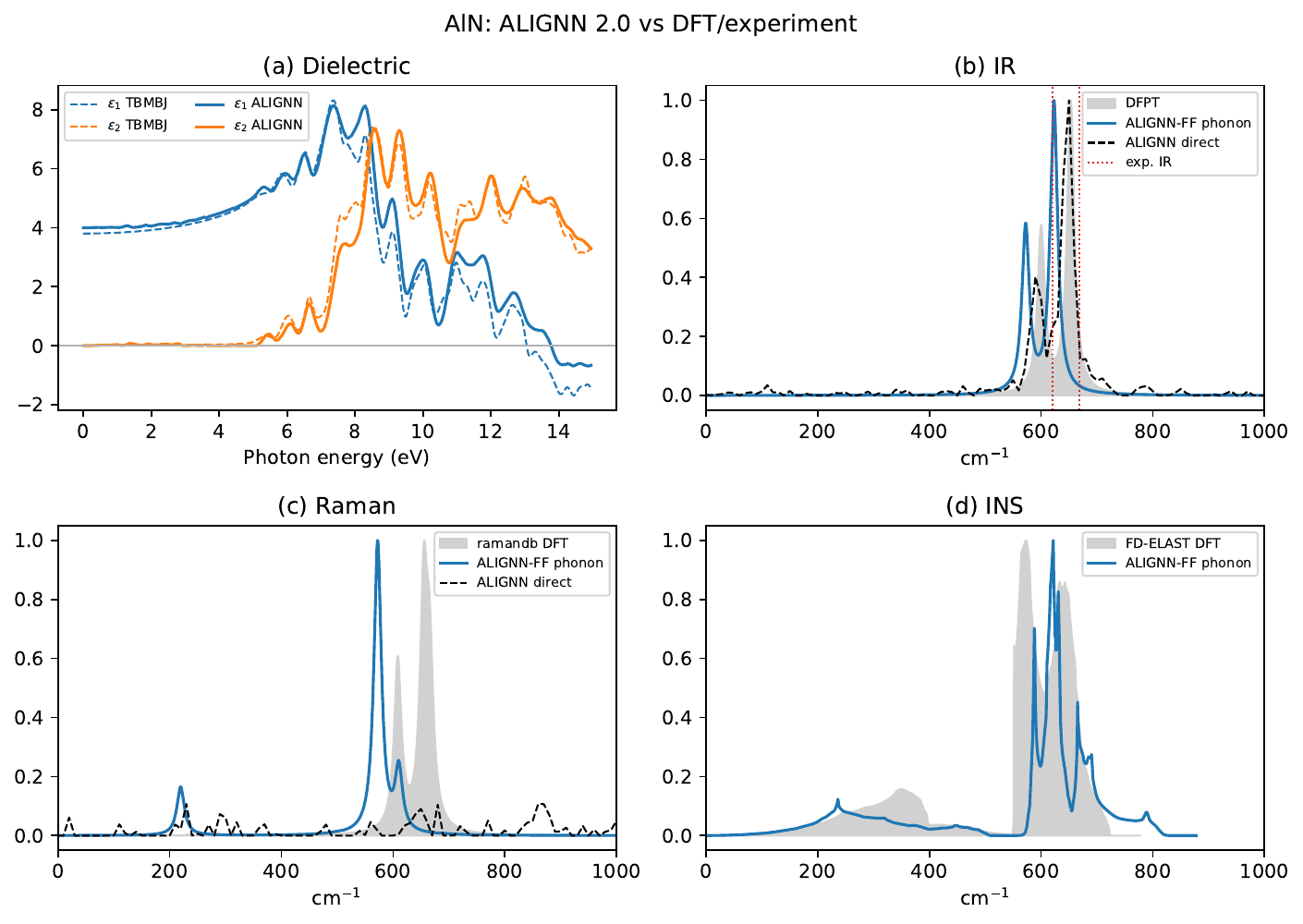}
\caption{Spectroscopic validation for AlN (wurtzite; IR JVASP-39, Raman mp-661),
panels as in Fig.~\ref{fig:spectra}.}
\label{fig:spec_aln}
\end{figure}

\begin{figure}[htbp]
\centering
\includegraphics[width=\textwidth]{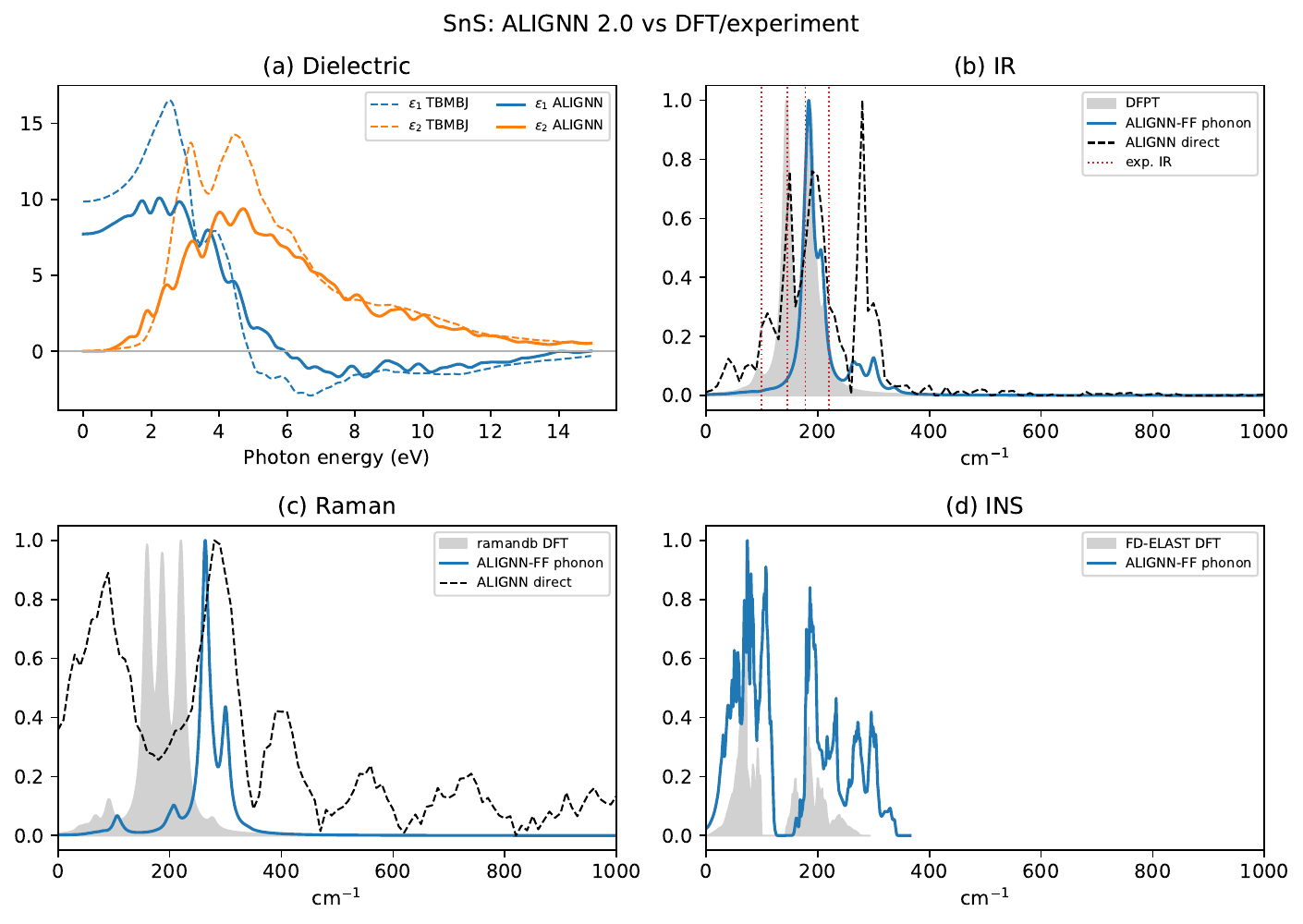}
\caption{Spectroscopic validation for SnS (\emph{Pnma}; IR JVASP-1109, Raman
mp-2231), panels as in Fig.~\ref{fig:spectra}.}
\label{fig:spec_sns}
\end{figure}

\begin{figure}[htbp]
\centering
\includegraphics[width=\textwidth]{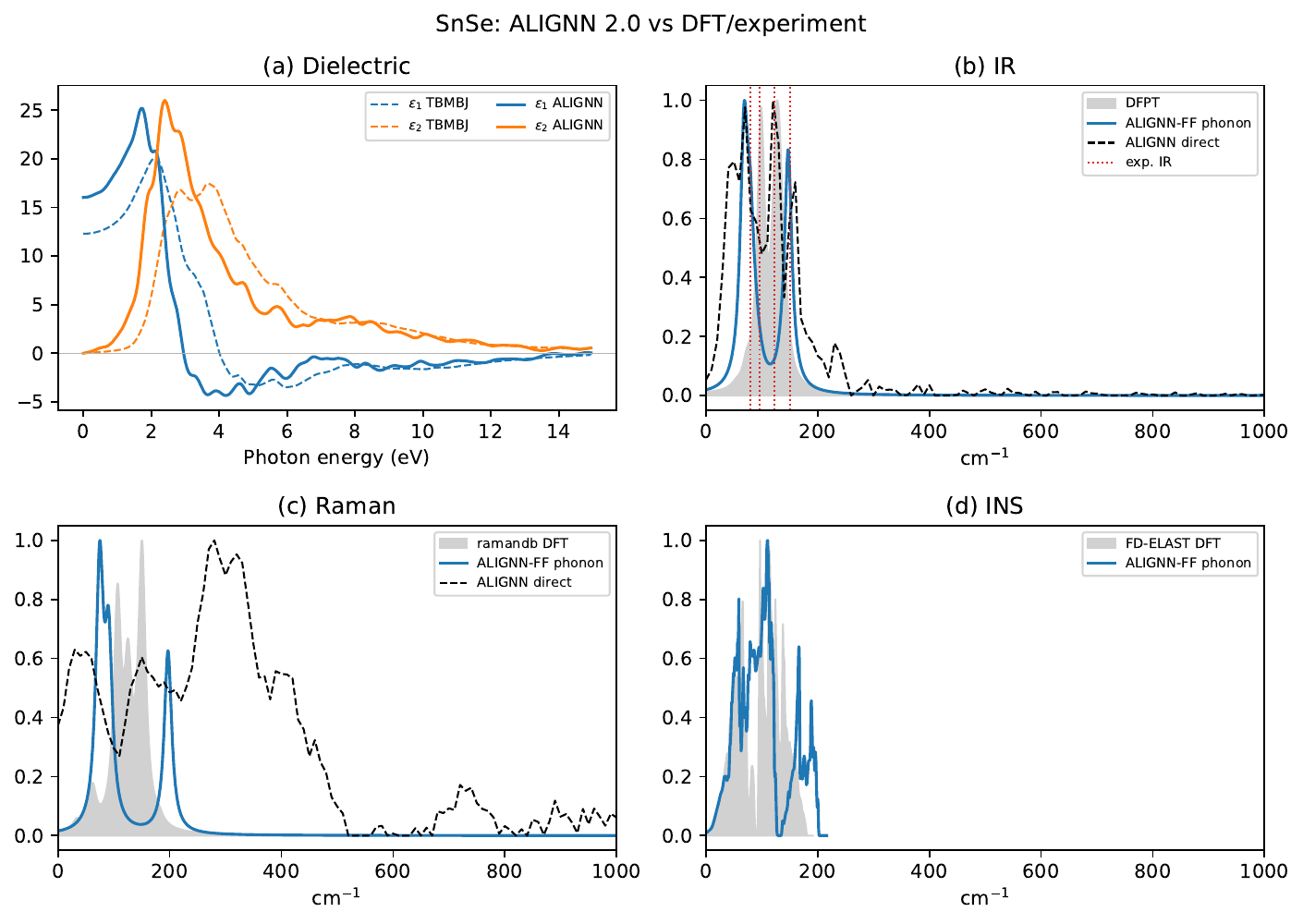}
\caption{Spectroscopic validation for SnSe (\emph{Pnma}; IR JVASP-299, Raman
mp-691), panels as in Fig.~\ref{fig:spectra}.}
\label{fig:spec_snse}
\end{figure}

\begin{figure}[htbp]
\centering
\includegraphics[width=\textwidth]{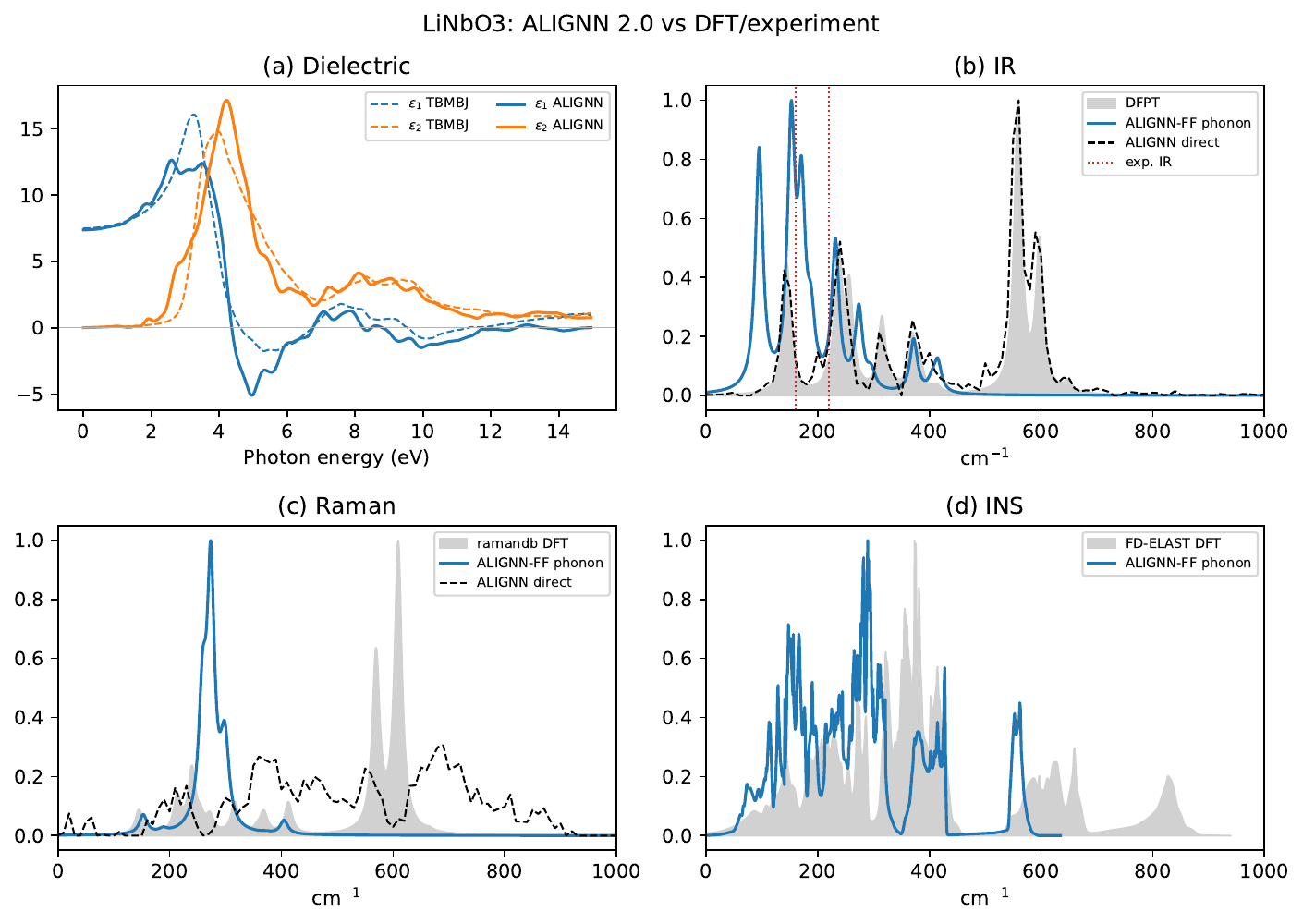}
\caption{Spectroscopic validation for LiNbO$_3$ (\emph{R3c}; IR JVASP-1240, Raman
mp-3731), panels as in Fig.~\ref{fig:spectra}.}
\label{fig:spec_linbo3}
\end{figure}

\begin{figure}[htbp]
\centering
\includegraphics[width=\textwidth]{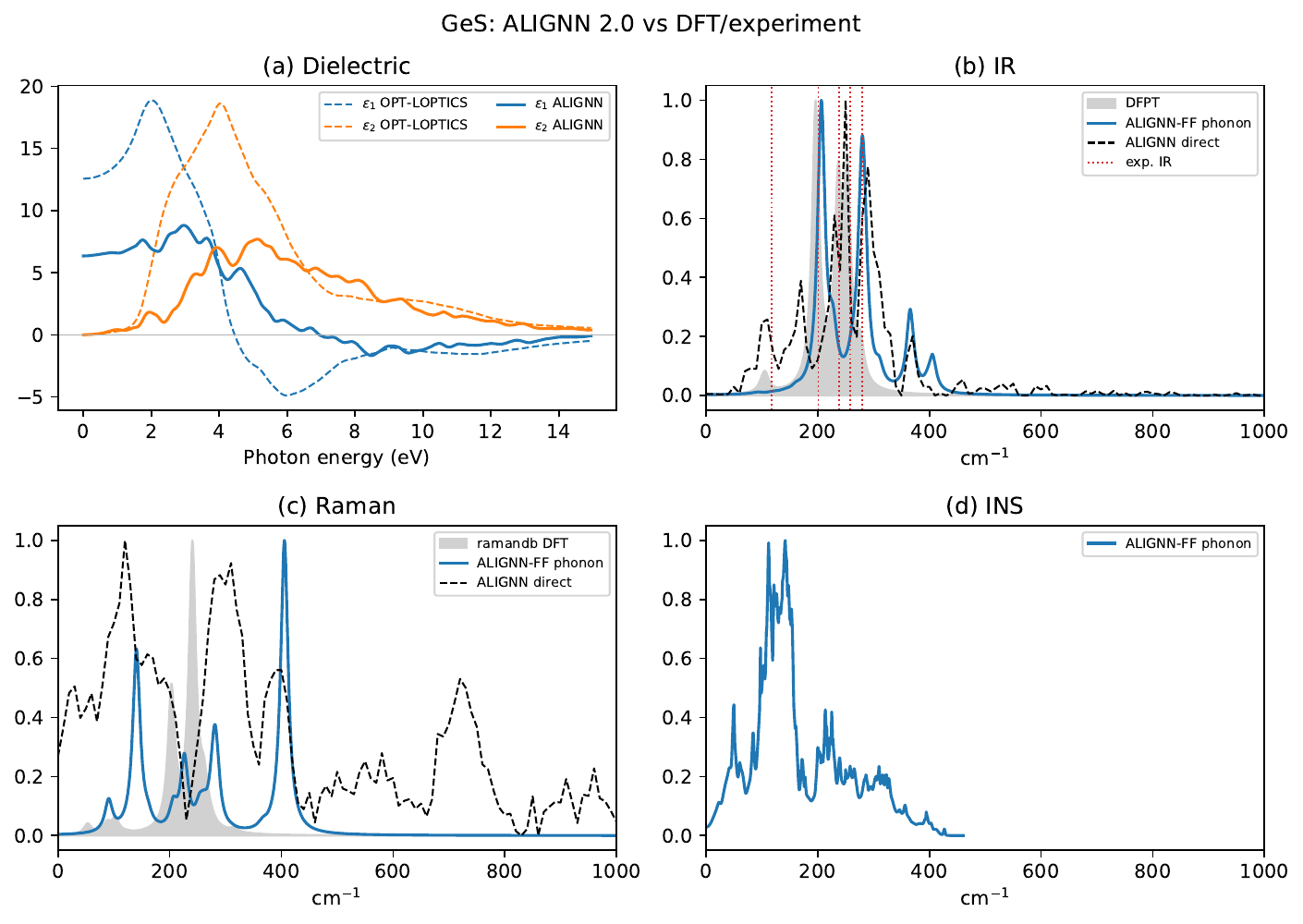}
\caption{Spectroscopic validation for GeS (\emph{Pnma}; IR JVASP-2169, Raman
mp-2242), panels as in Fig.~\ref{fig:spectra}.}
\label{fig:spec_ges}
\end{figure}

\begin{figure}[htbp]
\centering
\includegraphics[width=\textwidth]{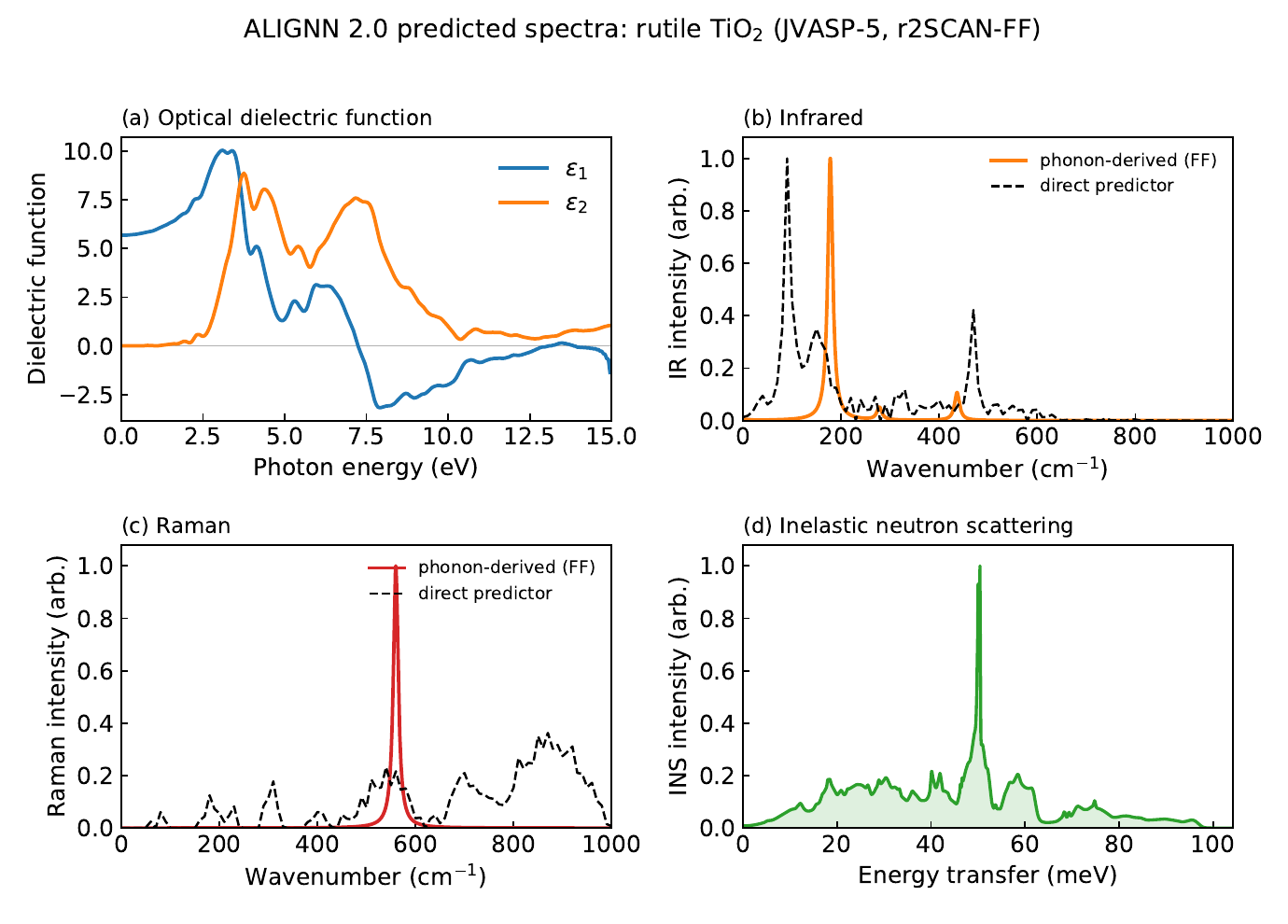}
\caption{Spectroscopic showcase for rutile TiO$_2$ (JVASP-5) predicted by ALIGNN 2.0
from a single relaxed structure: (a) optical dielectric ($\varepsilon_2$ from TBmBJ,
$\varepsilon_1$ from Kramers--Kronig); (b) infrared and (c) Raman computed two
independent ways, from r2SCAN ALIGNN-FF phonons (solid) and the direct end-to-end
predictor (dashed), agreeing on the principal E$_u$ IR band near $180$~cm$^{-1}$ and
resolving the E$_g$/A$_{1g}$ Raman lines at $\sim\!400$/$560$~cm$^{-1}$; and
(d) inelastic neutron scattering from the same force constants.}
\label{fig:spec_tio2}
\end{figure}

\begin{table}[htbp]
\centering
\caption{\textbf{Interface work of adhesion from the ALIGNN~2.0 force field (MATPES).}
Predicted $W_{\mathrm{ad}}=(E_{\mathrm{slab1}}+E_{\mathrm{slab2}}-E_{\mathrm{interface}})/A$,
with all three cells relaxed and the two slabs obtained by a self-consistent split of the
relaxed interface (same lateral cell and atom set), for lattice-matched interfaces spanning
metal, ceramic, semiconductor and 2D chemistries built with the \textsc{Intermat} protocol~\cite{choudhary2024intermat}.
Positive $W_{\mathrm{ad}}$ denotes adhesion; values are single lowest-energy stackings without
lateral-registry optimization, so weakly registered contacts can read near zero or slightly
negative.}
\label{tab:interface}
\setlength{\tabcolsep}{7pt}
\begin{tabular}{l c}
\toprule
Interface (film / substrate, Miller) & $W_{\mathrm{ad}}$ (J/m$^2$) \\
\midrule
\multicolumn{2}{l}{\emph{Metal\,/\,metal}} \\
\quad Cu\,(111)\,/\,W\,(001) & +6.61 \\
\quad Al\,(111)\,/\,W\,(001) & +7.15 \\
\multicolumn{2}{l}{\emph{Metal\,/\,nitride}} \\
\quad Cu\,(111)\,/\,TiN\,(111) & +1.79 \\
\quad Al\,(111)\,/\,TiN\,(111) & +1.45 \\
\quad Ti\,(001)\,/\,TiN\,(001) & +5.39 \\
\multicolumn{2}{l}{\emph{Metal\,/\,oxide}} \\
\quad Al\,(111)\,/\,Al$_2$O$_3$\,(001) & -0.06 \\
\quad Cu\,(111)\,/\,SiO$_2$\,(101) & +1.87 \\
\quad Al\,(111)\,/\,SiO$_2$\,(101) & +0.73 \\
\quad Cu\,(001)\,/\,ZrO$_2$\,(110) & +1.38 \\
\quad Cu\,(111)\,/\,Cu$_2$O\,(111) & +2.00 \\
\multicolumn{2}{l}{\emph{Ceramic\,/\,ceramic}} \\
\quad AlN\,(100)\,/\,Al$_2$O$_3$\,(001) & +0.03 \\
\quad TiO$_2$\,(001)\,/\,TiN\,(001) & +2.13 \\
\multicolumn{2}{l}{\emph{Semiconductor\,/\,oxide}} \\
\quad Si\,(111)\,/\,SiO$_2$\,(001) & +1.12 \\
\multicolumn{2}{l}{\emph{2D\,/\,3D}} \\
\quad MoS$_2$\,(001)\,/\,Al$_2$O$_3$\,(001) & -1.02 \\
\quad graphene\,(001)\,/\,SiO$_2$\,(101) & +0.57 \\
\bottomrule
\end{tabular}
\end{table}

\subsection*{Full results table}

\begin{scriptsize}
\setlength{\tabcolsep}{2.5pt}
\begin{longtable}{>{\raggedright\arraybackslash}p{3.15cm}ccccccc}
\caption{Unified ALIGNN~2.0 results (JARVIS-Leaderboard protocol), three sections;
rows are numbered for reference. \textbf{(a) Single-property prediction} (test MAE):
ALIGNN~2.0 on radius and \SI{8}{\angstrom} kNN graphs, vs original ALIGNN and CGCNN.
Baseline (MAD) is a predict-the-mean model's test MAE; Skill $=100(1-\mathrm{MAE_{2.0}}/\mathrm{MAD})$
for the better variant. \textbf{(b) Multi-property} (held-out MAE): spectral, per-atom,
tensor outputs; $D$ is output dimension; value in col.\ ``radius''. \textbf{(c) Force
fields}: \texttt{mlearn} per-element energy/force in col.\ ``radius''; large sets give
energy\,/\,force. $^{\dagger}$still training. $^{\ddagger}$\texttt{mlearn} MAE pending
verification (per-atom normalization); mlearn Skill therefore omitted. For the
large force-field sets, Skill is energy-only (the force MAD baseline is near zero,
so force Skill is undefined, shown ``--''). ``--'' otherwise marks a value that is
unavailable or a baseline that is ill-defined (e.g.\ Born, net charge). \best{Bold}:
row best.}
\label{tab:results}\\
\toprule
Task (unit) & $N_{\mathrm{tr}}/N_{\mathrm{val}}/N_{\mathrm{te}}$ & ALIGNN 2.0 & ALIGNN 2.0 & orig.\ & CGCNN & Baseline & Skill \\
 & & (radius) & (kNN) & ALIGNN & & (MAD) & (\%) \\
\midrule
\endfirsthead
\toprule
Task (unit) & $N_{\mathrm{tr}}/N_{\mathrm{val}}/N_{\mathrm{te}}$ & ALIGNN 2.0 & ALIGNN 2.0 & orig.\ & CGCNN & Baseline & Skill \\
 & & (radius) & (kNN) & ALIGNN & & (MAD) & (\%) \\
\midrule
\endhead
\multicolumn{8}{l}{\textbf{(a) Single-property prediction} — test MAE} \\
\midrule
1) formation\_energy (eV/atom) & 44569/5572/5572 & 0.0316 & \best{0.0307} & 0.0331 & 0.0551 & 0.876 & 96.5 \\
2) optb88vdw\_total\_energy (eV/atom) & 44569/5572/5572 & 0.0321 & \best{0.0314} & 0.0367 & 0.0584 & 1.786 & 98.2 \\
3) optb88vdw\_bandgap (eV) & 44569/5572/5572 & 0.1314 & \best{0.1306} & 0.1423 & 0.1857 & 0.999 & 86.9 \\
4) mbj\_bandgap (eV) & 14535/1817/1815 & \best{0.2721} & 0.2730 & 0.3104 & 0.3261 & 1.765 & 84.6 \\
5) QM9 HOMO--LUMO gap (eV) & 110{,}000/10{,}000/10{,}829 & \best{0.031} & & 0.0345 & & 0.834 & 96.3 \\
6) QMOF bandgap (eV) & 16{,}340/2042/2042 & 0.208 & & \best{0.202} & & 0.946 & 78.7 \\
7) ehull (eV/atom) & 44290/5537/5537 & \best{0.0576} & 0.0590 & 0.0763 & 0.0590 & 1.148 & 95.0 \\
8) bulk\_modulus\_kv (GPa) & 15744/1968/1968 & 9.885 & \best{9.302} & 10.399 & 11.015 & 53.76 & 82.7 \\
9) shear\_modulus\_gv (GPa) & 15744/1968/1968 & 9.063 & \best{8.825} & 9.476 & 10.079 & 27.06 & 67.4 \\
10) magmom\_oszicar ($\mu_B$) & 41766/5222/5222 & 0.2608 & \best{0.2567} & 0.2574 & 0.3065 & 1.254 & 79.5 \\
11) slme (\%) & 7250/906/906 & 4.493 & \best{4.447} & 4.521 & 5.014 & 11.21 & 60.3 \\
12) spillage & 9101/1137/1137 & 0.3527 & \best{0.3456} & 0.3510 & 0.3844 & 0.518 & 33.3 \\
13) kpoint\_length\_unit (\AA) & 44313/5540/5539 & 9.699 & \best{9.342} & 9.515 & 9.875 & 17.94 & 47.9 \\
14) encut (eV) & 44308/5539/5539 & 131.81 & \best{128.08} & 133.80 & 134.83 & 262.6 & 51.2 \\
15) epsx & 35592/4449/4449 & 20.705 & \best{20.139} & 20.394 & 22.199 & 57.45 & 64.9 \\
16) epsy & 35592/4449/4449 & 20.088 & \best{19.829} & 19.999 & 21.787 & 57.32 & 65.4 \\
17) epsz & 35592/4449/4449 & 19.633 & \best{19.453} & 19.568 & 21.121 & 55.79 & 65.1 \\
18) mepsx & 13447/1681/1681 & 24.646 & \best{23.847} & 24.046 & 26.929 & 63.39 & 62.4 \\
19) mepsy & 13447/1681/1681 & 23.823 & 24.044 & \best{23.648} & 26.556 & 63.68 & 62.6 \\
20) mepsz & 13447/1681/1681 & \best{23.247} & 23.531 & 23.731 & 26.629 & 60.71 & 61.7 \\
21) dfpt\_piezo\_max\_dij (pC/N) & 2677/334/334 & 12.603 & \best{12.498} & 20.570 & 18.392 & 22.69 & 44.9 \\
22) dfpt\_piezo\_max\_dielectric & 3764/470/470 & 26.823 & \best{24.305} & 28.151 & 30.961 & 43.91 & 44.7 \\
23) exfoliation\_energy (meV/atom) & 650/81/81 & 40.272 & \best{37.628} & 52.703 & 45.762 & 61.03 & 38.3 \\
24) max\_efg ($10^{21}$V/m$^2$) & 9493/1186/1186 & 19.802 & 19.248 & \best{19.121} & 22.957 & 44.46 & 56.7 \\
25) avg\_elec\_mass ($m_e$) & 14114/1764/1764 & 0.0837 & \best{0.0810} & 0.0853 & 0.0921 & 0.225 & 64.1 \\
26) avg\_hole\_mass ($m_e$) & 14114/1764/1764 & 0.1299 & 0.1240 & \best{0.1239} & 0.1406 & 0.399 & 68.9 \\
27) n\_Seebeck ($\mu$V/K) & 18568/2321/2321 & 41.524 & \best{40.346} & 40.921 & 45.660 & 111.5 & 63.8 \\
28) n\_powerfact ($\mu$W/mK$^2$) & 18568/2321/2321 & 469.07 & 451.90 & \best{442.30} & 485.59 & 709.2 & 36.3 \\
29) ph\_heat\_capacity (J/mol/K) & 9644/1205/1205 & \best{9.577} & -- & 9.606 & 12.936 & 40.16 & 76.2 \\
30) Thermal Cond.\ (log$_{10}\kappa_L$) & 3227/403/404 & 0.376 & \best{0.362} & -- & -- & 0.597 & 39.4 \\
31) Tc\_supercon (K) & 556/30/30 & 1.637 & \best{1.490} & 2.032 & -- & 2.723 & 45.3 \\
32) Tc\_supercon\_hydride (K) & 763/95/95 & 9.937 & \best{9.425} & -- & -- & 33.56 & 71.9 \\
33) Tc\_supercon\_\allowbreak hydride\_plus\_bulk (K) & 1595/199/199 & 8.670 & \best{8.407} & -- & -- & 22.33 & 62.3 \\
34) alex\_supercon Tc (K) & 6592/824/825 & 0.883 & \best{0.864} & & & 2.818 & 69.3 \\
35) alex\_supercon $N(E_F)$ (states/eV) & 6592/824/825 & 0.821 & \best{0.791} & & & 1.559 & 49.3 \\
36) alex\_supercon $\theta_D$ (K) & 6592/824/825 & 11.33 & \best{10.68} & & & 80.30 & 86.7 \\
37) alex\_supercon $\lambda$ & 6592/824/825 & 0.0707 & \best{0.0679} & & & 0.194 & 65.0 \\
38) alex\_supercon $\omega_{\log}$ (K) & 6592/824/825 & 20.31 & \best{20.08} & & & 55.37 & 63.7 \\
\midrule
\multicolumn{8}{l}{\textbf{(b) Multi-property} — spectra / per-atom / tensor; held-out MAE (col.\ ``radius'')} \\
\midrule
39) eDOS, electronic DOS ($D{=}300$) & 4103/227/229 & \best{0.0138} & & & & 0.0213 & 35.2 \\
40) pDOS, phonon DOS ($D{=}200$) & 4103/227/229 & 0.0819 & & & & 0.117 & 29.8 \\
41) Raman spectrum ($D{=}200$) & 4059/507/508 & 0.0378 & \best{0.0326} & & & 0.0497 & 34.4 \\
42) Bader charge, per atom ($e$) & 75{,}028/3000/3000 & 0.0192 & & & & 2.124 & 99.1 \\
43) Net charge, per atom ($e$) & 75{,}033/3000/3000 & 0.0167 & & & & -- & -- \\
44) Magnetic moment, per atom ($\mu_B$) & 89{,}231/3000/3000 & 0.0256 & & & & 2.063 & 98.8 \\
45) Dielectric tensor ($D{=}9$) & 4103/227/229 & 1.690 & & & & 3.401 & 50.3 \\
46) Born effective charge ($e$) & 4472/248/249 & 0.234 & & & & -- & -- \\
47) Piezoelectric tensor, C/m$^2$ ($D{=}18$) & 4513/250/252 & 0.077 & & & & 0.089 & 13.9 \\
48) Elastic $C_{ij}$ tensor, GPa ($D{=}36$) & 15{,}936/885/886 & 5.593 & & & & 18.73 & 70.1 \\
\midrule
\multicolumn{8}{l}{\textbf{(c) Interatomic force fields} — \texttt{mlearn} per-element energy/force; large sets energy\,/\,force} \\
\midrule
49) \texttt{mlearn}-Cu, energy (meV/atom) & 262/--/31 & 1.32$^{\ddagger}$ & & & & 146.13 & \\
50) \texttt{mlearn}-Cu, force (eV/\AA) & 262/--/31 & 0.0179$^{\ddagger}$ & & & & 0.4514 & \\
51) \texttt{mlearn}-Ni, energy (meV/atom) & 263/--/31 & 3.04$^{\ddagger}$ & & & & 167.48 & \\
52) \texttt{mlearn}-Ni, force (eV/\AA) & 263/--/31 & 0.0275$^{\ddagger}$ & & & & 0.4984 & \\
53) \texttt{mlearn}-Li, energy (meV/atom) & 241/--/29 & 5.86$^{\ddagger}$ & & & & 49.26 & \\
54) \texttt{mlearn}-Li, force (eV/\AA) & 241/--/29 & 0.0264$^{\ddagger}$ & & & & 0.2062 & \\
55) \texttt{mlearn}-Ge, energy (meV/atom) & 228/--/25 & 12.48$^{\ddagger}$ & & & & 221.34 & \\
56) \texttt{mlearn}-Ge, force (eV/\AA) & 228/--/25 & 0.0690$^{\ddagger}$ & & & & 0.4236 & \\
57) \texttt{mlearn}-Si, energy (meV/atom) & 214/--/25 & 13.88$^{\ddagger}$ & & & & -- & \\
58) \texttt{mlearn}-Si, force (eV/\AA) & 214/--/25 & 0.0872$^{\ddagger}$ & & & & -- & \\
59) \texttt{mlearn}-Mo, energy (meV/atom) & 194/--/23 & 9.86$^{\ddagger}$ & & & & 340.28 & \\
60) \texttt{mlearn}-Mo, force (eV/\AA) & 194/--/23 & 0.1121$^{\ddagger}$ & & & & 0.9496 & \\
61) ALIGNN-FF-DB (E/F) & 276{,}401/15{,}355/15{,}355 & 32.4$^{\dagger}$ / 0.0564$^{\dagger}$ & & & & 1808 / 0.039 & 98.2 / -- \\
62) MATPES-PBE (E/F) & 391{,}241/21{,}735/21{,}736 & 40.4 / 0.1475 & & & & 1508 / 0.467 & 97.3 / -- \\
63) FD-FF, 1.1\,M (E/F) & 1{,}097{,}227/60{,}957/60{,}958 & 28.9$^{\dagger}$ / 0.0445$^{\dagger}$ & & & & 1742 / 0.241 & 98.3 / -- \\
64) MPtrj (E/F) & 1{,}376{,}739/76{,}485/76{,}485 & 56.7$^{\dagger}$ / 0.0707$^{\dagger}$ & & & & 1478 / 0.151 & 96.2 / -- \\
\bottomrule
\end{longtable}
\end{scriptsize}

\subsection*{Data and model availability}
All trained ALIGNN~2.0 models are archived on figshare (project 279395) and,
where a matching benchmark exists, mirrored as JARVIS-Leaderboard contributions
(\texttt{atomgptlab/}\allowbreak\texttt{jarvis\_leaderboard}, teams \texttt{alignn2\_radius} and \texttt{alignn2\_knn}). Each ``figshare'' cell links to the archive DOI; each ``Leaderboard'' cell links to the benchmark page (live once the contributions are merged and the site is rebuilt). ``--'' marks a graph variant or benchmark not applicable to that model.
Tables~\ref{tab:avail_single}--\ref{tab:avail_ff} enumerate the archives for the
single-output, multi-output, and force-field models respectively.

\begin{scriptsize}
\begin{longtable}{>{\raggedright\arraybackslash}p{5.5cm} c c c}
\caption{Single-output property-prediction models: figshare archives (radius and kNN graphs) and JARVIS-Leaderboard benchmark pages.}\label{tab:avail_single}\\
\toprule Model / property & figshare (radius) & figshare (kNN) & Leaderboard \\ \midrule \endfirsthead
\toprule Model / property & figshare (radius) & figshare (kNN) & Leaderboard \\ \midrule \endhead
$\varepsilon_x$ (MBJ) & \href{https://doi.org/10.6084/m9.figshare.33135077}{33135077} & \href{https://doi.org/10.6084/m9.figshare.33135080}{33135080} & \href{https://atomgptlab.github.io/jarvis_leaderboard/AI/SinglePropertyPrediction/dft_3d_mepsx/}{link} \\
$\varepsilon_x$ (OPT) & \href{https://doi.org/10.6084/m9.figshare.33135017}{33135017} & \href{https://doi.org/10.6084/m9.figshare.33135020}{33135020} & \href{https://atomgptlab.github.io/jarvis_leaderboard/AI/SinglePropertyPrediction/dft_3d_epsx/}{link} \\
$\varepsilon_y$ (MBJ) & \href{https://doi.org/10.6084/m9.figshare.33135083}{33135083} & \href{https://doi.org/10.6084/m9.figshare.33135086}{33135086} & \href{https://atomgptlab.github.io/jarvis_leaderboard/AI/SinglePropertyPrediction/dft_3d_mepsy/}{link} \\
$\varepsilon_y$ (OPT) & \href{https://doi.org/10.6084/m9.figshare.33135023}{33135023} & \href{https://doi.org/10.6084/m9.figshare.33135026}{33135026} & \href{https://atomgptlab.github.io/jarvis_leaderboard/AI/SinglePropertyPrediction/dft_3d_epsy/}{link} \\
$\varepsilon_z$ (MBJ) & \href{https://doi.org/10.6084/m9.figshare.33135089}{33135089} & \href{https://doi.org/10.6084/m9.figshare.33135092}{33135092} & \href{https://atomgptlab.github.io/jarvis_leaderboard/AI/SinglePropertyPrediction/dft_3d_mepsz/}{link} \\
$\varepsilon_z$ (OPT) & \href{https://doi.org/10.6084/m9.figshare.33135029}{33135029} & \href{https://doi.org/10.6084/m9.figshare.33135032}{33135032} & \href{https://atomgptlab.github.io/jarvis_leaderboard/AI/SinglePropertyPrediction/dft_3d_epsz/}{link} \\
$T_c$ (JARVIS-SC) & \href{https://doi.org/10.6084/m9.figshare.33135209}{33135209} & \href{https://doi.org/10.6084/m9.figshare.33135212}{33135212} & -- \\
$T_c$ hydride & \href{https://doi.org/10.6084/m9.figshare.33135215}{33135215} & \href{https://doi.org/10.6084/m9.figshare.33135218}{33135218} & \href{https://atomgptlab.github.io/jarvis_leaderboard/AI/SinglePropertyPrediction/dft_3d_Tc_supercon_hydride/}{link} \\
2D-MatPedia band gap & \href{https://doi.org/10.6084/m9.figshare.33135242}{33135242} & \href{https://doi.org/10.6084/m9.figshare.33135245}{33135245} & \href{https://atomgptlab.github.io/jarvis_leaderboard/AI/SinglePropertyPrediction/twod_matpd_bandgap/}{link} \\
alex\_supercon\_debye & \href{https://doi.org/10.6084/m9.figshare.33135185}{33135185} & -- & -- \\
alex\_supercon\_dosef & \href{https://doi.org/10.6084/m9.figshare.33135182}{33135182} & -- & -- \\
alex\_supercon\_la & \href{https://doi.org/10.6084/m9.figshare.33135188}{33135188} & -- & -- \\
alex\_supercon\_Tc & \href{https://doi.org/10.6084/m9.figshare.33135179}{33135179} & -- & -- \\
alex\_supercon\_wlog & \href{https://doi.org/10.6084/m9.figshare.33135191}{33135191} & -- & -- \\
Avg. electron mass & \href{https://doi.org/10.6084/m9.figshare.33134975}{33134975} & \href{https://doi.org/10.6084/m9.figshare.33134978}{33134978} & \href{https://atomgptlab.github.io/jarvis_leaderboard/AI/SinglePropertyPrediction/dft_3d_avg_elec_mass/}{link} \\
Avg. hole mass & \href{https://doi.org/10.6084/m9.figshare.33134981}{33134981} & \href{https://doi.org/10.6084/m9.figshare.33134984}{33134984} & \href{https://atomgptlab.github.io/jarvis_leaderboard/AI/SinglePropertyPrediction/dft_3d_avg_hole_mass/}{link} \\
Bulk modulus $K_V$ & \href{https://doi.org/10.6084/m9.figshare.33134987}{33134987} & \href{https://doi.org/10.6084/m9.figshare.33134990}{33134990} & \href{https://atomgptlab.github.io/jarvis_leaderboard/AI/SinglePropertyPrediction/dft_3d_bulk_modulus_kv/}{link} \\
C2DB band gap & \href{https://doi.org/10.6084/m9.figshare.33135236}{33135236} & \href{https://doi.org/10.6084/m9.figshare.33135239}{33135239} & \href{https://atomgptlab.github.io/jarvis_leaderboard/AI/SinglePropertyPrediction/c2db_gap/}{link} \\
Energy above hull & \href{https://doi.org/10.6084/m9.figshare.33135005}{33135005} & \href{https://doi.org/10.6084/m9.figshare.33135008}{33135008} & \href{https://atomgptlab.github.io/jarvis_leaderboard/AI/SinglePropertyPrediction/dft_3d_ehull/}{link} \\
Exfoliation energy & \href{https://doi.org/10.6084/m9.figshare.33135041}{33135041} & \href{https://doi.org/10.6084/m9.figshare.33135044}{33135044} & \href{https://atomgptlab.github.io/jarvis_leaderboard/AI/SinglePropertyPrediction/dft_3d_exfoliation_energy/}{link} \\
Formation energy (JARVIS-DFT) & \href{https://doi.org/10.6084/m9.figshare.33135047}{33135047} & \href{https://doi.org/10.6084/m9.figshare.33135050}{33135050} & \href{https://atomgptlab.github.io/jarvis_leaderboard/AI/SinglePropertyPrediction/dft_3d_formation_energy_peratom/}{link} \\
Halide-perov. HSE decomp. & \href{https://doi.org/10.6084/m9.figshare.33135434}{33135434} & \href{https://doi.org/10.6084/m9.figshare.33135437}{33135437} & \href{https://atomgptlab.github.io/jarvis_leaderboard/AI/SinglePropertyPrediction/halide_peroskites_HSE_decomp_energy/}{link} \\
Halide-perov. HSE gap & \href{https://doi.org/10.6084/m9.figshare.33135401}{33135401} & \href{https://doi.org/10.6084/m9.figshare.33135404}{33135404} & \href{https://atomgptlab.github.io/jarvis_leaderboard/AI/SinglePropertyPrediction/halide_peroskites_HSE_gap/}{link} \\
Halide-perov. PBE decomp. & \href{https://doi.org/10.6084/m9.figshare.33135425}{33135425} & \href{https://doi.org/10.6084/m9.figshare.33135428}{33135428} & \href{https://atomgptlab.github.io/jarvis_leaderboard/AI/SinglePropertyPrediction/halide_peroskites_PBE_decomp_energy/}{link} \\
Halide-perov. PBE gap & \href{https://doi.org/10.6084/m9.figshare.33135395}{33135395} & \href{https://doi.org/10.6084/m9.figshare.33135398}{33135398} & \href{https://atomgptlab.github.io/jarvis_leaderboard/AI/SinglePropertyPrediction/halide_peroskites_PBE_gap/}{link} \\
Halide-perov. refr. index & \href{https://doi.org/10.6084/m9.figshare.33135416}{33135416} & \href{https://doi.org/10.6084/m9.figshare.33135419}{33135419} & \href{https://atomgptlab.github.io/jarvis_leaderboard/AI/SinglePropertyPrediction/halide_peroskites_Ref_ind/}{link} \\
hMOF CO$_2$ uptake & -- & \href{https://doi.org/10.6084/m9.figshare.33137036}{33137036} & \href{https://atomgptlab.github.io/jarvis_leaderboard/AI/SinglePropertyPrediction/hmof_co2/}{link} \\
k-point length & \href{https://doi.org/10.6084/m9.figshare.33135053}{33135053} & \href{https://doi.org/10.6084/m9.figshare.33135056}{33135056} & \href{https://atomgptlab.github.io/jarvis_leaderboard/AI/SinglePropertyPrediction/dft_3d_kpoint_length_unit/}{link} \\
Lattice therm. cond. & \href{https://doi.org/10.6084/m9.figshare.33135194}{33135194} & \href{https://doi.org/10.6084/m9.figshare.33135197}{33135197} & -- \\
Magnetic moment & \href{https://doi.org/10.6084/m9.figshare.33135059}{33135059} & \href{https://doi.org/10.6084/m9.figshare.33135062}{33135062} & \href{https://atomgptlab.github.io/jarvis_leaderboard/AI/SinglePropertyPrediction/dft_3d_magmom_oszicar/}{link} \\
Max EFG & \href{https://doi.org/10.6084/m9.figshare.33135065}{33135065} & \href{https://doi.org/10.6084/m9.figshare.33135068}{33135068} & \href{https://atomgptlab.github.io/jarvis_leaderboard/AI/SinglePropertyPrediction/dft_3d_max_efg/}{link} \\
MBJ band gap & \href{https://doi.org/10.6084/m9.figshare.33135071}{33135071} & \href{https://doi.org/10.6084/m9.figshare.33135074}{33135074} & \href{https://atomgptlab.github.io/jarvis_leaderboard/AI/SinglePropertyPrediction/dft_3d_mbj_bandgap/}{link} \\
MXene275 formation energy & \href{https://doi.org/10.6084/m9.figshare.33135221}{33135221} & \href{https://doi.org/10.6084/m9.figshare.33135224}{33135224} & \href{https://atomgptlab.github.io/jarvis_leaderboard/AI/SinglePropertyPrediction/mxene275_formation_energy/}{link} \\
n-power factor & \href{https://doi.org/10.6084/m9.figshare.33135101}{33135101} & \href{https://doi.org/10.6084/m9.figshare.33135104}{33135104} & \href{https://atomgptlab.github.io/jarvis_leaderboard/AI/SinglePropertyPrediction/dft_3d_n_powerfact/}{link} \\
n-Seebeck & \href{https://doi.org/10.6084/m9.figshare.33135095}{33135095} & \href{https://doi.org/10.6084/m9.figshare.33135098}{33135098} & \href{https://atomgptlab.github.io/jarvis_leaderboard/AI/SinglePropertyPrediction/dft_3d_n_Seebeck/}{link} \\
OMDB band gap & \href{https://doi.org/10.6084/m9.figshare.33135329}{33135329} & \href{https://doi.org/10.6084/m9.figshare.33135332}{33135332} & \href{https://atomgptlab.github.io/jarvis_leaderboard/AI/SinglePropertyPrediction/omdb_bandgap/}{link} \\
OptB88vdW band gap & \href{https://doi.org/10.6084/m9.figshare.33135107}{33135107} & \href{https://doi.org/10.6084/m9.figshare.33135110}{33135110} & \href{https://atomgptlab.github.io/jarvis_leaderboard/AI/SinglePropertyPrediction/dft_3d_optb88vdw_bandgap/}{link} \\
OptB88vdW total energy & \href{https://doi.org/10.6084/m9.figshare.33135113}{33135113} & \href{https://doi.org/10.6084/m9.figshare.33135116}{33135116} & \href{https://atomgptlab.github.io/jarvis_leaderboard/AI/SinglePropertyPrediction/dft_3d_optb88vdw_total_energy/}{link} \\
PDBbind binding affinity & \href{https://doi.org/10.6084/m9.figshare.33135335}{33135335} & \href{https://doi.org/10.6084/m9.figshare.33135338}{33135338} & \href{https://atomgptlab.github.io/jarvis_leaderboard/AI/SinglePropertyPrediction/pdbbind_binding_affinity/}{link} \\
Piezo. $d_{ij}$ (max) & \href{https://doi.org/10.6084/m9.figshare.33134999}{33134999} & \href{https://doi.org/10.6084/m9.figshare.33135002}{33135002} & \href{https://atomgptlab.github.io/jarvis_leaderboard/AI/SinglePropertyPrediction/dft_3d_dfpt_piezo_max_dij/}{link} \\
Piezo. dielectric (max) & \href{https://doi.org/10.6084/m9.figshare.33134993}{33134993} & \href{https://doi.org/10.6084/m9.figshare.33134996}{33134996} & \href{https://atomgptlab.github.io/jarvis_leaderboard/AI/SinglePropertyPrediction/dft_3d_dfpt_piezo_max_dielectric/}{link} \\
Plane-wave cutoff & \href{https://doi.org/10.6084/m9.figshare.33135011}{33135011} & \href{https://doi.org/10.6084/m9.figshare.33135014}{33135014} & \href{https://atomgptlab.github.io/jarvis_leaderboard/AI/SinglePropertyPrediction/dft_3d_encut/}{link} \\
Polymer-Genome GGA gap & \href{https://doi.org/10.6084/m9.figshare.33135230}{33135230} & \href{https://doi.org/10.6084/m9.figshare.33135233}{33135233} & \href{https://atomgptlab.github.io/jarvis_leaderboard/AI/SinglePropertyPrediction/polymer_genome_gga_gap/}{link} \\
QM9 HOMO--LUMO gap & -- & \href{https://doi.org/10.6084/m9.figshare.33135167}{33135167} & -- \\
QMOF band gap & -- & \href{https://doi.org/10.6084/m9.figshare.33135170}{33135170} & -- \\
Shear modulus $G_V$ & \href{https://doi.org/10.6084/m9.figshare.33135119}{33135119} & \href{https://doi.org/10.6084/m9.figshare.33135122}{33135122} & \href{https://atomgptlab.github.io/jarvis_leaderboard/AI/SinglePropertyPrediction/dft_3d_shear_modulus_gv/}{link} \\
SLME & \href{https://doi.org/10.6084/m9.figshare.33135125}{33135125} & \href{https://doi.org/10.6084/m9.figshare.33135128}{33135128} & \href{https://atomgptlab.github.io/jarvis_leaderboard/AI/SinglePropertyPrediction/dft_3d_slme/}{link} \\
SNUMAT HSE gap & \href{https://doi.org/10.6084/m9.figshare.33135974}{33135974} & \href{https://doi.org/10.6084/m9.figshare.33135977}{33135977} & \href{https://atomgptlab.github.io/jarvis_leaderboard/AI/SinglePropertyPrediction/snumat_Band_gap_HSE/}{link} \\
Spin--orbit spillage & \href{https://doi.org/10.6084/m9.figshare.33135131}{33135131} & \href{https://doi.org/10.6084/m9.figshare.33135134}{33135134} & \href{https://atomgptlab.github.io/jarvis_leaderboard/AI/SinglePropertyPrediction/dft_3d_spillage/}{link} \\
\bottomrule\end{longtable}\end{scriptsize}

\begin{scriptsize}
\begin{longtable}{>{\raggedright\arraybackslash}p{5.0cm} c c c c}
\caption{Multi-output (tensor/spectra/atomwise) models: figshare archives and JARVIS-Leaderboard pages. $D$ is the output dimension.}\label{tab:avail_multi}\\
\toprule Model / property & $D$ & figshare (radius) & figshare (kNN) & Leaderboard \\ \midrule \endfirsthead
\toprule Model / property & $D$ & figshare (radius) & figshare (kNN) & Leaderboard \\ \midrule \endhead
Bader charge (per atom) & 1 & \href{https://doi.org/10.6084/m9.figshare.33135137}{33135137} & -- & -- \\
Born effective charge & 1 & \href{https://doi.org/10.6084/m9.figshare.33135155}{33135155} & -- & -- \\
Dielectric function (mBJ spectrum) & 300 & \href{https://doi.org/10.6084/m9.figshare.33135254}{33135254} & \href{https://doi.org/10.6084/m9.figshare.33135266}{33135266} & \href{https://atomgptlab.github.io/jarvis_leaderboard/AI/Spectra/mbjdiel_dielectric/}{link} \\
Dielectric tensor & 9 & \href{https://doi.org/10.6084/m9.figshare.33135143}{33135143} & -- & -- \\
Elastic $C_{ij}$ tensor & 36 & \href{https://doi.org/10.6084/m9.figshare.33135152}{33135152} & -- & -- \\
Electronic DOS & 300 & \href{https://doi.org/10.6084/m9.figshare.33135158}{33135158} & -- & -- \\
IR spectrum & 200 & \href{https://doi.org/10.6084/m9.figshare.33135248}{33135248} & \href{https://doi.org/10.6084/m9.figshare.33135251}{33135251} & \href{https://atomgptlab.github.io/jarvis_leaderboard/AI/Spectra/irdb_ir/}{link} \\
Magn. moment (per atom) & 1 & \href{https://doi.org/10.6084/m9.figshare.33135140}{33135140} & -- & -- \\
Net charge (per atom) & 1 & \href{https://doi.org/10.6084/m9.figshare.33135227}{33135227} & -- & -- \\
Phonon DOS & 200 & \href{https://doi.org/10.6084/m9.figshare.33135164}{33135164} & -- & -- \\
Piezoelectric tensor & 18 & \href{https://doi.org/10.6084/m9.figshare.33135149}{33135149} & -- & -- \\
Raman spectrum & 200 & -- & \href{https://doi.org/10.6084/m9.figshare.33134963}{33134963} & -- \\
\bottomrule\end{longtable}\end{scriptsize}

\begin{scriptsize}
\begin{longtable}{>{\raggedright\arraybackslash}p{5.5cm} c c c}
\caption{Force-field models: figshare archives and JARVIS-Leaderboard MLFF benchmark pages.}\label{tab:avail_ff}\\
\toprule Model / property & figshare (radius) & figshare (kNN) & Leaderboard \\ \midrule \endfirsthead
\toprule Model / property & figshare (radius) & figshare (kNN) & Leaderboard \\ \midrule \endhead
JARVIS-FD & \href{https://doi.org/10.6084/m9.figshare.33135203}{33135203} & -- & -- \\
JARVIS-FD-EV & \href{https://doi.org/10.6084/m9.figshare.33134966}{33134966} & -- & -- \\
JARVIS-FF-DB & \href{https://doi.org/10.6084/m9.figshare.33135200}{33135200} & -- & \href{https://atomgptlab.github.io/jarvis_leaderboard/AI/MLFF/alignn_ff_db_energy/}{link} \\
MATPES-PBE & \href{https://doi.org/10.6084/m9.figshare.33134972}{33134972} & -- & \href{https://atomgptlab.github.io/jarvis_leaderboard/AI/MLFF/matpes_energy/}{link} \\
MATPES-PBE (smooth, default) & \href{https://doi.org/10.6084/m9.figshare.33148208}{33148208} & -- & \href{https://atomgptlab.github.io/jarvis_leaderboard/AI/MLFF/matpes_energy/}{link} \\
MATPES-R2SCAN & \href{https://doi.org/10.6084/m9.figshare.33190782}{33190782} & -- & -- \\
mlearn-Si & \href{https://doi.org/10.6084/m9.figshare.33135206}{33135206} & -- & \href{https://atomgptlab.github.io/jarvis_leaderboard/AI/MLFF/mlearn_Si_energy/}{link} \\
MPtrj & \href{https://doi.org/10.6084/m9.figshare.33134969}{33134969} & -- & \href{https://atomgptlab.github.io/jarvis_leaderboard/AI/MLFF/mptrj_energy/}{link} \\
\bottomrule\end{longtable}\end{scriptsize}

\subsection*{Supplementary methods}
ALIGNN 2.0 retains the alternating atom-graph and line-graph update structure of
ALIGNN~\cite{choudhary2021alignn} but implements every graph operation in native
PyTorch tensors~\cite{paszke2019pytorch} under a neighbor strategy we refer to as
pure-torch, with no external graph runtime. Atoms are represented by
ninety-two-dimensional CGCNN embedding vectors~\cite{xie2018cgcnn}; edges carry
radial-basis-function distance features; and triplets, which are the nodes of the
line graph, carry bond-angle features expanded in a radial basis. The model
alternates edge-gated graph convolutions on the atom graph with convolutions on
the line graph, so that angular information computed on the line graph modulates
the messages passed on the atom graph. The line graph is constructed on the fly
from the atom graph's edges, and, because training proceeds over batches of
structures of different sizes, both graphs are batched by concatenation with
consistent index offsetting, so that atoms, edges, and triplets belonging to
different structures never exchange messages; this batched line-graph
construction, expressed entirely in vectorized tensor operations, is what allows
the dependency-free implementation to train efficiently at the scale of more than
a million configurations. Two model sizes are used throughout: a compact
configuration with two atom-graph and two line-graph layers and a hidden width of
sixty-four, used for controlled comparisons, and a production configuration with
four atom-graph and four line-graph layers and a hidden width of two hundred and
fifty-six.

Two graph constructions are compared under otherwise identical settings. The
force-field-compatible radius graph uses a five-angstrom cutoff, retains at most
twelve nearest neighbors, and builds the line graph from a three-and-a-half
angstrom three-body cutoff; its neighbor set varies continuously with atomic
displacement, which is the property required for conservative dynamics. The
alternative kNN graph uses an eight-angstrom cutoff with twelve nearest neighbors
and the full angular line graph, providing a wider angular receptive field at the
cost of a neighbor list that changes discontinuously as atoms move. All other
architectural and training settings are held fixed between the two constructions
so that any difference in accuracy is attributable to the graph alone. For
force-field training the total energy is predicted as a per-atom energy multiplied
by the number of atoms; atomic forces are obtained analytically as the negative
gradient of the predicted energy with respect to atomic positions through
automatic differentiation; and, where trained, the stress is obtained from the
corresponding virial. The training loss combines a graph-level term for energy
with a gradient term for forces and, when enabled, a stress term, with fixed
relative weights. The same per-atom head that produces forces is reused, with the
gradient disabled, to fit non-gradient per-atom targets such as atomic charges and
magnetic moments, and a graph-level vector head of the appropriate output
dimension is used for spectral and tensorial targets.

The property datasets are drawn from JARVIS-DFT~\cite{choudhary2020jarvis}, in
which the underlying quantities are computed with the projector-augmented-wave
method~\cite{kresse1996vasp} using, for the \texttt{dft\_3d} properties, the
OptB88vdW van der Waals density functional~\cite{klimes2011optb88}. Formation and
total energies, band gaps, and derived electronic descriptors follow from
self-consistent calculations; elastic stiffness tensors are obtained from
finite strain-stress relations; dielectric, Born-effective-charge, and
piezoelectric tensors are obtained from density-functional perturbation
theory~\cite{baroni2001dfpt}; thermoelectric Seebeck coefficients and power
factors follow from semiclassical Boltzmann transport; and the lattice thermal
conductivity, reported on a base-ten logarithmic scale, quantifies phonon
transport. Electronic densities of states are taken directly from the
self-consistent calculations, while phonon densities of states derive from
force constants evaluated by the finite-displacement method and post-processed
with phonopy~\cite{togo2015phonopy}. Atomic charges are obtained by Bader
decomposition of the charge density~\cite{henkelman2006bader}, and site magnetic
moments from spin-polarized calculations. Superconducting critical temperatures,
including the separately distributed high-pressure hydride sets, follow the
electron-phonon-coupling screening workflow of Ref.~\cite{choudhary2022supercon}.
For force fields, the element-specific \texttt{mlearn} sets provide DFT molecular
dynamics snapshots for six elements assembled to benchmark machine-learning
potentials~\cite{zuo2020mlearn}; the ALIGNN-FF database aggregates JARVIS-DFT
relaxation trajectories~\cite{choudhary2023alignnff}; and the MATPES set provides
of order four hundred thousand structures carefully sampled from a very large pool
of molecular-dynamics snapshots to span equilibrium and near-equilibrium
configurations~\cite{kaplan2025matpes}. To broaden the configuration space sampled
by a finite-displacement potential, we additionally harvest OptB88vdW
configurations from energy-volume curves and from vacancy and surface relaxation
trajectories generated in JARVIS-DFT workflows, parsing every ionic step of each
converged calculation into an energy-per-atom-and-forces record consistent with
the finite-displacement data's reference, and merge these with the
finite-displacement set while holding the original test partition fixed so that
any change in accuracy is attributable to the added training data alone. All
property and force-field evaluations follow the JARVIS-Leaderboard protocol and
its fixed train, validation, and test splits and report the mean absolute error on
the test partition~\cite{choudhary2024leaderboard}.

Alongside the model comparisons we report, for each single-property task where it
is well defined, the mean absolute error of a baseline that predicts the mean of
the training targets, evaluated on the test partition, and a skill score defined
as one minus the ratio of the model error to this baseline error; the skill score
is positive when the model outperforms the naive predictor and equals the fraction
of the baseline error removed. For a small number of targets whose per-atom or
reference conventions are still being reconciled, the corresponding baseline or
skill entry is omitted rather than reported provisionally.

Models are trained with a one-cycle learning-rate
schedule~\cite{smith2019superconvergence} and the AdamW
optimizer~\cite{loshchilov2019adamw}. Because the schedule anneals the learning
rate to near zero only at the final epoch, all results are reported at full
convergence, with property models trained for one hundred and fifty to three
hundred epochs and force fields for up to several tens to a few hundred epochs
depending on dataset size. Warm starting from a converged checkpoint is used when
continuing a potential's training on an augmented dataset. Training was performed
on current-generation Blackwell-class GPUs and on Grace-Blackwell hardware; the
dependency-free implementation runs on these accelerators without modification.
Molecular-dynamics stability is assessed by micro-canonical integration with a
one-femtosecond time step, using structures handled through the atomic simulation
environment~\cite{larsen2017ase}, initializing supercells at finite temperature
and reporting the linear drift of the total energy per atom in
milli-electronvolts per atom per picosecond together with its maximum absolute
deviation. The pure-PyTorch ALIGNN implementation, the trained models, and the
associated LAMMPS-compatible~\cite{thompson2022lammps} ALIGNN-FF potential are
openly available at \url{https://github.com/atomgptlab/alignn}, with an
interactive web application at \url{https://atomgpt.org/alignn}, whose
agentic AtomGPT.org platform exposes these models alongside the wider
materials-design toolchain~\cite{lee2026agapi}; evaluations
follow the JARVIS-Leaderboard protocol~\cite{choudhary2024leaderboard} and the
contributions will be provided under the corresponding leaderboard entries.

\subsection*{Reproducibility, hyperparameter search, and a\texorpdfstring{\\}{ }zero-shot universal-potential baseline}

To test end-to-end reproducibility, the formation-energy and
exfoliation-energy benchmarks were retrained from a self-contained
bundle (data download, split construction verified against the
public benchmark files, four training configurations, and an automatic
reported-versus-retrained comparison) on two unrelated clusters with
different architectures (aarch64 Grace--Blackwell and x86 nodes). The kNN
formation-energy model reproduced the reported test error to within one
percent on both clusters (0.0305 and 0.0308 versus 0.0307~eV/atom); the
radius models reproduced 40.6 versus 40.3~meV/atom on exfoliation energy and
0.0330 versus 0.0316~eV/atom on formation energy, agreement within a few
percent. The campaign also independently detected, and confirmed the fix
for, an earlier pipeline defect in which the test split was featurized
without the three-body cutoff; radius-graph models required the corrected
test featurization to replicate, whereas kNN models are structurally immune
because their three-body cutoff equals the neighbor cutoff.

\begin{figure}[t]
\centering
\includegraphics[width=\textwidth]{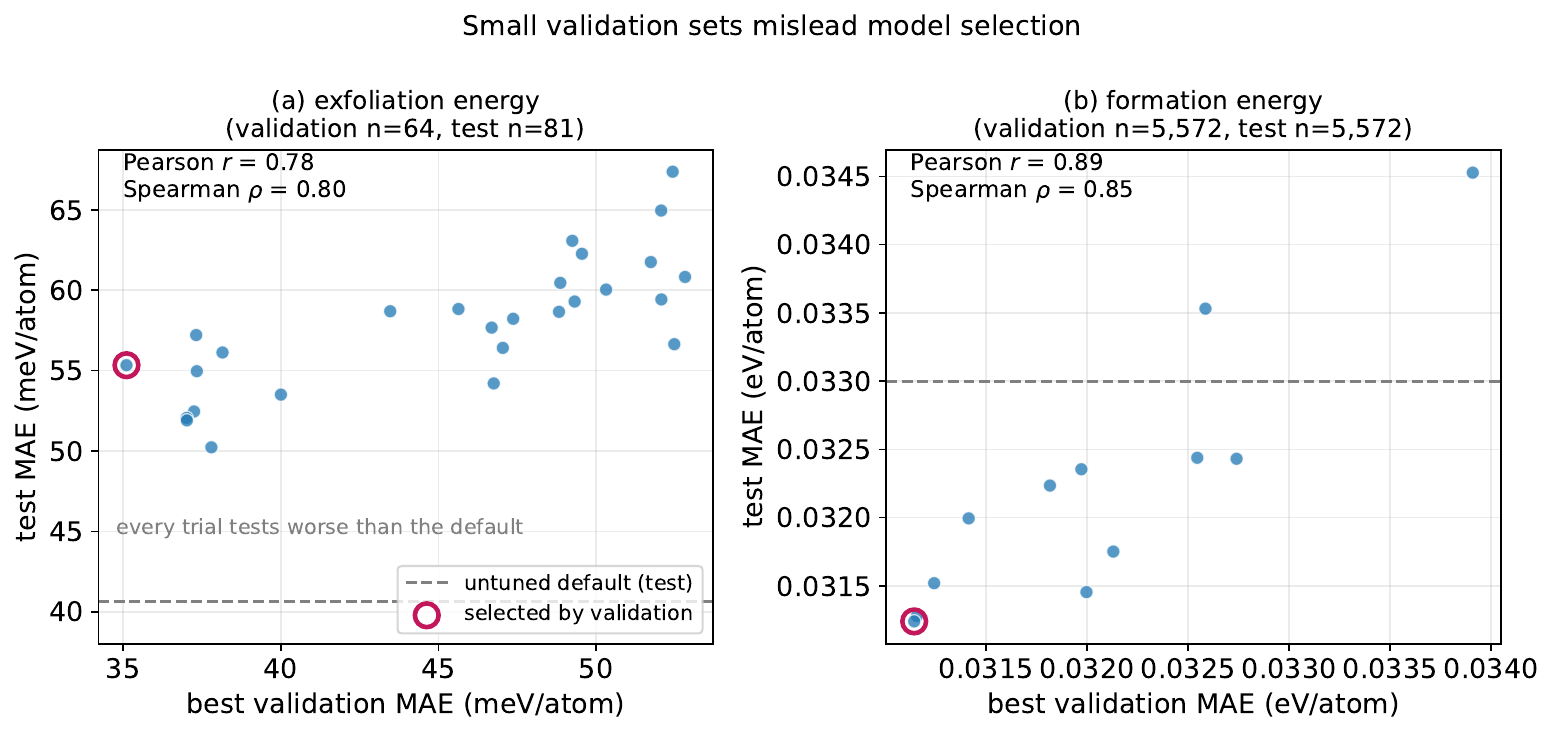}
\caption{Hyperparameter selection is only as reliable as the validation
set is large. Each point is one random-search trial, plotting the best
validation error used for model selection against the held-out test error;
the dashed line marks the untuned default configuration's test error and
the circled point marks the trial selected by validation. (a) On the
81-structure exfoliation benchmark the 64-structure validation subset is
systematically optimistic and every trial that beat the default on
validation lost to it on test, so the defaults are reported. (b) On the
5{,}572-structure formation-energy benchmark validation and test agree,
and the validation-selected winner is also the best on test, below the
default; its three-seed ensembles give the tuned results quoted in the
text.}
\label{fig:tuning}
\end{figure}

Hyperparameter sensitivity was probed with a validation-ranked random search
(learning rate, batch size, weight decay, width, depth, and convolution
variant), with the test split never entering model selection; final
evaluation used the best-validation checkpoint, optionally with an
exponential moving average of the weights. Figure~\ref{fig:tuning} plots,
for every trial of both searches, the validation error used for selection
against the held-out test error, with the untuned default marked. On the 5{,}572-structure
formation-energy validation set the search is well conditioned, and the
winning configuration (learning rate $3.5\times10^{-3}$, batch 128, hidden
width 384, five ALIGNN and four GCN layers, \texttt{n\_alignn}
convolution, weight decay $10^{-4}$, with weight-space exponential moving
averaging and best-validation-checkpoint evaluation), retrained at full
length with three training-seed replicas that share the identical
leaderboard split, gives a median single-model test error of
0.0312~eV/atom and a three-seed prediction ensemble of 0.0294~eV/atom on
the radius graph; the same configuration transferred to the kNN graph
gives single-model errors of 0.0281--0.0287~eV/atom (median 0.0283) and a
three-seed ensemble of 0.0266~eV/atom, a thirteen percent improvement over
the untuned single model. On the 81-structure exfoliation benchmark the same
procedure fails instructively: although validation and test errors
correlate across trials, the 64-structure validation subset is
systematically optimistic (every trial's held-out error exceeds its
validation error, by twenty meV/atom for the validation-selected trial),
and every configuration that outperformed the untuned default on
validation, including the selected one, went on to underperform it on
test, as did every other trial; the untuned defaults are therefore
reported there, and we regard this as a caution against tuning on
small-sample benchmarks generally.

The universal-potential baseline evaluates UMA (uma-s,
\texttt{omat} task head)~\cite{wood2025uma} zero-shot on the identical
held-out splits. Formation energies per atom are obtained from UMA total
energies with per-element reference energies fitted by least squares on the
training split only; exfoliation energies are computed physically, as the
difference between monolayer and bulk single-point energies per atom, with
each bulk structure paired to its DFT-relaxed monolayer counterpart in
JARVIS-DFT-2D through the shared source identifier. UMA reaches
0.112~eV/atom on formation energy and 82~meV/atom on exfoliation energy
against the OptB88vdW targets, versus 0.0305 and 38.6 for the retrained
task-trained models on the same splits; the exfoliation gap is expected, since the PBE-level data on which
universal potentials are trained lacks the interlayer van der Waals binding
that OptB88vdW captures and that exfoliation energy directly measures.

\end{document}